\documentclass[11pt]{article}
\usepackage[margin=1in]{geometry}

\usepackage{bm}
\usepackage{multirow}
\usepackage{amssymb,amsmath,amsthm,amsfonts}
\usepackage{hyperref}
\usepackage[capitalise,nameinlink]{cleveref}
\usepackage{aliascnt}
\usepackage{letltxmacro}
\usepackage{cancel}
\usepackage{bm}
\usepackage{mathtools}
\usepackage{multicol}
\usepackage{enumitem}
\usepackage{authblk}
\usepackage{graphicx}
\usepackage{subcaption}
\usepackage{tablefootnote} 
\usepackage{float}
\usepackage{physics}
\usepackage{footnote}
\usepackage{xcolor}
\usepackage{mathrsfs}
\usepackage{bbm}
\usepackage{makecell}
\usepackage{pbox}
\usepackage{multirow}%
\usepackage[title]{appendix}%
\usepackage{textcomp}%
\usepackage{manyfoot}%
\usepackage{booktabs}%
\usepackage{algpseudocode}%
\usepackage{listings}%

\usepackage{xcolor}
\usepackage{amsmath}
\usepackage[most]{tcolorbox}
\usepackage{lmodern}
\usepackage{dsfont}
\usepackage{tikz}
\usepackage{standalone}

\allowdisplaybreaks[4]

\newcommand{\im}{\mathrm{i}}

\newcommand{\lrb}[1]{\left(#1\right)}
\newcommand{\lrc}[1]{\left\{#1 \right \}}

\numberwithin{equation}{section}

\newtheorem{theorem_}{Theorem}[section]

\newtheorem{definition_}{Definition}[section]
\newtheorem{remark_}{Remark}[section]

\crefname{assumption_}{Assumption}{Assumptions}
\Crefname{assumption_}{Assumption}{Assumptions}
\crefname{lemma_}{Lemma}{Lemmas}
\Crefname{lemma_}{Lemma}{Lemmas}
\crefname{corollary_}{Corollary}{Corollaries}
\Crefname{corollary_}{Corollary}{Corollaries}
\crefname{proposition_}{Proposition}{Propositions}
\Crefname{proposition_}{Proposition}{Propositions}
\crefname{definition_}{Definition}{Definitions}
\Crefname{definition_}{Definition}{Definitions}
\crefname{remark_}{Remark}{Remarks}
\Crefname{remark_}{Remark}{Remarks}
\crefname{question_}{Question}{Questions}
\Crefname{question_}{Question}{Questions}
\crefname{fact}{Fact}{Facts}
\Crefname{fact}{Fact}{Facts}

\makeatletter
\g@addto@macro\appendix{
  \crefalias{section}{appendix}
  \crefalias{subsection}{appendix}
  \crefalias{subsubsection}{appendix}
}
\makeatother

\AtBeginDocument{

}
\renewcommand{\eqref}[1]{\cref{#1}}
\AtBeginDocument{\let\ref\cref}

\graphicspath{{Fig/}}

\renewenvironment{quote}{%
  \list{}{%
    \rightmargin=0.5cm 
    \leftmargin=0.5cm  
  }%
  \item\relax
}{%
  \endlist
}
\usetikzlibrary{arrows.meta}
\usepackage{needspace}
\usepackage{placeins}

\crefname{theorem_}{Theorem}{Theorems}
\Crefname{theorem_}{Theorem}{Theorems}
\newtheorem*{informaltheorem}{Theorem}

\begin{document}

\title{Metastability in Quantum Dynamics for Optimization}
\author{
  Yuchen Lu\textsuperscript{*}\quad
  Zherui Chen\textsuperscript{*}\quad
  Tongyang Li\textsuperscript{\(\dagger\)}
}
\date{}
\maketitle

\begingroup
\renewcommand{\thefootnote}{\fnsymbol{footnote}}
\footnotetext[1]{These authors contributed equally to this work.}
\footnotetext[2]{Corresponding author.}
\endgroup

\begin{abstract}
Metastability separates apparent equilibration from true convergence: a dynamical system can remain near an intermediate state long before reaching its stationary state. In quantum dynamics used for optimization, such a plateau can delay progress toward a target concentrated near the global minimizers of a nonconvex potential. In this paper, we initiate the study of the general mechanisms by which coherent evolution and dissipation shape metastability, using quantum Langevin dynamics (QLD) as a concrete example \cite{chen2023quantum}, where the objective function is encoded in the potential and coherent transfer between wells interacts with tunable dissipation. To quantify this interaction, we solve a two-state model precisely and derive a spectral-separation criterion in terms of tunneling frequency, dissipation strength, and damping imbalance. Extending the analysis to multi-state models, we show that sufficiently weak tunneling between wells can separate slow population transfer from fast relaxation within each well. In addition, 
we establish conditions under which changing the structure of dissipation shortens the time needed to approach a prescribed target state. Numerical experiments on double-well and multi-well potentials demonstrate these improvements while keeping the target and accuracy fixed and accounting for the time spent on control.
\end{abstract}

\vspace{0in}
\setlength{\columnsep}{25pt}
{\small \tableofcontents}

\section{Introduction} \label{sec:intro}

Metastability describes a system that remains in an apparently stable configuration for a long time before undergoing further relaxation.
After a quick initial transient, the density operator \(\rho(t)\) may
remain close to an intermediate state for a long interval while still
being far from the stationary state \(\rho_{\mathrm{ss}}\).
During this plateau, fast modes have largely decayed, while slow modes retain memory of the initial state \(\rho_0\) \cite{macieszczak2016towards,macieszczak2021classical}.
Small changes in \(\rho(t)\) over time can give the impression that
the dynamics has converged. During a metastable plateau, however,
slowly decaying modes can keep the state far from its stationary
state \(\rho_{\mathrm{ss}}\), so substantial further evolution
occurs only on a much longer timescale.
A plateau can therefore mislead a stopping rule or dominate the time
needed to reach a prescribed accuracy, even when the eventual
stationary state is known.

The central question is therefore what makes the slow modes decay so much more slowly than the fast ones. Previous studies have explored this questions in various settings, including dissipative spin models~\cite{olmos2012facilitated,rose2016metastability} and optical systems~\cite{fink2018signatures,minganti2018spectral,casteels2017critical,rodriguez2017probing}.
Spectral theory constructs a low-dimensional description from the
slow eigenmodes of the Lindbladian \(\mathcal L\)
\cite{macieszczak2016towards}, with extensions to approximately
classical dynamics~\cite{macieszczak2021classical} and individual
quantum trajectories~\cite{brown2024unravelling}. Quantum metastability has attracted increasing attention in recent years. People has explored mixing
bottlenecks~\cite{chen2024local,rakovszky2024bottlenecks},
the structure of thermal metastable states and learning from
them~\cite{bergamaschi2025structural,wang2026efficient},
and reduced models of metastable dynamics~\cite{li2026quantum}.
In the mean-field interchange model, further cooling can even
restore faster thermalization after a metastable
regime~\cite{escobar2026metastability}.
These results motivate identifying the physical couplings responsible for slow relaxation and determining how accessible controls can reduce the resulting delay.

In quantum dynamics for optimization, such slow relaxation can directly increase the time needed to obtain a useful solution. Here, the objective function \(V(x)\) is encoded as the potential energy of a Hamiltonian. The quantum state is then evolved with the aim of concentrating its position distribution near the global minimizers, so that a position measurement yields a candidate solution~\cite{leng2023quantum,chen2023quantum,catli2026efficient,abbas2024challenges,herman2023constrained,chakrabarti2026provable}. However, rapid equilibration of quantum dynamics within one basin need not imply rapid redistribution between competing minima~\cite{rakovszky2024bottlenecks}. Therefore, the {quantum} dynamics may exhibit metastable behavior, where the system appears to have equilibrated within a local minimum but has not yet explored the global landscape~\cite{macieszczak2016towards}. This can lead to long plateaus in the convergence to the desired target state, which is concentrated near the global minimizers of \(V(x)\).

To overcome the challenges of metastability, 
classical Langevin methods use fluctuations
to assist exploration~\cite{geman1986diffusions,chiang1987diffusion,
welling2011bayesian,raginsky2017nonconvex,zhang2017hitting,
gao2022global,chak2023generalized};
quantum tunneling supplies a distinct transfer mechanism, with
advantages established for particular landscapes and preparations
\cite{liu2023tunneling,denchev2016computational,muthukrishnan2016tunneling}. Quantum Hamiltonian descent and its
extensions provide further dynamical approaches to nonconvex
optimization~\cite{leng2023quantum,leng2025quantum,leng2025gradient}.
In an open quantum system, coherent transfer and dissipative relaxation act together \cite{plenio2008dephasing}. The presence of tunneling does not by itself exclude a metastable bottleneck, and stronger damping need not make every relevant process faster \cite{syassen2008strong}. These observations consolidate into two questions:
\begin{quote}
\emph{(i) How can tunneling and dissipation be leveraged to characterize metastable bottlenecks?}\\[0.5em]
\emph{(ii) Under what conditions can dissipative control bypass such bottlenecks and reduce the time required to reach a fixed target state at a prescribed accuracy?}
\end{quote}
We address these questions using quantum Langevin dynamics (QLD)~\cite{chen2023quantum} as our central model. {QLD evolves an open quantum system through a Lindblad equation~\cite{lindblad1976generators}. In one-dimensional QLD, the density operator evolves according to the following Lindbladian equation
\begin{align}
    \frac{\mathrm d\rho}{\mathrm dt} =-\frac{\im}{\hbar}[H,\rho]+\mathcal D(\rho),\qquad \mathcal D(\rho) =2L\rho L^\dagger-\left(L^\dagger L \rho + \rho L^\dagger L\right),
\end{align}
where the Hamiltonian term \(H=p^2/(2m)+V(x)\) encodes the objective function \(V(x)\), with mass \(m>0\), while the jump operator of Lindbladian \(L=\mu x+\im\nu p\) couples the position \(x\) and momentum \(p\) to dissipative evolution.} Previous QLD work establishes convergence results for convex and certain quasar-convex potentials and investigates schedules for the temperature \(T(t)\) and effective Planck constant \(\hbar(t)\). Its numerical experiments on more general nonconvex landscapes also reveal slow relaxation and trapping near local minima~\cite{chen2023quantum}.

In this paper, we develop a theoretical explanation for such metastable behavior and use the resulting understanding to guide dissipative control. An analytically solvable two-state reduction gives explicit relations between tunneling, dissipation, and the decay rates, while multi-state models reveal the transfer channels responsible for slow relaxation. These results connect the observed behavior to concrete physical parameters and guide the design of schedules for \(\mu(t),\nu(t)\) that can shorten relaxation to a prescribed target state while keeping \(H\) fixed.

\subsection{Origin of metastability}
Our first contribution connects the  metastable behavior to physical
transfer and damping parameters. For a relaxing Lindbladian
\(\mathcal L\) on a Hilbert space of dimension \(d_H\geq2\),
order the decay rates \(\Gamma_i=-\operatorname{Re}\lambda_i\)
as \(0=\Gamma_1<\Gamma_2\leq\cdots\leq\Gamma_{d_H^2}\), and set
\begin{align}
 \chi_i=\frac{\Gamma_{i+1}-\Gamma_i}{\Gamma_{i+1}+\Gamma_i},
 \qquad
 \chi_{\max}=\max_{2\leq i<d_H^2}\chi_i.
 \label{eq:intro_spectral_separation}
\end{align}
{We use the bounded relative indicator \(\chi_i\) because metastability concerns a separation of relaxation timescales, rather than an absolute duration. 
The aim is to detect a large ratio \(\Gamma_{i+1}/\Gamma_i\gg1\), separating the exponential decay times \(1/\Gamma_{i+1}\) and \(1/\Gamma_i\). The indicator \(\chi_i\in[0,1)\) increases monotonically with this ratio: it vanishes for equal decay rates \(\Gamma_i=\Gamma_{i+1}\) and approaches one as their ratio grows. Maximizing over \(i\) identifies the strongest separation. A uniform rescaling \(\mathcal L\mapsto c\mathcal L\), with \(c>1\), shortens every evolution time by \(1/c\) without changing \(\chi_i\). Any plateau is therefore shorter in absolute time but equally long relative to the relaxation timescales. For a chosen threshold \(0<\chi_c<1\), the spectral window \(\chi_{\max}>\chi_c\) records this relative separation, not a trajectory-level guarantee of a plateau (\cref{def:chi_separation}).}
Whether a particular trajectory
\(\rho(t)\) exhibits a plateau also depends on its slow-mode
weights and finite-time evolution.

To understand how tunneling and dissipation determine this spectral
separation, we first consider the symmetric two-state QLD model.
Here, \(\Omega_{\mathrm{tun}}^{(2)}>0\) is the frequency of coherent
transfer between the wells, \(\gamma_{\mathcal D}^{(2)}>0\)
measures the overall dissipation strength, and dissipative anisotropy
\(\delta_{\mathcal D}\in[-1,1]\) measures the normalized
imbalance between the direct damping rates of the real and imaginary
parts of the energy-basis coherence. Equal damping gives
\(\delta_{\mathcal D}=0\), while
\(|\delta_{\mathcal D}|\) close to one means a strong imbalance.
Precise definitions are given in
\cref{eq:two_state_tunneling_scale,eq:two_state_dissipation_norm,eq:two_state_dissipative_anisotropy}.

\begingroup
\renewenvironment{theorem_}{\begin{informaltheorem}[informal]}{\end{informaltheorem}}
\begin{theorem_}
For the symmetric two-state QLD model above and any prescribed threshold
\(1/3<\chi_c<1\), the following exact spectral-separation criterion holds:
\begin{align}
 \chi_{\max}>\chi_c
 \quad\Longleftrightarrow\quad
 \delta_{\mathcal D}^{2}>
 \chi_c^2+4(2-\delta_{\mathcal D}^{2})
 \left(\frac{\Omega_{\mathrm{tun}}^{(2)}}
            {\gamma_{\mathcal D}^{(2)}}\right)^2.
 \label{eq:intro_two_state_spectral_criterion}
\end{align}
\end{theorem_}
\endgroup

The criterion, proved in \cref{thm:two_state_lower_edge}, requires
both sufficiently unequal damping, \(|\delta_{\mathcal D}|>\chi_c\), and sufficiently weak tunneling relative to dissipation, as measured by \(\Omega_{\mathrm{tun}}^{(2)}/\gamma_{\mathcal D}^{(2)}\). Physically, rapid tunneling mixes the two coherence components, whereas sufficiently slow mixing allows their unequal damping to produce a large separation of decay rates.
For a finite potential bias \(E_\delta\),
\cref{thm:biased_two_state_window} gives necessary and sufficient
spectral-separation criteria at sufficiently weak or strong dissipation
away from threshold equality, showing that bias can permit separation
even at arbitrarily weak positive dissipation.

For models retaining more states, the physical picture is that
relaxation within each well can remain fast while redistribution
between the wells becomes slow. Under the local-relaxation and QLD
coupling assumptions of \cref{thm:qld_sector_origin}, we show that
sufficiently weak nonzero coupling between the two wells, with a
unique stationary state, creates one slow population mode while
the other modes retain finite decay rates. The coupling
must be small across all retained transfer channels: a small
matrix element for one selected pair of states is insufficient if
other states still provide an efficient route between the wells.
Conversely, \cref{thm:qld_strong_dissipation_exclusion} shows that sufficiently strong dissipation excludes the spectral window, if the dissipator $\mathcal{D}$ has a unique stationary state and satisfies \(\chi_{\max}(\mathcal D) < \chi_c\).

\subsection{Dissipative control for optimization}

To assess whether controlling a metastable bottleneck improves
relaxation-based optimization, we compare a baseline trajectory
\(\rho^{\mathrm{base}}(t)\) with a controlled trajectory
\(\rho^{\mathrm{ctrl}}(t)\). Both start from the same initial state
\(\rho_0\) and are evaluated against the same target
\(\rho_{\mathrm{tar}}\) and tolerance \(\varepsilon>0\).
We measure their performance by the convergence time
\begin{align}
 T_\varepsilon[\rho;\rho_{\mathrm{tar}}]
 &=\inf\{T\geq0:D_{\mathrm{tr}}(\rho(t),\rho_{\mathrm{tar}})
             \leq\varepsilon\ \text{for all }t\geq T\},
 \label{eq:intro_control_settling_time}
\end{align}
where \(D_{\mathrm{tr}}(\rho,\sigma)=\|\rho-\sigma\|_1/2\)
is the trace distance. If no such \(T\) exists, the infimum is
\(+\infty\). Denote the two times by
\(T_\varepsilon^{\mathrm{base}}\) and
\(T_\varepsilon^{\mathrm{ctrl}}\). An improvement means
\(T_\varepsilon^{\mathrm{ctrl}}<T_\varepsilon^{\mathrm{base}}\).
The controlled time includes all preparation and switching stages;
admissible controls and their strength limits are fixed before the
comparison.

To optimize a given potential function $V$, i.e., the Hamiltonian part \(H\) is fixed, dissipative control can change either
the structure of the dissipator \(\mathcal D\) or its overall
strength. We consider two routes:
\begin{itemize}
    \item \textbf{Slow-mode suppression.} Change the relative weights of
    the dissipative couplings to suppress the occupied slow component.
    In the symmetric two-state model, the damping contrast is
    quantified by the anisotropy \(\delta_{\mathcal D}\).
    \item \textbf{Dissipation rescaling.} Rescale the dissipative couplings
    together, preserving the normalized dissipator while changing
    its strength \(\gamma_{\mathcal D}\). Whether this change helps
    depends on the model and the target accuracy.
\end{itemize}
The first route is related to slow-mode preparation and temporary
dissipative control \cite{carollo2021mpemba,bao2025temporary}.
In QLD, both routes can be implemented by adjusting the nonnegative
position- and momentum-coupling coefficients \(\mu,\nu\) at fixed
\(H\): changing their ratio alters the dissipative structure,
whereas common rescaling changes its strength. For a continuous
strength-control protocol, the coefficients are rescaled over a
finite interval and then held fixed.

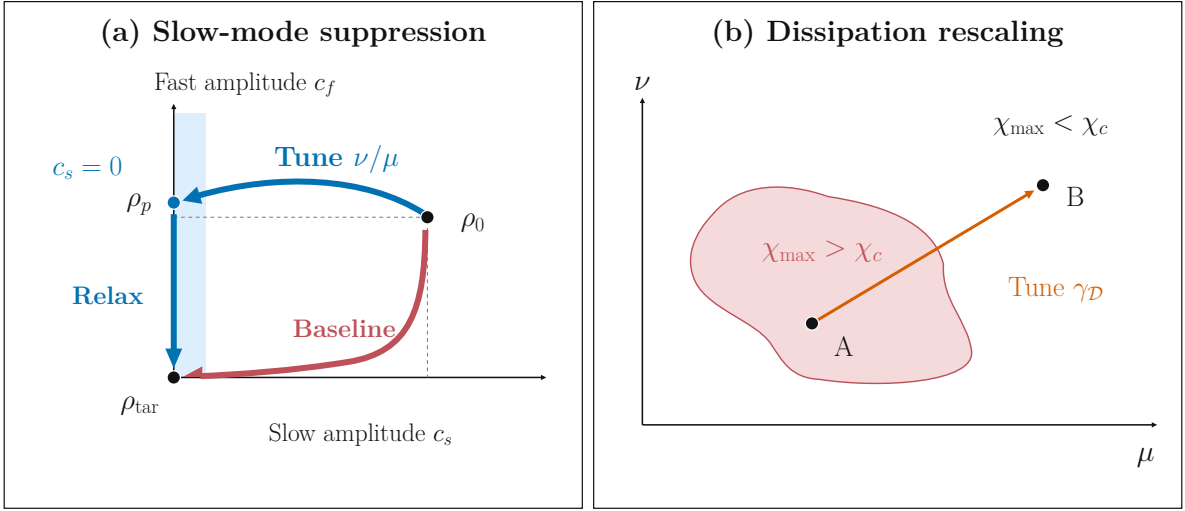
\begin{figure}[!htbp]
 \centering
 \resizebox{0.95\linewidth}{!}{
\begingroup%
\definecolor{controlBlack}{rgb}{0.0666504,0.0666504,0.0666504}%
\definecolor{controlBlue}{rgb}{0,0.447021,0.697266}%
\definecolor{controlBlueLight}{rgb}{0.867188,0.9375,0.988281}%
\definecolor{controlRed}{rgb}{0.744141,0.313721,0.345215}%
\definecolor{controlRedLight}{rgb}{0.957031,0.855469,0.859375}%
\definecolor{controlOrange}{rgb}{0.835938,0.368652,0}%
\begin{tikzpicture}[x=1bp,y=-1bp,
    every node/.style={inner sep=0pt,outer sep=0pt,text=controlBlack,
      font=\normalfont\fontsize{28.488bp}{34.186bp}\selectfont},
    control axis/.style={draw=controlBlack,line width=1.729bp},
    control curve/.style={line width=6.746bp},
    control point/.style={draw=white,line width=1.361bp}]
  \path[use as bounding box] (0,0) rectangle (1332,627);

  \draw[controlBlack,line width=1.361bp]
    (4.493,53.202) rectangle (653.967,622.427)
    (666.723,53.202) rectangle (1327.45,622.427);
  \fill[controlBlueLight] (194.981,178.465) rectangle (231,475.451);

  \draw[control axis] (194.981,475.451)--(607.932,475.451);
  \fill[controlBlack] (613.488,475.451)--(607.535,478.427)
    --(607.535,472.474)--cycle;
  \draw[control axis] (194.981,475.451)--(194.981,173.561);
  \fill[controlBlack] (194.981,168.005)--(197.957,173.958)
    --(192.005,173.958)--cycle;
  \draw[gray,line width=1.191bp,dash pattern=on 4.25bp off 4bp]
    (194.981,295.451)--(479.976,295.451)--(479.976,475.451);

  \draw[controlBlue,control curve] (475.498,290.972)
    .. controls (410.981,249.700) and (311.910,245.930) .. (222.676,271.158);
  \fill[controlBlue] (204.761,276.714)--(221.117,261.095)
    --(227.069,280.427)--cycle;
  \draw[controlBlue,control curve] (194.981,292)--(194.981,445);
  \fill[controlBlue] (194.981,465)--(185.481,443)
    --(204.481,443)--cycle;

  \draw[controlRed,control curve] (477.709,309.681)
    .. controls (477,399) and (460,447) .. (391,458)
    .. controls (329,468) and (268,472) .. (225,474);
  \fill[controlRed] (205,475)--(227,463)
    --(227,475)--cycle;
  \filldraw[control point,fill=controlBlack] (194.981,475.479) circle[radius=7.498bp];
  \filldraw[control point,fill=controlBlack] (479.976,295.479) circle[radius=7.498bp];
  \filldraw[control point,fill=controlBlue] (194.981,278.982) circle[radius=7.498bp];

  \draw[control axis] (721.488,528.714)--(1294.170,528.714);
  \fill[controlBlack] (1299.730,528.714)--(1293.780,531.690)
    --(1293.780,525.737)--cycle;
  \draw[control axis] (721.460,528.742)--(721.460,166.786);
  \fill[controlBlack] (721.460,161.230)--(724.436,167.183)
    --(718.483,167.183)--cycle;

  \filldraw[fill=controlRedLight,draw=controlRed,line width=1.587bp]
    (800.972,291.681)
    .. controls (830.962,255.709) and (883.460,255.709) .. (929.976,272.207)
    .. controls (984.713,280.456) and (1048.460,299.193) .. (1059.720,348.685)
    .. controls (1057.480,389.930) and (1086.730,410.934) .. (1091.210,445.432)
    .. controls (1096.480,481.432) and (989.957,488.207) .. (920.962,477.690)
    .. controls (887.967,476.188) and (878.981,457.451) .. (865.460,437.183)
    .. controls (848.225,412.437) and (797.230,398.944) .. (781.469,368.953)
    .. controls (767.239,340.437) and (780.732,312.685) .. (800.972,291.681)
    --cycle;
  \draw[controlOrange,line width=3.373bp] (912.005,414.733)--(1153.660,270.166);
  \fill[controlOrange] (1161.770,265.319)--(1155.670,274.843)
    --(1150.480,266.169)--cycle;
  \filldraw[control point,fill=controlBlack] (912.005,414.733) circle[radius=7.498bp];
  \filldraw[control point,fill=controlBlack] (1171.520,259.479) circle[radius=7.498bp];

  \node[font=\normalfont\bfseries\fontsize{32.995bp}{39.594bp}\selectfont]
    at (329.23,89.43) {(a) Slow-mode suppression};
  \node[font=\normalfont\bfseries\fontsize{32.995bp}{39.594bp}\selectfont]
    at (997.10,89.43) {(b) Dissipation rescaling};
  \node at (275,146.5) {Fast amplitude $c_f$};
  \node[text=controlBlue,font=\normalfont\fontsize{31.493bp}{37.792bp}\selectfont]
    at (98,240) {$c_s=0$};
  \node[text=controlBlue,font=\normalfont\bfseries\fontsize{31.493bp}{37.792bp}\selectfont]
    at (378,230) {Tune $\nu/\mu$};
  \node[text=controlBlue,font=\normalfont\bfseries\fontsize{28.488bp}{34.186bp}\selectfont]
    at (121,383) {Relax};
  \node[text=controlRed,font=\normalfont\bfseries\fontsize{28.488bp}{34.186bp}\selectfont]
    at (386.60,420) {Baseline};
  \node at (405,540) {Slow amplitude $c_s$};
  \node[font=\normalfont\fontsize{32.287bp}{38.744bp}\selectfont]
    at (156,281) {$\rho_p$};
  \node[font=\normalfont\fontsize{32.287bp}{38.744bp}\selectfont]
    at (531,302) {$\rho_0$};
  \node[font=\normalfont\fontsize{32.287bp}{38.744bp}\selectfont]
    at (157,507) {$\rho_{\mathrm{tar}}$};
  \node[font=\normalfont\fontsize{33.789bp}{40.547bp}\selectfont]
    at (721.46,145) {$\nu$};
  \node[font=\normalfont\fontsize{33.789bp}{40.547bp}\selectfont]
    at (1287,565) {$\mu$};
  \node[text=controlRed,font=\normalfont\fontsize{32.995bp}{39.594bp}\selectfont]
    at (923,334) {$\chi_{\max}>\chi_c$};
  \node[font=\normalfont\fontsize{31.493bp}{37.792bp}\selectfont]
    at (1180,192) {$\chi_{\max}<\chi_c$};
  \node[text=controlOrange,font=\normalfont\fontsize{31.493bp}{37.792bp}\selectfont]
    at (1186,376) {Tune $\gamma_{\mathcal D}$};
  \node[font=\normalfont\fontsize{31.493bp}{37.792bp}\selectfont]
    at (948,441) {A};
  \node[font=\normalfont\fontsize{31.493bp}{37.792bp}\selectfont]
    at (1208,272) {B};
\end{tikzpicture}%
\endgroup}
 \caption[Two routes of dissipative control for optimization]{Two routes of dissipative control.
(a) The horizontal and vertical axes show \(c_s\) and \(c_f\), respectively.
The blue band indicates a small residual slow contribution;
its left edge \(c_s=0\) represents exact cancellation.
The illustrated change in \(\nu/\mu\) prepares \(\rho_p\) on this edge.
After restoring the baseline generator, the remaining contribution
\(c_f\rho_{\mathrm f}\) decays as \(\rho\) approaches the stationary
target \(\rho_{\mathrm{tar}}\) along the vertical axis.
(b) Common rescaling \((\mu,\nu)\mapsto c(\mu,\nu)\) changes
\(\gamma_{\mathcal D}\) at fixed ratio \(\nu/\mu\).
The illustrated exit from \(\chi_{\max}>\chi_c\) is model-dependent;
it alone does not guarantee a shorter \(T_\varepsilon\).}
 \label{fig:intro_dissipative_control_routes}
\end{figure}

Consider the following decomposition of the state \(\rho\):
\begin{align}
 \rho&=\rho_{\mathrm{tar}}+c_s\xi_{\mathrm s}
                              +c_f\xi_{\mathrm f},
 \label{eq:intro_slow_fast_decomposition}
\end{align}
where \(c_s,c_f\geq0\) are the trace-norm sizes of the baseline
slow and fast contributions, respectively. The corresponding operator \(\xi_{\mathrm s}\) or
\(\xi_{\mathrm f}\) is traceless and normalized to unit trace norm. These operators may depend on
\(\rho\) and contain multiple modes.
The first route aims to reduce \(c_s\), with exact cancellation
of the entire slow contribution corresponding to \(c_s=0\).
\Cref{fig:intro_dissipative_control_routes} illustrates
slow-mode suppression in state space and strength tuning in the
coefficient plane \((\mu,\nu)\).

{Let \(\tau_{\mathrm{slow}}\) denote the lifetime of the occupied
baseline slow mode. For structure tuning, \(\tau_{\mathrm{fast}}\)
describes decay of the remaining fast components after slow-mode
cancellation; for strength tuning, it is the inverse of a certified
controlled contraction rate, not necessarily a baseline fast-mode
lifetime. An exponential decay \(e^{-\Gamma t}\) has timescale
\(1/\Gamma\). Cancelling a well-separated slow mode can yield
\(\tau_{\mathrm{fast}}\ll\tau_{\mathrm{slow}}\), whereas strength
tuning only requires \(\tau_{\mathrm{fast}}<\tau_{\mathrm{slow}}\).
For a fixed protocol, both timescales are independent of the accuracy
\(\varepsilon\).}

\begingroup
\renewenvironment{theorem_}{\begin{informaltheorem}[informal]}{\end{informaltheorem}}
\begin{theorem_}
{Consider a QLD baseline with an occupied, simple slow mode
separated from all other nonstationary modes. Suppose admissible
coefficients \(\mu(t),\nu(t)\geq0\) permit either exact cancellation
of this component before restoring the baseline, or common rescaling
that yields a contraction bound with
\(\tau_{\mathrm{fast}}<\tau_{\mathrm{slow}}\).
If both trajectories converge to the same target
\(\rho_{\mathrm{tar}}\), a fixed control protocol gives, schematically,}
\begin{align}
 T_\varepsilon^{\mathrm{base}}
 &\sim \tau_{\mathrm{slow}}\log(1/\varepsilon),
 \label{eq:informal_baseline_timescale_summary}\\
 T_\varepsilon^{\mathrm{ctrl}}
 &\lesssim T_{\mathrm{ctrl}}
       +\tau_{\mathrm{fast}}\log(1/\varepsilon),
 \label{eq:informal_control_timescale_summary}\\
 T_\varepsilon^{\mathrm{ctrl}}
 &<T_\varepsilon^{\mathrm{base}}
 \qquad(0<\varepsilon<\varepsilon_0).
 \label{eq:informal_control_accuracy_window}
\end{align}
Here \(T_{\mathrm{ctrl}}\) is the duration of a fixed
preparation or rescaling stage, independent of \(\varepsilon\),
and \(\varepsilon_0>0\) depends on the chosen protocol.
\end{theorem_}
\endgroup

{Under these conditions, the shorter relaxation timescale
\(\tau_{\mathrm{fast}}\) eventually outweighs the fixed control
duration \(T_{\mathrm{ctrl}}\) as \(\varepsilon\) decreases.
If strength tuning instead leaves a stationary-state bias relative
to \(\rho_{\mathrm{tar}}\), the comparison applies only at tolerances
\(\varepsilon\) above that bias, not as \(\varepsilon\rightarrow 0^{+}\).
In either case, the relevant benefit is a shorter convergence time
\(T_\varepsilon^{\mathrm{ctrl}}\), not merely a smaller
spectral-separation indicator \(\chi_{\max}\).}

For the symmetric two-state QLD model, we give explicit choices of
\(\mu,\nu\) and duration estimates for both routes
(\cref{thm:two_state_dissipative_pulse,thm:two_state_strength_control}).
For general potentials \(V(x)\), the first route requires
accessible coefficients \(\mu,\nu\) that remove the occupied
slow component. The second requires stronger contraction while
keeping the limiting state within the prescribed tolerance
\(\varepsilon\) of the target \(\rho_{\mathrm{tar}}\)
(\cref{thm:qld_dissipative_pulse,thm:qld_strength_control}).
These conditions depend on the model; tunable coefficients alone do
not guarantee them. The finite-accuracy analysis in
\cref{rem:control_robustness} also accounts for residual slow-mode
weights and switching errors while retaining the same target
\(\rho_{\mathrm{tar}}\) and counting the full control time
\(T_{\mathrm{ctrl}}\).

\subsection{Organization}

\begin{table}[!tbp]
 \centering
 \small
 \setlength{\tabcolsep}{6pt}
 \renewcommand{\arraystretch}{1.08}
 \begin{tabular}{@{}>{\centering\arraybackslash}m{0.17\textwidth}
                     >{\centering\arraybackslash}m{0.14\textwidth}
                     >{\raggedright\arraybackslash}m{\dimexpr0.69\textwidth-4\tabcolsep\relax}@{}}
  \toprule
  Model & Theorem & \multicolumn{1}{c@{}}{Main conclusion} \\
  \midrule
  \multicolumn{3}{@{}l}{\textbf{Section 3: Origin of metastability}}\\
  \addlinespace[2pt]
  \multirow{3}{=}{\centering Two-state QLD}
  & \Cref{thm:two_state_lower_edge}
  & Symmetric wells: an explicit necessary-and-sufficient criterion for the spectral window.\\
  \addlinespace[2pt]
  & \Cref{thm:biased_two_state_window}
  & {Biased wells: necessary-and-sufficient criteria in the weak- and strong-dissipation regimes.}\\
  \midrule
  \addlinespace[3pt]
  \multirow{3}{=}{\centering Multi-state QLD}
  & \Cref{thm:qld_sector_origin}
  & Weak tunneling: a sufficient criterion for spectral separation, with quadratically slow population transfer between wells.\\
  \addlinespace[2pt]
  & \Cref{thm:qld_strong_dissipation_exclusion}
  & Strong dissipation: a sufficient exclusion criterion when the dissipator lies strictly below the separation threshold.\\
  \midrule
  \addlinespace[3pt]
  \multirow{3}{=}{\centering General Lindbladian}
  & \Cref{thm:weak_sector_certificate}
  & Weak perturbations: a general sufficient criterion for spectral separation near a degenerate stationary space.\\
  \addlinespace[2pt]
  & \Cref{thm:general_strong_dissipation_exclusion}
  & Strong dissipation: a general sufficient exclusion criterion determined by the dissipator.\\
  \midrule
  \multicolumn{3}{@{}l}{\textbf{Section 4: Dissipative control for optimization}}\\
  \addlinespace[2pt]
  \multirow{2}{=}[-1pt]{\centering Two-state QLD}
  & \Cref{thm:two_state_dissipative_pulse}
  & Convergence time reduction by slow-mode suppression.\\
  \addlinespace[2pt]
  & \Cref{thm:two_state_strength_control}
  & Convergence time reduction by using dissipation rescaling.\\
  \midrule
  \addlinespace[3pt]
  \multirow{2}{=}[-1pt]{\centering General-potential\\QLD}
  & \Cref{thm:qld_dissipative_pulse}
  & Convergence time reduction by slow-component cancellation.\\
  \addlinespace[2pt]
  & \Cref{thm:qld_strength_control}
  & Convergence time reduction by dissipation rescaling.\\
  \bottomrule
 \end{tabular}
 \caption{Theorems underlying the spectral analysis and dissipative
 control. The conclusions are subject to the assumptions of the
 cited statements. Spectral exclusion concerns
 \(\chi_{\max}\); the control results compare convergence times
 \(T_\varepsilon\) at a fixed target accuracy \(\varepsilon\).}
 \label{tab:intro_theorem_summary}
\end{table}

\Cref{sec:preliminaries} introduces the QLD model and the spectral
and dynamical notions of metastability.
\Cref{sec:origin_metastability} studies bottleneck formation and
exclusion, progressing from exact two-state boundaries to multi-state
QLD and general Lindbladians. \Cref{sec:metastability_control}
connects occupied slow components to finite-time delay and develops
QLD controls that can bypass it. \Cref{sec:numerical_simulations} numerically supports our analysis of the origin of metastability and demonstrates shorter fixed-target convergence times under dissipative control.
\Cref{sec:conclusion_future_work} summarizes the conclusions and
future directions. In appendices,  \Cref{app:non_diagonalizable_dynamics} treats
non-diagonalizable dynamics; \cref{app:proof_of_theorems,app:proof_dissipative_control}
collect the proofs of the theorems in
\cref{sec:origin_metastability,sec:metastability_control}, respectively.
The formal theorems and their main conclusions are summarized in
\cref{tab:intro_theorem_summary}.

\section{Preliminaries}
\label{sec:preliminaries}
We introduce the concrete dynamics used to resolve metastable
bottlenecks, then the spectral and trajectory-level concepts used to
analyze them. The QLD parameters make these mechanisms computable;
the Lindbladian framework distinguishes features specific to this
model from statements valid for more general open quantum systems.

\subsection{Quantum Langevin dynamics (QLD)}

To examine how coherent transfer and dissipation shape relaxation,
we use quantum Langevin dynamics (QLD), which encodes an objective function
\(V\colon\mathbb R^d\to\mathbb R\) in a Hamiltonian \(H\), with the
aim of concentrating the density operator \(\rho(t)\) near the global minimum
of \(V\) through dissipative evolution \cite{chen2023quantum}.
For spatial dimension \(d\geq1\), the density operator \(\rho(t)\)
evolves under the generator \(\mathcal L\) as
\begin{align}
 \frac{\mathrm d\rho}{\mathrm dt}
 =\mathcal L(\rho)
 =-\frac{\im}{\hbar}[H,\rho]
 +\sum_{j=1}^{d}\left(2L_j\rho L_j^\dagger
                  -\{L_j^\dagger L_j,\rho\}\right).
 \tag{QLD}\label{eq:QLD}
\end{align}
The Hamiltonian \(H\) and jump operators \(L_j\) are
\begin{align}
 H&=\frac{1}{2m}\sum_{j=1}^{d}p_j^2+V(x),
 \label{eq:qld_hamiltonian}\\
 L_j&=\mu x_j+\im\nu p_j,\qquad 1\leq j\leq d.
 \label{eq:qld_jump_operators}
\end{align}
The position operators \(x_j\) and momentum operators \(p_j\)
satisfy \([x_j,p_k]=\im\hbar\delta_{jk}I\) on their common
domain, and the anticommutator \(\{A,B\}=AB+BA\) fixes our
normalization of the dissipative terms. The thermal parametrization
of the coefficients \(\mu,\nu\geq0\) in \cite{chen2023quantum} is
\begin{align}
 \mu^2&=\frac{\eta m\Omega}{2\hbar}
       \coth\!\left(\frac{\hbar\Omega}{4k_BT}\right),
 \label{eq:qld_thermal_mu}\\
 \nu^2&=\frac{\eta}{2\hbar m\Omega}
       \tanh\!\left(\frac{\hbar\Omega}{4k_BT}\right),
 \qquad 2\hbar\mu\nu=\eta.
 \label{eq:qld_thermal_nu}
\end{align}
Here the mass \(m\), reduced Planck constant \(\hbar\), Boltzmann
constant \(k_B\), bath temperature \(T\), and reference frequency
\(\Omega\) are positive, and the damping rate \(\eta\) is nonnegative.
In the control analysis, the coefficients \(\mu,\nu\) may instead be
varied within a specified admissible set. In a finite-dimensional
discretization, a nonnegative pair \((\mu,\nu)\) defines a Lindblad
generator \(\mathcal L\), but need not be realizable with fixed bath
parameters \(\eta,T,\Omega\). In the continuum, the unbounded
operators \(H,x_j,p_j\) additionally require domains and conditions
ensuring that \(\mathcal L\) generates a trace-preserving semigroup;
the formal Lindblad expression alone does not establish this \cite{chebotarev1998sufficient}.
This distinction matters for the constrained protocols in
\cref{sec:metastability_control}.

For an observable \(B\), write its expectation as
\(\langle B\rangle_t=\operatorname{Tr}(B\rho(t))\).
The convex convergence result of \cite{chen2023quantum} applies to
a continuously differentiable convex potential \(V\) with a unique
minimizer \(x^\star\), under its finite-Laplacian-expectation and
high-temperature assumptions. In coordinates centered at
\(x^\star=0\), with \(V(0)=0\),
\(\hbar\Omega\ll k_BT\), and the stated initial-energy condition
\(\langle H\rangle_0\geq\zeta d k_BT\), it gives
\begin{align}
 \langle V\rangle_t-V(x^\star)
 &\leq \zeta d k_BT+O(e^{-\eta t}),
 \qquad \zeta=2+O\!\left(\frac{\hbar\Omega}{k_BT}\right).
 \label{eq:qld_convex_energy_bound}
\end{align}
The high-temperature correction \(\zeta-2\) is subject to the
regularity and approximation conditions of that result; it is not a
uniform estimate for arbitrary nonconvex potentials \(V\).
The centering is part of this comparison: translating \(V\) while
keeping the jump operators \(L_j=\mu x_j+\im\nu p_j\) unchanged
need not preserve the dynamics \(\mathcal L\). None of the
nonconvex results below assumes knowledge of the minimizer
\(x^\star\) or uses \cref{eq:qld_convex_energy_bound} as an escape
guarantee.
The same work studies quasar-convex potentials \(V\) and reports
numerical benefits of time-dependent cooling on selected nonconvex
benchmarks, including comparisons with classical optimization,
Quantum Hamiltonian Descent \cite{leng2023quantum}, and adiabatic
evolution. Those experiments do not establish a universal escape or
acceleration guarantee. The spectral mechanisms and conditional
guarantees studied here address this distinction.

Allowing the coefficients \(\mu(t),\nu(t)\) to depend on time
defines a time-dependent generator \(\mathcal L_t\) by
\cref{eq:QLD,eq:qld_jump_operators}. Its propagator
\(\mathcal U(t,0)\) solves
\begin{align}
 \partial_t\mathcal U(t,0)&=\mathcal L_t\mathcal U(t,0),
 \qquad \mathcal U(0,0)=\mathcal I,\qquad
 \rho(t)=\mathcal U(t,0)\rho_0.
 \label{eq:qld_time_dependent_propagator}
\end{align}
Here \(\mathcal I\) is the identity superoperator and
\(\rho_0=\rho(0)\) is the initial state. For finite-dimensional,
piecewise-continuous Lindblad generators \(\mathcal L_t\), the
propagator \(\mathcal U(t,0)\) is completely positive and trace
preserving. The spectral formulas below concern fixed generators
\(\mathcal L\), or individual constant stages of a control protocol,
not a single eigendecomposition of a time-dependent propagator
\(\mathcal U(t,0)\).

\subsection{Metastability of Lindbladian dynamics}

The spectral framework of \cite{macieszczak2016towards} distinguishes
rapid relaxation toward a low-dimensional set from slower evolution
within that set. We formulate the spectral diagnostic first, without
assuming diagonalizability. Let the generator \(\mathcal L\) act on
\(\mathcal B(\mathbb C^{d_H})\), where the Hilbert-space dimension
\(d_H\geq2\) differs from the spatial dimension \(d\) in
\cref{eq:qld_hamiltonian}. Unless stated otherwise, \(\mathcal L\)
is relaxing: it has a simple stationary eigenvalue \(0\), a stationary
state \(\rho_{\mathrm{ss}}\), and strictly negative real parts for
all remaining eigenvalues \(\lambda_k\). Order these eigenvalues
\(\lambda_k\), counting algebraic multiplicities, by
\begin{align}
 0=\lambda_1>\operatorname{Re}\lambda_2
 \geq\cdots\geq\operatorname{Re}\lambda_{d_H^2}.
 \label{eq:ordered_eigenvalues}
\end{align}
These finite-dimensional assumptions concern the chosen discretization
or reduction. They are not an assertion about every continuum
generator \(\mathcal L\).

\begin{definition_}[Spectral metastability and candidate intermediate times \(\tau\)]
\label{def:metastability}
For a relaxing generator \(\mathcal L\), define the decay rates
\(\Gamma_j\) by
\begin{align}
 \Gamma_j&=-\operatorname{Re}\lambda_j,\qquad
 0=\Gamma_1<\Gamma_2\leq\cdots\leq\Gamma_{d_H^2}.
 \label{eq:decay_rates_Gamma}
\end{align}
A separated slow sector with \(M\) operator directions, including
the stationary direction \(\rho_{\mathrm{ss}}\), is characterized by
\begin{align}
 \Gamma_M&\ll\Gamma_{M+1},\qquad 2\leq M<d_H^2.
 \label{eq:metastable_rate_separation}\\
 \Gamma_{M+1}^{-1}&\ll\tau\ll\Gamma_M^{-1}.
 \label{eq:metastable_candidate_window}
\end{align}
The candidate intermediate times \(\tau\) in
\cref{eq:metastable_candidate_window} lie between the exponential
decay scales. Their separation does not, by itself, guarantee an approximately stationary
trajectory \(\rho(t)\) away from \(\rho_{\mathrm{ss}}\).
\end{definition_}

To make the scale comparison in \cref{eq:metastable_rate_separation}
operational, we use a bounded ratio and record which cut is selected.
This convention also applies at repeated eigenvalues \(\lambda_k\)
and will be used consistently in the phase maps.
\begin{definition_}[Adjacent separations \(\chi_i\) and the maximizing cut \(i^\ast\)]
\label{def:chi_separation}
For the decay rates \(\Gamma_i\) in \cref{eq:decay_rates_Gamma}, set
\begin{align}
 \chi_i&=\frac{\Gamma_{i+1}-\Gamma_i}{\Gamma_{i+1}+\Gamma_i},
 \qquad 2\leq i<d_H^2,
 \label{eq:adjacent_decay_rate_separation}\\
 \chi_{\max}&=\max_{2\leq i<d_H^2}\chi_i,\qquad
 i^\ast=\min\{i:\chi_i=\chi_{\max}\}.
 \label{eq:maximizing_spectral_cut}
\end{align}
The index \(i\) labels the cut after \(i-1\) nonstationary operator
directions. The stationary rate \(\Gamma_1=0\) is excluded from
the comparison. Ties are resolved by choosing the smallest index
\(i\).
\end{definition_}

For a fixed threshold \(0<\chi_c<1\), the operational spectral
window is \(\chi_{\max}>\chi_c\). Inverting
\cref{eq:adjacent_decay_rate_separation} gives
\begin{align}
 \chi_i>\chi_c
 \quad\Longleftrightarrow\quad
 \frac{\Gamma_{i+1}}{\Gamma_i}>\frac{1+\chi_c}{1-\chi_c}.
 \label{eq:spectral_threshold_rate_ratio}
\end{align}
We choose a twentyfold rate separation as the numerical screening
convention: substituting the ratio \(20\) into
\cref{eq:adjacent_decay_rate_separation} gives
\(\chi_c=(20-1)/(20+1)=19/21\).
Thus \(\chi_{\max}>19/21\) means that at least one adjacent
rate ratio exceeds \(20\). This chosen resolution of timescales
is not a universal constant or a thermodynamic phase boundary.
Equal decay rates \(\Gamma_i=\Gamma_{i+1}\) give \(\chi_i=0\);
a nonzero cut therefore does not divide an exactly conjugate pair.
If only part of the spectrum \(\sigma(\mathcal L)\) is resolved,
the maximum over those rates must be reported as a windowed diagnostic,
not as the full-spectrum quantity \(\chi_{\max}\).

When \(\mathcal L\) is diagonalizable, its right eigenmatrices
\(\mathbf R_k\) and left eigenmatrices \(\mathbf L_k\) give an
explicit interpretation of this separation. We use the bilinear trace
pairing, rather than silently conjugating the coefficients:
\begin{align}
 \mathcal L(\mathbf R_k)&=\lambda_k\mathbf R_k,\qquad
 \mathcal L^\dagger(\mathbf L_k)=\lambda_k\mathbf L_k,
 \label{eq:biorthogonal_eigenmatrix_equations}\\
 \operatorname{Tr}(\mathbf L_k\mathbf R_j)&=\delta_{kj},\qquad
 \mathbf R_1=\rho_{\mathrm{ss}},\qquad \mathbf L_1=I.
 \label{eq:biorthogonal_normalization}
\end{align}
Here the adjoint \(\mathcal L^\dagger\) is the Hilbert--Schmidt
adjoint. Since \(\mathcal L\) preserves Hermiticity,
\(\operatorname{Tr}(B\mathcal L(C))
=\operatorname{Tr}(\mathcal L^\dagger(B)C)\);
equivalently, \(\mathbf L_k^\dagger\) is an adjoint eigenmatrix
with eigenvalue \(\overline{\lambda_k}\) in the sesquilinear
convention. Defining the initial coefficients
\(c_k=\operatorname{Tr}(\mathbf L_k\rho_0)\) gives
\begin{align}
 \rho(t)&=\sum_{k=1}^{d_H^2}e^{\lambda_kt}c_k\mathbf R_k
 =\rho_{\mathrm{ss}}+
   \sum_{k=2}^{d_H^2}e^{\lambda_kt}c_k\mathbf R_k.
 \label{eq:diagonalizable_state_expansion}
\end{align}
The second equality uses \(c_1=\operatorname{Tr}\rho_0=1\).
The trace norm \(\|B\|_1\), Hilbert--Schmidt norm
\(\|B\|_{\mathrm{HS}}\), and trace distance \(D_{\mathrm{tr}}\)
used here and below are
\begin{align}
 \|B\|_1&=\operatorname{Tr}\sqrt{B^\dagger B},\qquad
 \|B\|_{\mathrm{HS}}=\sqrt{\operatorname{Tr}(B^\dagger B)},
 \label{eq:prelim_operator_norms}\\
 D_{\mathrm{tr}}(\rho,\sigma)&=\frac12\|\rho-\sigma\|_1.
 \label{eq:prelim_trace_distance}
\end{align}
For vectors \(v\), \(\|v\|_2\) is the Euclidean norm; for
matrices \(B\), \(\|B\|_2=\|B\|_{\mathrm{op}}\) denotes its
induced operator norm, not \(\|B\|_{\mathrm{HS}}\). Superoperator
norms \(\|\mathcal A\|_{2\to2}\) instead use the
Hilbert--Schmidt norm on both their input and output, as specified
in \cref{eq:two_state_dissipation_norm} below.
For a conjugation-closed slow sector, the spectral projector
\(\mathcal P\) and its discarded contribution satisfy
\begin{align}
 \mathcal P(B)&=\sum_{k=1}^{M}\operatorname{Tr}(\mathbf L_kB)
                    \mathbf R_k,
 \label{eq:diagonalizable_slow_projector}\\
 \mathcal P\rho(t)&=\rho_{\mathrm{ss}}+
                 \sum_{k=2}^{M}e^{\lambda_kt}c_k\mathbf R_k,
 \label{eq:diagonalizable_slow_expansion}\\
 \|\rho(t)-\mathcal P\rho(t)\|_1
 &\leq\sum_{k=M+1}^{d_H^2}|c_k|\|\mathbf R_k\|_1e^{-\Gamma_kt}.
 \label{eq:diagonalizable_fast_error}
\end{align}
The remainder bound in \cref{eq:diagonalizable_fast_error} displays
the modal amplitudes \(|c_k|\|\mathbf R_k\|_1\) that a rate ratio
alone does not control. Moreover, small decay rates \(\Gamma_k\)
need not make \(e^{\lambda_kt}c_k\mathbf R_k\) nearly constant:
the imaginary parts \(\operatorname{Im}\lambda_k\) can generate
coherent motion. If the slow coefficients \(c_k\), \(2\leq k\leq M\),
vanish, there is no corresponding plateau for that initial state
\(\rho_0\). The projector \(\mathcal P\) also need not preserve
positivity. We therefore use \(\chi_{\max}\) as a spectral
diagnostic and assess its effect on convergence for a specified
initial state and target in \cref{sec:metastability_control}.

Diagonalizability of \(\mathcal L\) is an assumption for
\cref{eq:biorthogonal_eigenmatrix_equations,eq:diagonalizable_state_expansion},
not for \cref{def:metastability,def:chi_separation}. It is a common
simplification in metastability analyses \cite{brown2024unravelling},
but neither finite-dimensional discretization nor noise guarantees it
at every parameter value. In particular, the exact two-state calculation
below contains a defective critical point. Generalized eigenmatrices
replace \(\mathbf R_k\) there, and polynomial factors multiply the
exponentials \(e^{\lambda_kt}\). The peripheral eigenvalues
\(\operatorname{Re}\lambda_k=0\) of a finite-dimensional Lindbladian
\(\mathcal L\) are semisimple (their Jordan blocks have size one),
because its quantum channels
\(e^{t\mathcal L}\) are bounded. For decaying Jordan blocks,
the relevant requirement is quantitative suppression of the fast
contribution and control of slow evolution
\cite{macieszczak2021classical,szehr2015spectral}.
\Cref{app:non_diagonalizable_dynamics} gives the full expansion and
the corresponding bounds.
\section{Origin of metastability}
\label{sec:origin_metastability}

The first question is when relaxation develops a bottleneck: which
degrees of freedom equilibrate rapidly, and which remain slow?
We answer it through the competition between coherent transfer and
dissipative relaxation, using QLD to make the relevant couplings
explicit before extracting conditions for general Lindbladians.
We distinguish the formation of a spectral window
\(\chi_{\max}>\chi_c\) from the appearance of a long-lived plateau
in a particular trajectory \(\rho(t)\). Throughout this section,
\(\chi_c\in(0,1)\) is a chosen diagnostic threshold, not a
thermodynamic phase boundary. The stationary eigenvalue \(\lambda_1=0\)
is included in the indexing of \cref{def:chi_separation}, so the cut
\(i=2\) separates one nonstationary slow mode from the remaining
modes. A change in the maximizing cut \(i^\ast\) records a change
of spectral ordering; it is not itself a dynamical cause.

The competition is first resolved exactly for the two-state generator
\(\mathcal L_2^{\mathrm{TP}}\) in \cref{sec:double_well_model},
giving conditions for spectral separation and its exclusion.
To identify which mechanisms persist with additional Hamiltonian
eigenstates \(\ket{\phi_n}\), \cref{sec:qld_general_origin}
expresses sufficient conditions through computable couplings for
arbitrary one-dimensional potentials \(V(x)\). Finally,
\cref{sec:general_origin} formulates the stationary-sector mechanism
for a general finite-dimensional Lindbladian \(\mathcal L\) and
gives complementary exclusion conditions. Both formation and
exclusion of the spectral window
\(\chi_{\max}>\chi_c\) belong to this section. Whether an admissible
control reduces the convergence time \(T_\varepsilon\) at a fixed
target accuracy \(\varepsilon\) is the distinct question of
\cref{sec:metastability_control}.

\subsection{Two-state QLD model for a double-well potential}

\label{sec:double_well_model}

To isolate the competition that produces a spectral bottleneck, we
consider a one-dimensional potential \(V(x)\) having two wells.
Retaining two Hamiltonian eigenstates \(\ket{\phi_0},\ket{\phi_1}\)
makes the separation condition \(\chi_{\max}>\chi_c\) for the
resulting generator \(\mathcal L_2^{\mathrm{TP}}\) exactly calculable.
This reduction isolates coherent and dissipative effects; it does
not assert that two states
accurately describe every trajectory of the full generator
\(\mathcal L\). The later comparison with larger projections \(P_r\)
tests that approximation. For mass \(m>0\), curvature frequency
\(\omega>0\), separation \(\ell>0\), plateau energy \(E_b>0\),
and well offset \(0\leq E_\delta<E_b\), set
\begin{align}
V(x)
=
\begin{cases}
\dfrac12m\omega^2
\left(x+\dfrac{\ell}{2}\right)^2+E_\delta,
&
x\in\left[
        -\frac{\ell}{2}-R_L,
        -\frac{\ell}{2}+R_L
    \right],
\\[1.2em]
\dfrac12m\omega^2
\left(x-\dfrac{\ell}{2}\right)^2,
&
x\in \left[
        \frac{\ell}{2}-R_R,
        \frac{\ell}{2}+R_R
    \right],
\\[1.2em]
E_b,
&
x \in \mathbb{R} \setminus 
    \lrb{
        \left[
        -\frac{\ell}{2}-R_L,
       -  \frac{\ell}{2}+R_L
    \right] 
    \cup \left[
        \frac{\ell}{2}-R_R,
        \frac{\ell}{2}+R_R
    \right]},
\end{cases} 
\label{eq:piecewise_double_well_potential_0710}
\end{align}
where 
\begin{align}
    R_L 
    =
    \sqrt{\frac{2(E_b-E_\delta)}{m\omega^2}}\,, 
    \qquad 
    R_R 
    =
    \sqrt{\frac{2E_b}{m\omega^2}},
    \label{eq:piecewise_turning_radii_0710}
\end{align}
are the truncation radii \(R_L,R_R\), chosen so that each quadratic
branch joins the plateau \(V(x)=E_b\) continuously. The minima
\(x=-\ell/2\) and \(x=\ell/2\) have energies \(E_\delta\) and
zero, respectively. Thus the barrier heights are \(E_b-E_\delta\)
from the left and \(E_b\) from the right.

We take \(0\leq E_\delta<E_b\) and \(\ell>R_L+R_R\), so the two well intervals are disjoint and separated by a nonempty flat barrier {of width \(d_b:=\ell-R_L-R_R>0\)}. The radii \(R_L,R_R\) are determined by \cref{eq:piecewise_turning_radii_0710}, rather than varied independently of the curvature parameter \(\omega\) and energy parameters \(E_b,E_\delta\).

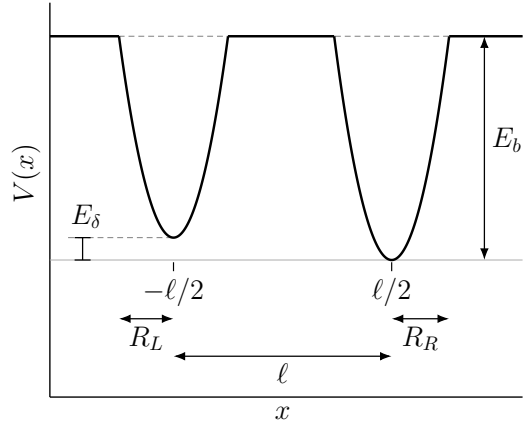
\begin{figure}[htbp]
    \centering
    \resizebox{0.42\textwidth}{!}{
\begin{tikzpicture}[x=0.65cm,y=1.0cm,
    every node/.style={font=\normalfont\fontsize{14.4}{17.28}\selectfont,text=black},
    dimension/.style={Latex-Latex,line width=0.65pt},
    guide/.style={gray,densely dashed,line width=0.4pt}]
    \pgfmathsetmacro{\rr}{sqrt(4/1.6)}
    \draw[line width=0.7pt] (-6.4,-2.45)--(6.6,-2.45);
    \draw[line width=0.7pt] (-6.4,-2.45)--(-6.4,4.6);
    \draw[gray!65,line width=0.45pt] (-6.4,0)--(6.6,0);
    \draw[guide] (-6.4,4)--(6.6,4);
    \draw[guide] (-5.9,0.4)--(-3,0.4);
    \draw[line width=1.25pt] (-6.4,4)--(-4.5,4);
    \draw[line width=1.25pt,domain=-4.5:-1.5,samples=120]
        plot (\x,{1.6*(\x+3)^2+0.4});
    \draw[line width=1.25pt] (-1.5,4)--({3-\rr},4);
    \draw[line width=1.25pt,domain={3-\rr}:{3+\rr},samples=120]
        plot (\x,{1.6*(\x-3)^2});
    \draw[line width=1.25pt] ({3+\rr},4)--(6.6,4);
    \draw[line width=0.6pt] (-3,-0.04)--(-3,-0.20);
    \draw[line width=0.6pt] (3,-0.04)--(3,-0.20);
    \node[below] at (-3,-0.18) {$-\ell/2$};
    \node[below] at (3,-0.18) {$\ell/2$};
    \draw[dimension] (-4.5,-1.05)--(-3,-1.05);
    \node[below] at (-3.75,-1.05) {$R_L$};
    \draw[dimension] (3,-1.05)--({3+\rr},-1.05);
    \node[below] at ({3+\rr/2},-1.05) {$R_R$};
    \draw[dimension] (-3,-1.72)--(3,-1.72);
    \node[below] at (0,-1.72) {$\ell$};
    \draw[dimension] (5.55,0)--(5.55,4);
    \node[right] at (5.55,2.15) {$E_b$};
    \draw[line width=0.65pt] (-5.5,0)--(-5.5,0.4);
    \draw[line width=0.65pt] (-5.72,0)--(-5.28,0);
    \draw[line width=0.65pt] (-5.72,0.4)--(-5.28,0.4);
    \node[above] at (-5.4,0.43) {$E_\delta$};
    \node[below] at (0,-2.48) {$x$};
    \node[rotate=90] at (-7.05,1.55) {$V(x)$};
\end{tikzpicture}%
}
    \caption{Schematic of the piecewise quadratic double-well potential \(V(x)\) in \cref{eq:piecewise_double_well_potential_0710}. The minima at \(x=-\ell/2\) and \(x=\ell/2\) have energies \(E_\delta\) and zero, respectively. The left and right truncation radii \(R_L\) and \(R_R\) are given in \cref{eq:piecewise_turning_radii_0710}. The plateau energy \(E_b\) is measured from the right-well minimum, so the barrier height \(E_b-E_\delta\) is measured from the left-well minimum. The drawing illustrates the geometry rather than a particular numerical parameter set.}
    \label{fig:piecewise_double_well_potential_0710}
\end{figure}

The geometry of \(V(x)\) is shown in
\cref{fig:piecewise_double_well_potential_0710}. Symmetry corresponds
to \(E_\delta=0\); for \(E_\delta>0\), the right minimum has lower
energy. Because \(V(x)\) is constant outside the wells, the
continuum Hamiltonian \(H\) need not have arbitrarily many bound
states. The two-state construction below requires the stated discrete
eigenstates \(\ket{\phi_0},\ket{\phi_1}\), or a specified finite
spatial discretization.

 We restrict both the operators
\(H,L\) and the evolving state \(\rho_2(t)\) to the two lowest
Hamiltonian eigenstates \(\ket{\phi_0},\ket{\phi_1}\).
This defines an autonomous model \(\mathcal L_2^{\mathrm{TP}}\),
rather than merely compressing the output of \(\mathcal L\).

Let the Hamiltonian eigenstates \(\ket{\phi_n}\) and corresponding energies \(E_n\) satisfy \(H\ket{\phi_n}=E_n\ket{\phi_n}\), with $n\in \lrc{0,1}$ and \(E_0<E_1\), and define the projector \(P_2\) by
\begin{align}
    P_2=\sum_{n=0}^{1}\ket{\phi_n}\!\bra{\phi_n}.
    \label{eq:two_state_projector}
\end{align}
For an initial density operator \(\rho_0\) with
\(\operatorname{Tr}(P_2\rho_0P_2)>0\), initialize the reduced state
\(\rho_2(0)\) by
\begin{align}
    \rho_2(0)=\frac{P_2\rho_0P_2}{\operatorname{Tr}(P_2\rho_0P_2)},
    \qquad \rho_2(0)\succeq0,\qquad \operatorname{Tr}\rho_2(0)=1.
    \label{eq:two_state_normalized_projection}
\end{align}
For the one-dimensional jump operator \(L=\mu x+\im\nu p\), define the projected Hamiltonian \(H_2\), projected jump operator \(L^{(2)}\), projected position operator \(x_2\), and projected momentum operator \(p_2\) by
\begin{align}
    H_2&=P_2HP_2,
    & L^{(2)}&=P_2LP_2=\mu x_2+\im\nu p_2,
    \label{eq:two_state_projected_hamiltonian}\\
    x_2&=P_2xP_2,
    & p_2&=P_2pP_2.
    \label{eq:two_state_projected_position}
\end{align}
The autonomous two-state Lindbladian \(\mathcal L_2^{\mathrm{TP}}\) is
\begin{align}
    \frac{d\rho_2}{dt}
    =\mathcal L_2^{\mathrm{TP}}(\rho_2)
    =-\frac{\im}{\hbar}[H_2,\rho_2]
     +2L^{(2)}\rho_2L^{(2)\dagger}
     -\{L^{(2)\dagger}L^{(2)},\rho_2\},
    \label{eq:two_state_tp_generator}
\end{align}
where all operators and products act within the retained space. The reduced Lindbladian \(\mathcal L_2^{\mathrm{TP}}\) in \cref{eq:two_state_tp_generator} preserves positivity and trace. Throughout this subsection, the eigenvalues \(\lambda_i\), decay rates \(\Gamma_i\), adjacent separations \(\chi_i\), maximum separation \(\chi_{\max}\), and maximizing index \(i^\ast\) refer to the two-state Lindbladian \(\mathcal L_2^{\mathrm{TP}}\), with the ordering and definitions in \cref{eq:ordered_eigenvalues,eq:decay_rates_Gamma,def:chi_separation}.

The state \(\rho_2(t)\) subsequently solves
\cref{eq:two_state_tp_generator}; it is not defined by renormalizing
\(P_2\rho(t)P_2\) at every time.

\subsubsection{Exact two-state dynamics}
\label{sec:exact_two_state}

The decay rates \(\Gamma_i\) reveal whether a slow mode separates
from the remaining relaxation processes. To compute them and the
maximum separation \(\chi_{\max}\) exactly, we express
\cref{eq:two_state_tp_generator} through a real \(3\times3\)
Bloch matrix \(A\). Its entries and spectral classification below
provide the inputs to the window criteria in
\cref{thm:two_state_lower_edge,thm:biased_two_state_window}.

Choose real energy eigenfunctions \(\phi_0,\phi_1\) and, for
\(n,m\in\{0,1\}\), define
\begin{align}
 x_{nm}\coloneqq\langle\phi_n|x|\phi_m\rangle,\qquad
 p_{nm}\coloneqq\langle\phi_n|p|\phi_m\rangle,\qquad
 \omega_{01}\coloneqq\frac{E_1-E_0}{\hbar}>0.
 \label{eq:energy_basis_matrix_elements}
\end{align}
For the continuum Hamiltonian \(H=p^2/(2m)+V(x)\), assume these
eigenstates lie in the domains of the commutator \([H,x]\) and
matrix elements \(x_{nm},p_{nm}\). Then
\([H,x]=-\im\hbar p/m\) gives
\begin{align}
 p_{nm}=\frac{\im m}{\hbar}(E_n-E_m)x_{nm}.
 \label{eq:projected_momentum_identity}
\end{align}
A spatial discretization must satisfy the same commutator identity
to use this relation exactly. Otherwise, the projected jump operator
\(L^{(2)}\) must be built from its actual momentum matrix elements
\(p_{nm}\), and formulas using \cref{eq:projected_momentum_identity}
must be rederived.

Define the real coefficients \(a_0,a_x,a_z,b\) by
\begin{align}
 a_0\coloneqq\frac{\mu(x_{00}+x_{11})}{2},\qquad
 a_z\coloneqq\frac{\mu(x_{00}-x_{11})}{2},\qquad
 a_x\coloneqq\mu x_{01},\qquad b\coloneqq\nu m\omega_{01}x_{01}.
 \label{eq:bloch_coefficients}
\end{align}
The projected jump operator \(L^{(2)}\) then takes the form
\begin{align}
 \begin{aligned}[b]
 L^{(2)}
 =\begin{pmatrix}a_0+a_z&a_x+b\\a_x-b&a_0-a_z\end{pmatrix}
  =a_0I_2+a_x\sigma_x+a_z\sigma_z+\im b\sigma_y.
 \end{aligned}
 \label{eq:two_state_jump_pauli}
\end{align}
Here the Pauli matrices \(\sigma_x,\sigma_y,\sigma_z\), collected
in \(\boldsymbol\sigma=(\sigma_x,\sigma_y,\sigma_z)\), act in the
ordered energy basis \((\phi_0,\phi_1)\). Define the dissipator
\(\mathcal D_2\) by
\begin{align}
 \mathcal D_2(B)\coloneqq
 2L^{(2)}B L^{(2)\dagger}
 -\{L^{(2)\dagger}L^{(2)},B\}.
 \label{eq:two_state_dissipator}
\end{align}
Write the density operator \(\rho_2\) using the Bloch vector
\(\boldsymbol u=(u_x,u_y,u_z)^{\mathsf T}\in\mathbb R^3\),
\(\|\boldsymbol u\|_2\leq1\):
\begin{align}
 \begin{aligned}[b]
 \rho_2
 =\frac12\bigl(I_2+u_x\sigma_x+u_y\sigma_y+u_z\sigma_z\bigr)
 =\frac12
 \begin{pmatrix}
  1+u_z&u_x-\im u_y\\
  u_x+\im u_y&1-u_z
 \end{pmatrix}.
 \end{aligned}
 \label{eq:two_state_density_bloch_matrix}
\end{align}
Thus \(u_z=\rho_{2,00}-\rho_{2,11}\) is the population difference
and \(\rho_{2,01}=(u_x-\im u_y)/2\) is the coherence, where
\(\rho_{2,nm}=\langle\phi_n|\rho_2|\phi_m\rangle\).
Substitution into \cref{eq:two_state_tp_generator} gives
\begin{align}
 \dot{\boldsymbol u}=A\boldsymbol u+\boldsymbol f,\quad
 A=\begin{pmatrix}
 -4(b^2+a_z^2)&\omega_{01}&4(a_xa_z-ba_0)\\
 -\omega_{01}&-4(a_x^2+a_z^2)&0\\
 4(a_xa_z+ba_0)&0&-4(b^2+a_x^2)
 \end{pmatrix},\quad
 \boldsymbol f=\begin{pmatrix}-8ba_z\\0\\8ba_x\end{pmatrix}.
 \label{eq:exact_bloch_affine}
\end{align}
The affine source \(\boldsymbol f\) fixes the stationary state
\(\rho_{2,\mathrm{ss}}\), whereas the Bloch matrix \(A\) determines
the remaining eigenvalues:
\begin{align}
 \sigma(\mathcal L_2^{\mathrm{TP}})=\{0\}\cup\sigma(A).
 \label{eq:two_state_bloch_spectral_reduction}
\end{align}
Algebraic multiplicities are included. If \(A\) is invertible,
\begin{align}
 \boldsymbol u_{\mathrm{ss}}=-A^{-1}\boldsymbol f,\qquad
 \boldsymbol u(t)=\boldsymbol u_{\mathrm{ss}}
      +e^{tA}\bigl(\boldsymbol u(0)-\boldsymbol u_{\mathrm{ss}}\bigr).
 \label{eq:exact_bloch_solution}
\end{align}
These identities do not require diagonalizability of \(A\).

\paragraph{Symmetric double well.}
At \(E_\delta=0\), parity gives \(x_{00}=x_{11}=0\), hence
\(a_0=a_z=0\). Assume \(\omega_{01}>0\) and \(a_x^2+b^2>0\).
The longitudinal component \(u_z\) decouples from the transverse
components \(u_x,u_y\), with eigenvalues and stationary vector
\begin{align}
 \lambda_z&=-4(a_x^2+b^2), \label{eq:symmetric_longitudinal_eigenvalue}\\
 \lambda_\pm&=-2(a_x^2+b^2)\pm\sqrt{4(a_x^2-b^2)^2-\omega_{01}^2}, \label{eq:symmetric_transverse_eigenvalues}\\
 \boldsymbol u_{\mathrm{ss}}&=\left(0,0,\frac{2a_xb}{a_x^2+b^2}\right)^{\mathsf T}.
 \label{eq:symmetric_stationary_bloch}
\end{align}
Here \(\lambda_z\) governs population relaxation and
\(\lambda_\pm\) govern coherence relaxation in the energy basis.
To express their ordered separations, set
\begin{align}
 \Delta_B\coloneqq4(a_x^2-b^2)^2-\omega_{01}^2.
 \label{eq:symmetric_square_root_argument}
\end{align}
For \(\Delta_B>0\), define
\begin{align}
  \xi\coloneqq\frac{\sqrt{\Delta_B}}{2(a_x^2+b^2)}.
  \label{eq:symmetric_normalized_splitting}
\end{align}
Then \(0<\xi<1\), and the complete spectral diagnostic is
\begin{align}
 (\chi_2,\chi_3)
 &=\begin{cases}
 (0,\frac13),&\Delta_B\leq0,\\[1ex]
 \left(\xi,\dfrac{1-\xi}{3+\xi}\right),&\Delta_B>0,
 \end{cases} \label{eq:symmetric_adjacent_separation_summary}\\
 \chi_{\max}&=\max\{\chi_2,\chi_3\}.
 \label{eq:symmetric_separation_summary}
\end{align}
For \(\Delta_B<0\), the transverse components \(u_x,u_y\)
oscillate with frequency \(\sqrt{-\Delta_B}\) and decay with
rate \(2(a_x^2+b^2)\). For \(\Delta_B>0\), their distinct real
decay rates can separate; since \(\chi_3\leq1/3\), any window
with \(\chi_c>1/3\) must have \(\chi_2>\chi_c\) and \(i^\ast=2\).
The decay-rate ordering used for this criterion is established in
\cref{app:two_state_lower_edge}.

At \(\Delta_B=0\), the transverse block
\(A_\perp=(A_{ij})_{i,j\in\{x,y\}}\)
has a size-two Jordan block because \(\omega_{01}>0\).
Its propagator \(e^{tA_\perp}\) contains a term proportional to
\(t e^{-2(a_x^2+b^2)t}\), but its asymptotic exponential decay
rate remains \(2(a_x^2+b^2)\); hence
\(\chi_{\max}=1/3\) still holds. Indeed,
\(N:=A_\perp+2(a_x^2+b^2)I_2\) satisfies \(N\ne0\) and
\(N^2=\Delta_B I_2=0\), so
\(e^{tA_\perp}=e^{-2(a_x^2+b^2)t}(I_2+tN)\).
The finite-time effect of this polynomial factor is discussed in
\cref{app:non_diagonalizable_dynamics}.

\paragraph{Biased double well.}
For \(E_\delta\ne0\), the coefficients \(a_0,a_z\) can couple
population and coherence relaxation. Define the damping coefficients
\(\alpha_A,\beta_A,\gamma_A\) and couplings \(h_\pm\) by
\begin{align}
 \alpha_A&=4(b^2+a_z^2),\qquad
 \beta_A=4(a_x^2+a_z^2), \label{eq:biased_alpha_beta_coefficients}\\
 \gamma_A&=4(b^2+a_x^2),\qquad
 h_\pm=4(a_xa_z\pm ba_0).
 \label{eq:biased_bloch_coefficients}
\end{align}
The Bloch matrix \(A\) is equivalently
\begin{align}
 A=
 \begin{pmatrix}
  -\alpha_A & \omega_{01} & h_-\\
  -\omega_{01} & -\beta_A & 0\\
  h_+ & 0 & -\gamma_A
 \end{pmatrix}.
 \label{eq:biased_bloch_matrix_abbreviated}
\end{align}
Its eigenvalues are the roots of
\begin{align}
 \begin{aligned}[b]
 P_A(\lambda)&=\det(\lambda I_3-A)\\
 &=(\lambda+\gamma_A)
   [(\lambda+\alpha_A)(\lambda+\beta_A)+\omega_{01}^2]
   -h_-h_+(\lambda+\beta_A)\\
 &=\lambda^3+c_2\lambda^2+c_1\lambda+c_0=0,
 \end{aligned}
 \label{eq:biased_characteristic_equation}
\end{align}
where
\begin{align}
 c_2&=\alpha_A+\beta_A+\gamma_A, \label{eq:general_bloch_cubic_c2}\\
 c_1&=\alpha_A\beta_A+\alpha_A\gamma_A+\beta_A\gamma_A
       +\omega_{01}^2-h_-h_+, \label{eq:general_bloch_cubic_c1}\\
 c_0&=\alpha_A\beta_A\gamma_A+\gamma_A\omega_{01}^2
       -\beta_Ah_-h_+.
 \label{eq:general_bloch_cubic_coefficients}
\end{align}
Define the depressed-cubic coefficients \(p_A,q_A\) and Cardano
parameter \(\Delta_A\) by
\begin{align}
 p_A&\coloneqq c_1-\frac{c_2^2}{3},\qquad
 q_A\coloneqq\frac{2c_2^3}{27}-\frac{c_2c_1}{3}+c_0.
 \label{eq:biased_cardano_coefficients}
\end{align}
\begin{align}
 \Delta_A=\left(\frac{q_A}{2}\right)^2
          +\left(\frac{p_A}{3}\right)^3.
 \label{eq:cardano_parameter}
\end{align}
For \(\Delta_A<0\), all three roots are real and distinct.
For \(\Delta_A>0\), there is one real root \(\lambda_R\) and
one nonreal conjugate pair \(\lambda_C,\overline{\lambda_C}\).
For \(\Delta_A=0\), at least two real roots coincide; the triple
root occurs exactly when \(p_A=q_A=0\).
Thus \(A\) is diagonalizable over \(\mathbb C\) for
\(\Delta_A\ne0\), whereas repeated roots require a separate
eigenspace check. These cases follow from the cubic discriminant
\(\operatorname{disc}P_A=-4p_A^3-27q_A^2=-108\Delta_A\).

When all three roots have strictly negative real parts, sort
\(0<\Gamma_2\leq\Gamma_3\leq\Gamma_4\). The two candidate
separations \(\chi_2,\chi_3\) and maximizing cut \(i^\ast\) are
\begin{align}
 \chi_2&=\frac{\Gamma_3-\Gamma_2}{\Gamma_3+\Gamma_2},\qquad
 \chi_3=\frac{\Gamma_4-\Gamma_3}{\Gamma_4+\Gamma_3}, \label{eq:biased_adjacent_separations}\\
 \chi_{\max}&=\max\{\chi_2,\chi_3\},\qquad
 i^\ast=\begin{cases}
 2,&\Gamma_3^2\geq\Gamma_2\Gamma_4,\\
 3,&\Gamma_3^2<\Gamma_2\Gamma_4.
 \end{cases}
 \label{eq:biased_real_separations}
\end{align}
In the complex-pair regime \(\Delta_A>0\), put
\(\Gamma_R=-\lambda_R>0\) and
\(\Gamma_C=-\operatorname{Re}\lambda_C>0\). Then
\begin{align}
 \chi_{\max}
 =\frac{|\Gamma_R-\Gamma_C|}{\Gamma_R+\Gamma_C}.
 \label{eq:biased_pair_separation}
\end{align}
The cut is \(i^\ast=2\) when \(\Gamma_R<\Gamma_C\) and
\(i^\ast=3\) when \(\Gamma_C<\Gamma_R\); at equality,
\(\chi_{\max}=0\) and the tie-breaking rule selects \(i^\ast=2\).
Thus the root type determined by \(\Delta_A\) and the strongest
decay-rate separation located by \(i^\ast\) are distinct pieces of
information. Comparing the two fractions in
\cref{eq:biased_adjacent_separations} gives the sign of
\(\Gamma_3^2-\Gamma_2\Gamma_4\), including the tie convention;
a nonreal pair contributes two equal decay rates \(\Gamma_C\).

\subsubsection{Spectral criteria for metastability}

To identify what creates the spectral window \(\chi_{\max}>\chi_c\),
we now relate it to the coherent coupling frequency
\(\Omega_{\mathrm{tun}}^{(2)}\), dissipator norm
\(\gamma_{\mathcal D}^{(2)}\), and dissipative anisotropy
\(\delta_{\mathcal D}\). These quantities distinguish coherent
transfer from dissipative damping, allowing the exact spectrum
\(\sigma(A)\) to identify when a slow mode separates.

\paragraph{Coupling frequency \(\Omega_{\mathrm{tun}}^{(2)}\).}
The energy eigenstates \(\ket{\phi_0},\ket{\phi_1}\) need not
be localized in individual wells. We therefore diagonalize the
projected position operator \(x_2\) in \(\operatorname{Ran}P_2\)
and restrict attention to parameter values for which its eigenvalues
lie on opposite sides of \(x=0\):
\begin{align}
 x_2\ket{\psi_\alpha^{(2)}}
 &=x_\alpha^{(2)}\ket{\psi_\alpha^{(2)}},
 \qquad \alpha\in\{L,R\},
 \qquad x_L^{(2)}<0<x_R^{(2)}.
 \label{eq:two_state_localized_basis}
\end{align}
The orthonormal states \(\ket{\psi_L^{(2)}},\ket{\psi_R^{(2)}}\)
minimize and maximize the mean position within the retained space.
This is a precise localization convention, not a claim that either
state has support exclusively in one well.

The off-diagonal matrix element of \(H_2\) in this basis generates
coherent transfer. Define its magnitude \(J^{(2)}\) and associated
frequency \(\Omega_{\mathrm{tun}}^{(2)}\) by
\begin{align}
 J^{(2)}
 &\coloneqq
 \bigl|\langle\psi_L^{(2)}|H_2|\psi_R^{(2)}\rangle\bigr|,
 \qquad
 \Omega_{\mathrm{tun}}^{(2)}
 \coloneqq\frac{2J^{(2)}}{\hbar}.
 \label{eq:two_state_tunneling_scale}
\end{align}
The magnitude \(J^{(2)}\) is independent of basis phases.
Define the localized energy imbalance \(\epsilon^{(2)}\) and
mean energy \(\bar E\) by
\begin{align}
 \epsilon^{(2)}
 &\coloneqq
 \langle\psi_R^{(2)}|H_2|\psi_R^{(2)}\rangle
 -
 \langle\psi_L^{(2)}|H_2|\psi_L^{(2)}\rangle,
 \qquad
 \bar E\coloneqq\frac{E_0+E_1}{2}.
 \label{eq:two_state_localized_energies}
\end{align}
The imbalance \(\epsilon^{(2)}\) includes kinetic and potential
contributions, so it need not equal \(E_\delta\). Choosing the
relative phase so that the off-diagonal entry is nonnegative gives
\begin{align}
 [H_2]_{\mathrm{localized}}
 =
 \bar E I_2+
 \begin{pmatrix}
 -\epsilon^{(2)}/2 & J^{(2)}\\
 J^{(2)} & \epsilon^{(2)}/2
 \end{pmatrix}.
 \label{eq:two_state_localized_hamiltonian}
\end{align}
For coherent evolution under \(H_2\) alone, define
\begin{align}
 P_{L\to R}^{(2)}(t)
 =\bigl|\langle\psi_R^{(2)}|e^{-\im H_2t/\hbar}
                         |\psi_L^{(2)}\rangle\bigr|^2,
 \qquad t\geq0.
 \label{eq:two_state_coherent_transfer_definition}
\end{align}
\begingroup
\emergencystretch=2em
Diagonalizing and exponentiating
\cref{eq:two_state_localized_hamiltonian} yields
\((\hbar\omega_{01})^2=(\epsilon^{(2)})^2+4(J^{(2)})^2\),
\(P_{L\to R}^{(2)}(t)=\left({\Omega_{\mathrm{tun}}^{(2)}}/{\omega_{01}}\right)^2
\sin^2\!\left({\omega_{01}t}/{2}\right)\), and
\(P_{\max}^{(2)}\coloneqq\max_{t\geq0}P_{L\to R}^{(2)}(t)
=\left({\Omega_{\mathrm{tun}}^{(2)}}/{\omega_{01}}\right)^2\).
Complete transfer is possible exactly when \(\epsilon^{(2)}=0\),
whereas \(\Omega_{\mathrm{tun}}^{(2)}=0\) gives no coherent transfer.
With bias, the coupling frequency \(\Omega_{\mathrm{tun}}^{(2)}\)
and Bohr frequency \(\omega_{01}\) are distinct: their ratio
determines the maximum transfer probability \(P_{\max}^{(2)}\),
while \(\omega_{01}\) sets the oscillation frequency.
\par\endgroup

\paragraph{Dissipation strength \(\gamma_{\mathcal D}^{(2)}\).}
The dissipator \(\mathcal D_2\) acts on all operator directions,
rather than a single pair of states. We measure its overall action
with the induced Hilbert--Schmidt norm
\begin{align}
 \gamma_{\mathcal D}^{(2)}
 &=\|\mathcal D_2\|_{2\to2}
 =\sup_{B\ne0}\frac{\|\mathcal D_2(B)\|_{\mathrm{HS}}}
                       {\|B\|_{\mathrm{HS}}},
 \qquad
 \|B\|_{\mathrm{HS}}=\sqrt{\operatorname{Tr}(B^\dagger B)}.
 \label{eq:two_state_dissipation_norm}
\end{align}
Here \(2\to2\) uses the Hilbert--Schmidt norm for both input
and output, and the supremum ranges over all nonzero operators
\(B\) on \(\operatorname{Ran}P_2\), not only density operators.
Equivalently, \(\gamma_{\mathcal D}^{(2)}\) is the largest singular
value of a Hilbert--Schmidt orthonormal matrix representation of
\(\mathcal D_2\).

In the symmetric model, the transverse components \(u_x,u_y\)
are damped with coefficients \(4b^2,4a_x^2\). Their relative
contrast defines the dissipative anisotropy
\begin{align}
 \delta_{\mathcal D}=\frac{a_x^2-b^2}{a_x^2+b^2},
 \qquad a_x^2+b^2>0,\qquad -1\leq\delta_{\mathcal D}\leq1.
 \label{eq:two_state_dissipative_anisotropy}
\end{align}
Thus balanced damping gives \(\delta_{\mathcal D}=0\);
\(|\delta_{\mathcal D}|\) near one means that one transverse
component is much less directly damped than the other.

\begin{theorem_}[Exact symmetric spectral window \(\chi_{\max}>\chi_c\)]
\label{thm:two_state_lower_edge}
Fix a symmetric two-state model \(\mathcal L_2^{\mathrm{TP}}\) with
\(\Omega_{\mathrm{tun}}^{(2)}=\omega_{01}>0\),
\(\gamma_{\mathcal D}^{(2)}>0\),
\(\delta_{\mathcal D}\in[-1,1]\), and \(1/3<\chi_c<1\).
The following condition is necessary and sufficient for the spectral
window \(\chi_{\max}>\chi_c\):
\begin{align}
 \chi_{\max}>\chi_c
 \quad\Longleftrightarrow\quad
 \delta_{\mathcal D}^{2}>
 \chi_c^2+4(2-\delta_{\mathcal D}^{2})
 \left(\frac{\Omega_{\mathrm{tun}}^{(2)}}
            {\gamma_{\mathcal D}^{(2)}}\right)^2.
 \label{eq:two_state_spectral_iff}
\end{align}
Equality in the parameter inequality in \cref{eq:two_state_spectral_iff}
gives
\(\chi_{\max}=\chi_2=\chi_c\). Along common positive rescaling
of the jump coefficients \(\mu,\nu\), the strict spectral window
\(\chi_{\max}>\chi_c\), once entered, has no finite upper exit.
\end{theorem_}

The criterion expresses the competition between Hamiltonian rotation
of \(u_x,u_y\) at frequency \(\Omega_{\mathrm{tun}}^{(2)}\)
and their unequal damping. Fast rotation prevents their decay rates
from separating; sufficient damping contrast and dissipation relative
to tunneling produce one slow and one fast transverse mode.
Increasing \(\gamma_{\mathcal D}^{(2)}\) cannot produce the
window if \(|\delta_{\mathcal D}|\leq\chi_c\).
Because \cref{eq:two_state_spectral_iff} is both
necessary and sufficient for \(\mathcal L_2^{\mathrm{TP}}\), it
identifies both the presence and absence of the spectral window
\(\chi_{\max}>\chi_c\), rather than only a sufficient regime.
The proof is given in \cref{app:two_state_lower_edge}.

The limits of \cref{eq:two_state_spectral_iff} provide useful checks.
Balanced damping \(\delta_{\mathcal D}=0\) gives
\(\chi_{\max}=1/3\) at every positive dissipation strength.
At fixed positive \(\Omega_{\mathrm{tun}}^{(2)},
\gamma_{\mathcal D}^{(2)}\), the condition fails as
\(|\delta_{\mathcal D}|\to\chi_c^{+}\): the damping contrast
can no longer exceed the threshold \(\chi_c\) and the coherent-mixing
contribution. Conversely, for fixed \(|\delta_{\mathcal D}|>\chi_c\)
and \(\gamma_{\mathcal D}^{(2)}>0\), it holds for every
sufficiently small \(\Omega_{\mathrm{tun}}^{(2)}>0\), because
the coherent-mixing contribution tends to zero. The relevant
comparison is the ratio
\(\Omega_{\mathrm{tun}}^{(2)}/\gamma_{\mathcal D}^{(2)}\),
not either magnitude alone. The limit
\(\Omega_{\mathrm{tun}}^{(2)}\to0^{+}\) is taken through positive frequencies;
an exactly vanishing frequency can change the stationary-sector
dimension and is excluded from the theorem.

A finite potential offset \(E_\delta\) introduces the coefficients
\(a_0,a_z\) and couplings \(h_\pm\), which mix \(u_z\) with
\(u_x\) in \cref{eq:biased_bloch_matrix_abbreviated}. We retain
the anisotropy \(\delta_{\mathcal D}\) in
\cref{eq:two_state_dissipative_anisotropy}, but for \(a_z\ne0\)
it no longer describes all transverse damping: both
\(\alpha_A,\beta_A\) contain \(4a_z^2\).
We therefore hold the entire normalized dissipator
\(\mathcal D_2/\gamma_{\mathcal D}^{(2)}\) fixed when varying
its magnitude. {The next result gives necessary and sufficient
tests within sufficiently weak or strong dissipation, provided the
corresponding limiting separation differs strictly from \(\chi_c\).}

\begin{theorem_}[Biased two-state spectral-window criteria]
\label{thm:biased_two_state_window}
Fix a two-state QLD model with potential \(V(x)\),
\(\omega_{01}>0\), and coefficients \(a_x^2+a_z^2>0\), \(b\ne0\).
Vary the dissipation strength \(\gamma_{\mathcal D}^{(2)}>0\)
by common rescaling of \(\mu,\nu\), keeping
\(\mathcal D_2/\gamma_{\mathcal D}^{(2)}\) fixed.
Fix \(1/3<\chi_c<1\). Each assertion below excludes equality
in its right-hand parameter comparison.

\emph{(i) Weak dissipation.} There exists \(\varepsilon_{\mathrm w}>0\)
such that, for every
\(0<\gamma_{\mathcal D}^{(2)}/\omega_{01}<\varepsilon_{\mathrm w}\),
\begin{align}
 \chi_{\max}>\chi_c
 \quad\Longleftrightarrow\quad
 (1+\delta_{\mathcal D})
 \bigl[(\omega_{01}/\Omega_{\mathrm{tun}}^{(2)})^2-1\bigr]
 >\frac{1+3\chi_c}{1-\chi_c}.
 \label{eq:biased_weak_physical_criterion}
\end{align}

\emph{(ii) Strong dissipation.} There exists \(\varepsilon_{\mathrm s}>0\)
such that, for every
\(0<\omega_{01}/\gamma_{\mathcal D}^{(2)}<\varepsilon_{\mathrm s}\),
\begin{align}
 \chi_{\max}>\chi_c
 \quad\Longleftrightarrow\quad
 \chi_{\max}(\mathcal D_2/\gamma_{\mathcal D}^{(2)})>\chi_c.
 \label{eq:biased_strong_physical_criterion}
\end{align}
The thresholds \(\varepsilon_{\mathrm w},\varepsilon_{\mathrm s}\)
depend on the fixed model and \(\chi_c\), not on the varying
strength \(\gamma_{\mathcal D}^{(2)}\). At every
\(\gamma_{\mathcal D}^{(2)}>0\), \(\chi_3\leq1/3\), so any
spectral window \(\chi_{\max}>\chi_c\) has \(\chi_2>\chi_c\)
and \(i^\ast=2\).
\end{theorem_}

Here \(\chi_{\max}(\mathcal D_2/\gamma_{\mathcal D}^{(2)})\)
is computed from the three nonstationary decay rates of the
normalized dissipator \(\mathcal D_2/\gamma_{\mathcal D}^{(2)}\),
using \cref{def:chi_separation}. Its eigenvalues are given explicitly
in \cref{eq:biased_pure_dissipator_real_eigenvalue,eq:biased_pure_dissipator_spectrum}.
We prove both criteria in \cref{app:two_state_biased_limits}
by controlling the errors in the weak- and strong-dissipation limits,
which makes
\cref{eq:biased_weak_physical_criterion,eq:biased_strong_physical_criterion}
valid at finite positive \(\gamma_{\mathcal D}^{(2)}\).
Thus each parameter test is necessary and sufficient within its
specified weak- or strong-dissipation interval, and the reversed
strict parameter inequality excludes the spectral window there.
Neither criterion classifies
intermediate strengths, implies monotonicity, or covers a critical
threshold equality.

The weak-dissipation criterion depends on the normalized coupling frequency
\(\Omega_{\mathrm{tun}}^{(2)}/\omega_{01}\) and anisotropy
\(\delta_{\mathcal D}\), not the common scale
\(\gamma_{\mathcal D}^{(2)}\), because that
scale cancels from the leading decay-rate ratio.
A small ratio \(\Omega_{\mathrm{tun}}^{(2)}/\omega_{01}\) means large
diagonal position contrast \(|x_{00}-x_{11}|\) relative to
\(|x_{01}|\), as shown in \cref{eq:biased_tunneling_position_contrast}.
At fixed \(\delta_{\mathcal D}\), this increases coherence
damping relative to population relaxation and can create a window
even at arbitrarily weak positive dissipation, although both
absolute rates vanish in that limit.
As checks, \(\Omega_{\mathrm{tun}}^{(2)}=\omega_{01}\) makes the
left-hand parameter expression in
\cref{eq:biased_weak_physical_criterion} zero, excluding the weak-dissipation
window. In contrast,
\(\Omega_{\mathrm{tun}}^{(2)}/\omega_{01}\to0^{+}\) at fixed
\(\delta_{\mathcal D}>-1\) makes that expression diverge, so the
weak-dissipation criterion is satisfied.
This conditional parameter limit does not imply that varying \(V(x)\)
at fixed \(\mu,\nu\) preserves \(\delta_{\mathcal D}\).
The strong-dissipation criterion instead retains the full
normalized dissipative structure.

Together, these results identify two mechanisms: anisotropic damping
overcoming coherent mixing in the symmetric model, and unequal
population and coherence damping in the biased model.
The strong-dissipation behavior is set by
\(\mathcal D_2/\gamma_{\mathcal D}^{(2)}\), not its magnitude
alone. We next ask which connections remain valid when more
Hamiltonian eigenstates \(\ket{\phi_n}\) are retained.

\subsection{Multi-state QLD model for general potentials}
\label{sec:qld_general_origin}

Additional transfer channels can preserve or remove a spectral
bottleneck. Although \cref{thm:two_state_lower_edge} is exact within
\(\mathcal L_2^{\mathrm{TP}}\), a single coherent matrix element
\(J^{(2)}\) cannot describe all paths through additional Hamiltonian
eigenstates \(\ket{\phi_n}\). To track these channels for a general
one-dimensional potential \(V(x)\), we define the localized coupling
\(J^{(r)}\) and dissipator magnitude \(\gamma_{\mathcal D}^{(r)}\)
in the energy-projected model \(\mathcal L_r^{\mathrm{TP}}\).
Couplings between the position sectors \(P_L^{(r)},P_R^{(r)}\)
then give a sufficient condition for spectral separation. The
complementary exclusion condition determines when increasing
the dissipation strength \(\gamma_{\mathcal D}^{(r)}\) removes the
spectral window \(\chi_{\max}>\chi_c\). The exclusion criterion
concerns relative decay rates \(\Gamma_i\); its implications for
target accuracy and acceleration are considered in
\cref{sec:qld_general_control}.
The aim is a conditional extension, not a universal classification
by two scalar strengths.
No discrete spectrum for every potential \(V(x)\), or accuracy of every
finite projection \(P_r\), is presumed.

For the double-well potential \(V(x)\) in
\cref{eq:piecewise_double_well_potential_0710}, the distinction between
\(E_\delta=0\) and \(E_\delta\ne0\) remains important.
\Cref{thm:two_state_lower_edge} excludes a finite upper exit from
the spectral window \(\chi_{\max}>\chi_c\) in the symmetric
two-state generator \(\mathcal L_2^{\mathrm{TP}}\), whereas
\cref{thm:biased_two_state_window} permits such an exit for more general
coefficients \(a_0,a_x,a_z,b\). Retaining additional
Hamiltonian eigenstates \(\ket{\phi_n}\) therefore introduces further
transitions, but their participation is not a necessary explanation for
every upper exit from \(\chi_{\max}>\chi_c\).

\subsubsection{Multi-state QLD model}
\label{sec:higher_state_qld}

Let the Hamiltonian \(H\) have at least \(r\) discrete orthonormal
eigenstates \(\ket{\phi_n}\), or let \(H\) be a finite spatial
discretization of the Hamiltonian \(H\) in \cref{eq:qld_hamiltonian}. Order their energies
\(E_n\) as \(E_0<E_1\leq\cdots\leq E_{r-1}\), with integer \(r\geq2\)
not exceeding the available dimension, and assume all displayed matrix
elements are well defined. As in
\cref{eq:two_state_projector,eq:two_state_normalized_projection}, define
the energy projector \(P_r\) and restrict the density operator \(\rho_r\)
itself to \(\operatorname{Ran}P_r\):
\begin{align}
 P_r&=\sum_{n=0}^{r-1}\ket{\phi_n}\bra{\phi_n},
 \qquad H\ket{\phi_n}=E_n\ket{\phi_n},
 \label{eq:r_state_energy_projector}\\
 \rho_r&=P_r\rho_rP_r,\qquad \rho_r\succeq0,\qquad
 \operatorname{Tr}\rho_r=1.
 \label{eq:r_state_density_restriction}
\end{align}
For the jump operator \(L=\mu x+\im\nu p\), the projected Hamiltonian
\(H_r\), position operator \(x_r\), momentum operator \(p_r\), and jump
operator \(L^{(r)}\) are
\begin{align}
 H_r&=P_rHP_r,\qquad x_r=P_rxP_r,\qquad p_r=P_rpP_r,
 \label{eq:r_state_projected_operators}\\
 L^{(r)}&=P_rLP_r=\mu x_r+\im\nu p_r.
 \label{eq:r_state_projected_jump}
\end{align}
All products below act within \(\operatorname{Ran}P_r\). Replacing the
projected operators in \cref{eq:two_state_tp_generator} by
\cref{eq:r_state_projected_operators,eq:r_state_projected_jump} defines the
autonomous generator \(\mathcal L_r^{\mathrm{TP}}\) and dissipator
\(\mathcal D_r\):
\begin{align}
 \mathcal L_r^{\mathrm{TP}}(\rho_r)
 &=-\frac{\im}{\hbar}[H_r,\rho_r]+\mathcal D_r(\rho_r), \label{eq:r_state_tp_generator}\\
 \mathcal D_r(\rho_r)
 &=2L^{(r)}\rho_rL^{(r)\dagger}
   -\{L^{(r)\dagger}L^{(r)},\rho_r\}.
 \label{eq:r_state_dissipator}
\end{align}
The generator \(\mathcal L_r^{\mathrm{TP}}\) preserves positivity and trace
by its Lindblad form.
In general, \(\rho_r(t)\) denotes the
autonomous solution of \cref{eq:r_state_tp_generator}, not the normalized
compression of the full solution \(\rho(t)\) at each time. Increasing
\(r\) retains more Hamiltonian eigenstates \(\ket{\phi_n}\), but does not
by itself prove monotone convergence of the decay rates \(\Gamma_i\) or
accuracy of the full density operator \(\rho(t)\). For a fixed spatial
discretization of dimension \(N\), taking \(r=N\) gives \(P_r=I\)
and \(\mathcal L_r^{\mathrm{TP}}=\mathcal L\).

To extend the coupling frequency \(\Omega_{\mathrm{tun}}^{(2)}\) from
\cref{eq:two_state_tunneling_scale}, we first specify which localized states
\(\ket{\psi_L^{(r)}},\ket{\psi_R^{(r)}}\) are compared. We retain the
dividing point \(x=0\) from \cref{eq:two_state_localized_basis} and use
the spectral subspaces of the projected position operator \(x_r\) on its
two sides. For a multiwell potential \(V(x)\), this choice specifies two
groups of wells, rather than a unique description of all possible
transitions.

\begin{definition_}[Localized-state coupling frequency \(\Omega_{\mathrm{tun}}^{(r)}\)]
\label{def:r_state_tunneling}
Define the spectral projectors \(P_L^{(r)},P_R^{(r)}\) within
\(\operatorname{Ran}P_r\) by
\begin{align}
 P_L^{(r)}&=\mathbf1_{(-\infty,0)}(x_r),\qquad
 P_R^{(r)}=\mathbf1_{[0,\infty)}(x_r),\qquad
 P_L^{(r)}+P_R^{(r)}=P_r.
 \label{eq:r_state_spatial_projectors}
\end{align}
Suppose both ranges \(\operatorname{Ran}P_\alpha^{(r)}\) are nonempty
and each restricted Hamiltonian
\(P_\alpha^{(r)}H_rP_\alpha^{(r)}|_{\operatorname{Ran}P_\alpha^{(r)}}\)
has a nondegenerate normalized ground state \(\ket{\psi_\alpha^{(r)}}\),
\(\alpha\in\{L,R\}\). The coupling energy \(J^{(r)}\) and coupling
frequency \(\Omega_{\mathrm{tun}}^{(r)}\) are
\begin{align}
 J^{(r)}
 &=\bigl|\langle\psi_L^{(r)}|H_r|\psi_R^{(r)}\rangle\bigr|,
 \qquad \Omega_{\mathrm{tun}}^{(r)}\coloneqq\frac{2J^{(r)}}{\hbar}.
 \label{eq:r_state_coupling_frequency}
\end{align}
\end{definition_}

The projectors \(P_L^{(r)},P_R^{(r)}\) assign
\(\ker x_r\) to \(\operatorname{Ran}P_R^{(r)}\) by convention. The local
ground states \(\ket{\psi_\alpha^{(r)}}\) minimize the mean energy
\(\langle\psi|H_r|\psi\rangle\) among normalized vectors
\(\ket\psi\in\operatorname{Ran}P_\alpha^{(r)}\); they are not, in
general, eigenstates of \(H_r\). When both ranges
\(\operatorname{Ran}P_L^{(r)},\operatorname{Ran}P_R^{(r)}\) are
one-dimensional and the eigenvalues of \(x_2\) have opposite signs,
\cref{def:r_state_tunneling} recovers
\cref{eq:two_state_localized_basis,eq:two_state_tunneling_scale}. For
\(r>2\), the states \(\ket{\psi_L^{(r)}},\ket{\psi_R^{(r)}}\) depend
on \(r\) and need not span an invariant subspace of \(H_r\). Thus
\(\Omega_{\mathrm{tun}}^{(r)}\) measures their direct
coupling, but does not determine the full transfer dynamics: other
retained states can also participate. At fixed potential \(V(x)\), retained
dimension \(r\), and projectors \(P_L^{(r)},P_R^{(r)}\), the coupling
\(J^{(r)}\) is independent of \(\mu,\nu\), although changing the jump
coefficients \(\mu,\nu\) changes the generator
\(\mathcal L_r^{\mathrm{TP}}\) and hence its population transfer.

We next extend the dissipator norm \(\gamma_{\mathcal D}^{(2)}\) from
\cref{eq:two_state_dissipation_norm}. Unlike the coupling \(J^{(r)}\),
which selects the matrix element
\(\langle\psi_L^{(r)}|H_r|\psi_R^{(r)}\rangle\), the norm
\(\gamma_{\mathcal D}^{(r)}\) must include the
action of \(\mathcal D_r\) on all populations
\(\langle\phi_n|\rho_r|\phi_n\rangle\) and coherences
\(\langle\phi_n|\rho_r|\phi_m\rangle\), \(n\ne m\), in
\(\operatorname{Ran}P_r\).

\begin{definition_}[Reduced dissipation strength \(\gamma_{\mathcal D}^{(r)}\)]
\label{def:r_state_dissipation}
For the dissipator \(\mathcal D_r\) in \cref{eq:r_state_dissipator}, define
\begin{align}
 \gamma_{\mathcal D}^{(r)}
 &\coloneqq\|\mathcal D_r\|_{2\to2}
 =\sup_{B\ne0}\frac{\|\mathcal D_r(B)\|_{\mathrm{HS}}}
                         {\|B\|_{\mathrm{HS}}}, \qquad 
 \frac{\gamma_{\mathcal D}^{(r)}}{\omega_{01}}
 =\frac{\|\mathcal D_r\|_{2\to2}}{(E_1-E_0)/\hbar}.
 \label{eq:r_state_dissipation_intensity}
\end{align}
Here the Hilbert--Schmidt norm
\(\|B\|_{\mathrm{HS}}=\sqrt{\operatorname{Tr}(B^\dagger B)}\)
is used at both input and output, and the supremum is over all nonzero
operators \(B=P_rBP_r\). The relative dissipation strength
\(\gamma_{\mathcal D}^{(r)}/\omega_{01}\) uses the
Bohr frequency \(\omega_{01}=(E_1-E_0)/\hbar>0\) from
\cref{eq:energy_basis_matrix_elements}.
\end{definition_}

In the Hilbert--Schmidt orthonormal basis
\(\{\ket{\phi_n}\bra{\phi_m}:0\leq n,m<r\}\), the norm
\(\gamma_{\mathcal D}^{(r)}\) is the largest singular value of the
\(r^2\times r^2\) matrix representing \(\mathcal D_r\). It bounds
\(\|\mathcal D_r(B)\|_{\mathrm{HS}}\) by
\(\gamma_{\mathcal D}^{(r)}\|B\|_{\mathrm{HS}}\), but is not an
individual decay rate \(\Gamma_i\) of \(\mathcal L_r^{\mathrm{TP}}\)
or a transfer rate between \(\operatorname{Ran}P_L^{(r)}\) and
\(\operatorname{Ran}P_R^{(r)}\). Its dependence on \(r\) also reflects
the change in \(\operatorname{Ran}P_r\), not necessarily a change in
the jump coefficients \(\mu,\nu\).

Here \(\|\cdot\|_{\mathrm{op}}\) is the operator norm induced by the
Hilbert-space vector norm \(\|\cdot\|_2\). Define the Hamiltonian
generator \(\mathcal K_r\) by
\begin{align}
 \mathcal K_r(B)&=-\frac{\im}{\hbar}[H_r,B],\qquad
 \mathcal L_r^{\mathrm{TP}}=\mathcal K_r+\mathcal D_r.
 \label{eq:r_state_hamiltonian_generator}
\end{align}

Spectral formation requires control of all
relevant operator directions, as addressed in
\cref{thm:qld_sector_origin}. We first record the Hamiltonian bandwidth
\(\omega_{\mathrm{bw}}^{(r)}\), which distinguishes a total spectral
scale from the single coupling \(J^{(r)}\).

{For the spectral comparison, substitute
\(H_r\ket{\phi_n}=E_n\ket{\phi_n}\) into
\cref{eq:r_state_hamiltonian_generator}. The matrix units
\(\ket{\phi_n}\bra{\phi_m}\), \(0\leq n,m<r\), form a
Hilbert--Schmidt orthonormal basis, so the norm of the diagonal
action of \(\mathcal K_r\) is the largest magnitude of its entries:}
\begin{align}
 \mathcal K_r(\ket{\phi_n}\bra{\phi_m})
 &=-\frac{\im(E_n-E_m)}{\hbar}\ket{\phi_n}\bra{\phi_m},
 \label{eq:r_state_hamiltonian_matrix_units}\\
 \|\mathcal K_r\|_{2\to2}
 &=\max_{0\leq n,m<r}\frac{|E_n-E_m|}{\hbar}
 =\frac{E_{r-1}-E_0}{\hbar}
 \coloneqq\omega_{\mathrm{bw}}^{(r)}>0.
 \label{eq:r_state_hamiltonian_bandwidth_norm}
\end{align}
The frequency \(\omega_{\mathrm{bw}}^{(r)}\) is the full retained
energy bandwidth divided by \(\hbar\). To compare it with the coupling
frequency \(\Omega_{\mathrm{tun}}^{(r)}\), use
\(\langle\psi_L^{(r)}|P_r|\psi_R^{(r)}\rangle=0\) and the spectral
bounds \(E_0P_r\preceq H_r\preceq E_{r-1}P_r\):
\begin{align}
 &\begin{aligned}[b]
 J^{(r)}
 &=\left|\left\langle\psi_L^{(r)}\left|
 H_r-\frac{E_{r-1}+E_0}{2}P_r
 \right|\psi_R^{(r)}\right\rangle\right|\\
 &\leq\left\|H_r-\frac{E_{r-1}+E_0}{2}P_r\right\|_{\mathrm{op}}
 =\frac{E_{r-1}-E_0}{2}.
 \end{aligned}
 \label{eq:r_state_centered_coupling_bound}
\end{align}
Combining \cref{eq:r_state_coupling_frequency,eq:r_state_hamiltonian_bandwidth_norm,eq:r_state_centered_coupling_bound}
yields
\begin{align}
 \|\mathcal K_r\|_{2\to2}
 =\omega_{\mathrm{bw}}^{(r)}
 =\frac{E_{r-1}-E_0}{\hbar},
 \qquad \Omega_{\mathrm{tun}}^{(r)}\le\omega_{\mathrm{bw}}^{(r)}.
 \label{eq:r_state_bandwidth}
\end{align}
Thus a small coupling frequency \(\Omega_{\mathrm{tun}}^{(r)}\) does
not imply a small bandwidth \(\omega_{\mathrm{bw}}^{(r)}\).
{For \(r=2\), \cref{eq:r_state_hamiltonian_bandwidth_norm}
reduces to \(\|\mathcal K_2\|_{2\to2}=\omega_{01}\); for
\(r>2\), the norm \(\|\mathcal K_r\|_{2\to2}\) also accounts
for the higher Bohr frequencies \(|E_n-E_m|/\hbar\).}
Accordingly, the spectral-exclusion estimate in
\cref{thm:qld_strong_dissipation_exclusion} uses
\(\omega_{\mathrm{bw}}^{(r)}\), rather than only
\(\Omega_{\mathrm{tun}}^{(r)}\). The formation condition below instead
compares intersector coupling with internal relaxation.

\subsubsection{Weak tunneling and strong dissipation}
\label{sec:qld_origin_certificate}

A spectral window \(\chi_{\max}>\chi_c\) concerns the entire generator
\(\mathcal L_r^{\mathrm{TP}}\). A small matrix element \(J^{(r)}\)
cannot control transitions through the other retained states
\(\ket{\phi_n}\). We therefore use the total intersector coupling
frequency \(\Omega_{\mathrm{tun}}^{\mathrm{op}}\), which includes
all coherent paths between the position sectors
\(P_L^{(r)},P_R^{(r)}\) from \cref{eq:r_state_spatial_projectors}:
\begin{align}
 \Omega_{\mathrm{tun}}^{\mathrm{op}}
 &\coloneqq\frac{2\|P_L^{(r)}H_rP_R^{(r)}\|_{\mathrm{op}}}{\hbar}
 \geq\Omega_{\mathrm{tun}}^{(r)}.
 \label{eq:qld_total_intersector_frequency}
\end{align}
For \(r=2\), each sector is one-dimensional and
\(\Omega_{\mathrm{tun}}^{\mathrm{op}}=\Omega_{\mathrm{tun}}^{(2)}\).
For \(r>2\), controlling this total frequency is stronger than
suppressing only the selected-pair frequency
\(\Omega_{\mathrm{tun}}^{(r)}\).

To distinguish transfer between wells from local relaxation, define
\(H_{0,r}=\sum_{\alpha=L,R}P_\alpha^{(r)}H_rP_\alpha^{(r)}\) and
\(L_{0,r}=\sum_{\alpha=L,R}P_\alpha^{(r)}L^{(r)}P_\alpha^{(r)}\).
For any jump operator \(B\), write
\(\mathcal D[B](\rho)=2B\rho B^\dagger-\{B^\dagger B,\rho\}\).
The reference generator \(\mathcal L_{0,r}\) removes all
Hamiltonian and dissipative transfer between the two sectors:
\begin{align}
 \mathcal L_{0,r}(\rho)
 &=-\frac{\im}{\hbar}[H_{0,r},\rho]+\mathcal D[L_{0,r}](\rho).
 \label{eq:qld_decoupled_reference}
\end{align}
The reference generator \(\mathcal L_{0,r}\) conserves each population
\(\operatorname{Tr}(P_\alpha^{(r)}\rho)\). It therefore has at least
two stationary operator directions. Exactly two directions require
equilibration within each sector \(P_\alpha^{(r)}\) and decay of the
intersector coherences \(P_L^{(r)}\rho P_R^{(r)}\); neither property
follows merely from a double-well shape of \(V(x)\).

In the following comparison, fix \(H_{0,r},x_r,\mu,\nu\), vary only
the Hamiltonian couplings between the sectors, and impose
\(p_r=\im m[H_r,x_r]/\hbar\), as in
\cref{eq:projected_momentum_identity}. This keeps \(L_{0,r}\) and
\(\mathcal L_{0,r}\) fixed and ties dissipative intersector transfer
to \(\Omega_{\mathrm{tun}}^{\mathrm{op}}\).
Thus weak tunneling suppresses both transfer mechanisms without
requiring weak local dissipation.
The commutator assumption must also hold for a spatial discretization;
an arbitrary change of potential \(V(x)\) need not preserve the fixed
local dynamics \(\mathcal L_{0,r}\).

Assume that the disconnected generator \(\mathcal L_{0,r}\)
relaxes within each sector \(P_\alpha^{(r)}\) to a unique normalized
state \(\rho_\alpha^{(r)}\), and damps all intersector coherences
\(P_L^{(r)}\rho P_R^{(r)}\). Thus
\(\ker\mathcal L_{0,r}
=\operatorname{span}\{\rho_L^{(r)},\rho_R^{(r)}\}\), and its
minimum relaxation gap \(\Delta_{\min}^{(r)}\) is
\begin{align}
 \Delta_{\min}^{(r)}
 =\min_{z\in\sigma(\mathcal L_{0,r})\setminus\{0\}}
       [-\operatorname{Re}z]>0.
 \label{eq:qld_minimum_relaxation_gap}
\end{align}
This is a decay-rate gap, not a Hamiltonian energy gap: it measures
the slowest remaining local relaxation or intersector-coherence decay.
The assumptions imply \(\|\mathcal D[L_{0,r}]\|_{2\to2}>0\),
but do not require diagonalizability of \(\mathcal L_{0,r}\).

\Needspace{15\baselineskip}
\begin{theorem_}[Weak tunneling relative to local relaxation]
\label{thm:qld_sector_origin}
Under the fixed-data, commutator, and local-relaxation assumptions
above, for every \(0<\chi_c<1\) there exist dimensionless constants
\(\varepsilon,C>0\) such that every admissible generator
\(\mathcal L_r^{\mathrm{TP}}\) with a simple stationary eigenvalue
\(0\) satisfies
\begin{align}
 0<\frac{\Omega_{\mathrm{tun}}^{\mathrm{op}}}{\Delta_{\min}^{(r)}}
 <\varepsilon
 \quad\Longrightarrow\quad \chi_{\max}\geq\chi_2>\chi_c.
 \label{eq:qld_weak_tunneling_epsilon_window}
\end{align}
In this range, the decay rates \(\Gamma_i\) obey
\begin{align}
 0<\Gamma_2\leq C\frac{(\Omega_{\mathrm{tun}}^{\mathrm{op}})^2}
                            {\Delta_{\min}^{(r)}},\qquad
 \Gamma_i\geq\frac{\Delta_{\min}^{(r)}}2,
 \quad i=3,\ldots,r^2.
 \label{eq:qld_weak_tunneling_fast_rate_bound}
\end{align}
\end{theorem_}

The physical mechanism is slow population redistribution between
otherwise relaxing sectors, as in \cite{macieszczak2016towards}.
The intersector coherences \(P_L^{(r)}\rho P_R^{(r)}\) respond at
first order in \(\Omega_{\mathrm{tun}}^{\mathrm{op}}\), whereas
population redistribution enters at second order. The proof in
\cref{app:proof_qld_sector_origin} establishes this second-order
mechanism. Consequently,
\(\Gamma_2\) becomes quadratically small while the remaining
decay rates \(\Gamma_i\), \(i\geq3\), stay bounded below by
\(\Delta_{\min}^{(r)}/2\). This is a transfer bottleneck, not
uniformly slow dissipation: the strength \(\gamma_{\mathcal D}^{(r)}\)
remains positive. At exactly \(\Omega_{\mathrm{tun}}^{\mathrm{op}}=0\),
the two stationary directions persist instead of producing a
relaxing system. For nonzero coupling, whether the spectral separation
produces an observable plateau in \(\rho_r(t)\) further depends on
the initial-state and finite-time conditions in
\cref{sec:metastability_control}.

The formation mechanism in \cref{thm:qld_sector_origin} weakens
intersector transfer while retaining local relaxation. Increasing
the dissipation strength \(\gamma_{\mathcal D}^{(r)}\) at fixed
Hamiltonian \(H_r\) is a different operation: both local damping
and dissipative intersector transfer can grow. To determine whether
this operation removes the spectral window \(\chi_{\max}>\chi_c\),
separate the dissipator magnitude \(\gamma_{\mathcal D}^{(r)}>0\)
from the unit-norm dissipator \(\widehat{\mathcal D}_r\):
\begin{align}
 \widehat{\mathcal D}_r
 &=\frac{\mathcal D_r}{\gamma_{\mathcal D}^{(r)}},\qquad
 \|\widehat{\mathcal D}_r\|_{2\to2}=1.
 \label{eq:r_state_unit_norm_dissipator}
\end{align}
Common rescaling of the jump coefficients \(\mu,\nu\) varies
\(\gamma_{\mathcal D}^{(r)}\) while keeping
\(H_r,\widehat{\mathcal D}_r\) fixed. It also leaves the
coupling frequency \(\Omega_{\mathrm{tun}}^{(r)}\)
unchanged.

Assume the normalized dissipator \(\widehat{\mathcal D}_r\) has
a simple stationary eigenvalue \(\zeta_1=0\), and that all its
other eigenvalues \(\zeta_i\) have strictly negative real parts.
Order its dimensionless decay rates \(d_i\) with algebraic
multiplicity as
\begin{align}
 0=d_1<d_2\leq\cdots\leq d_{r^2},\qquad
 d_i=-\operatorname{Re}\zeta_i.
 \label{eq:r_state_normalized_dissipator_rates}
\end{align}
Unlike the three rates \(d_i\) in
\cref{eq:biased_normalized_decay_rates}, this list includes the
stationary rate \(d_1=0\), consistently with
\cref{eq:decay_rates_Gamma}. The spectral separation
\(\chi_{\max}(\widehat{\mathcal D}_r)\) is therefore defined
by \cref{def:chi_separation} using the nonstationary rates
\(d_2,\ldots,d_{r^2}\). No diagonalizability is assumed.

\begin{theorem_}[Strong-dissipation exclusion of \(\chi_{\max}>\chi_c\)]
\label{thm:qld_strong_dissipation_exclusion}
Fix the Hamiltonian \(H_r\), with \(\omega_{\mathrm{bw}}^{(r)}>0\),
and normalized dissipator
\(\widehat{\mathcal D}_r\) satisfying the above spectral
assumptions. If
\(\chi_{\max}(\widehat{\mathcal D}_r)<\chi_c<1\), then there
exists \(\varepsilon>0\), independent of the dissipation strength
\(\gamma_{\mathcal D}^{(r)}\), such that
\begin{align}
 0<\frac{\omega_{\mathrm{bw}}^{(r)}}
           {\gamma_{\mathcal D}^{(r)}}<\varepsilon
 \quad\Longrightarrow\quad
 \chi_{\max}(\mathcal L_r^{\mathrm{TP}})<\chi_c.
 \label{eq:qld_strong_dissipation_epsilon_exit}
\end{align}
Here \(\omega_{\mathrm{bw}}^{(r)}\) is the Hamiltonian bandwidth
frequency in \cref{eq:r_state_hamiltonian_bandwidth_norm}.
Thus strong dissipation excludes the spectral window when the
normalized dissipator \(\widehat{\mathcal D}_r\) itself has no
separation above the threshold \(\chi_c\).
\end{theorem_}

This QLD criterion specializes
\cref{thm:general_strong_dissipation_exclusion} to the retained
space \(\operatorname{Ran}P_r\). A proof of this specialization
appears in \cref{app:proof_qld_strong_dissipation}.
The physical distinction from \cref{thm:qld_sector_origin} is that
the dissipator \(\widehat{\mathcal D}_r\) now governs the entire
decay spectrum, including intersector redistribution. If its
nonstationary rates \(d_i\) have no large adjacent separation,
the Hamiltonian term \(\mathcal K_r\) cannot sustain a spectral
window \(\chi_{\max}>\chi_c\) once its relative norm is below
the threshold \(\varepsilon\). The bandwidth frequency
\(\omega_{\mathrm{bw}}^{(r)}\), rather than only the tunneling
frequency \(\Omega_{\mathrm{tun}}^{(r)}\), enters because
\(\mathcal K_r\) acts on all retained operator directions.
If the spectrum of \(\mathcal L_r^{\mathrm{TP}}\) lies inside
the window at some finite \(\gamma_{\mathcal D}^{(r)}\), the
theorem guarantees its eventual exit, but not monotonicity or a
unique crossing. Faster convergence to the same target
\(\rho_{\mathrm{tar}}\) at a fixed accuracy \(\varepsilon\)
requires the additional tests in
\cref{sec:qld_general_control}.

This mechanism explains what must change beyond the symmetric
two-state model \(\mathcal L_2^{\mathrm{TP}}\). In its nonempty
spectral window, \(1/3<\chi_c<|\delta_{\mathcal D}|\),
\cref{thm:two_state_lower_edge} excludes an upper exit. When
\(\widehat{\mathcal D}_2\) has a simple stationary eigenvalue,
its limiting separation is
\(\chi_{\max}(\widehat{\mathcal D}_2)=|\delta_{\mathcal D}|
>\chi_c\); at \(|\delta_{\mathcal D}|=1\), its stationary
eigenvalue is degenerate instead. Neither case satisfies the
strong-dissipation exclusion hypotheses. Additional states
\(\ket{\phi_n}\) change the dissipator
\(\widehat{\mathcal D}_r\), so a model with \(r>2\) can
satisfy these hypotheses, but increasing \(r\) alone does not
guarantee them. A biased two-state model can also admit an upper
exit, as shown by \cref{thm:biased_two_state_window}.

{Strong dissipation need not remove the spectral window.
If \(\chi_{\max}(\widehat{\mathcal D}_r)>\chi_c\),
\cref{eq:projected_dissipative_limit} instead preserves that window
at large \(\gamma_{\mathcal D}^{(r)}\); equality requires a separate
analysis.} A degenerate stationary
space \(\ker\widehat{\mathcal D}_r\) can retain slow dynamics,
as in strongly dissipative steady-state-manifold constructions
\cite{zanardi2014coherent}. The theorem is valid for each fixed
finite projection \(P_r\), or a fixed full spatial discretization,
and does not assert a cutoff-independent threshold
\(\varepsilon\) in the continuum limit.

\subsection{Metastability in general Lindbladian dynamics}
\label{sec:general_origin}

Weak coupling between independently relaxing sectors is an established
source of metastability \cite{macieszczak2016towards,macieszczak2021classical}.
We now isolate the assumptions behind the QLD mechanisms that do not
depend on its particular jump operators. The stationary-sector
formulation gives a general condition for the spectral window
\(\chi_{\max}>\chi_c\), with \cref{thm:qld_sector_origin}
as a concrete realization. The strong-dissipation criterion then
identifies when a fixed dissipator \(\mathcal D\) excludes this
window, {providing the general result underlying
\cref{thm:qld_strong_dissipation_exclusion}}.
Throughout, the Lindbladian
\(\mathcal L\) acts on the finite-dimensional operator space
\(\mathcal B(\mathbb C^{d_H})\), and the decomposition
\(\mathcal L=-\im[H,\cdot]/\hbar+\mathcal D\) is fixed as part of
the physical model.

Let the reference Lindbladian \(\mathcal L_0\) have a stationary space
\(\ker\mathcal L_0\) of dimension \(2\leq M<d_H^2\), with every
nonzero eigenvalue \(z\) in the open left half-plane. Define the
minimum relaxation gap \(\Delta_{\min}\) by
\begin{align}
 \Delta_{\min}
 &=\min_{z\in\sigma(\mathcal L_0)\setminus\{0\}}
       [-\operatorname{Re}z]>0.
 \label{eq:general_minimum_relaxation_gap}
\end{align}
This extends the relaxation gap \(\Delta_{\min}^{(r)}\) in
\cref{eq:qld_minimum_relaxation_gap}. For example, two independently
relaxing sectors with stationary states \(\rho_L,\rho_R\) have
\(M=2\) if all other operator directions decay. Their mixtures
\(p\rho_L+(1-p)\rho_R\), \(0\leq p\leq1\), are stationary.
Thus \(M\) counts stationary operator directions, rather than
Hamiltonian energy levels.

Write \(\mathcal L=\mathcal L_0+\mathcal E\), where both
\(\mathcal L_0,\mathcal L\) are Lindbladians but the perturbation
\(\mathcal E\) need not be. With
\(\mathcal L_0=-\im[H_0,\cdot]/\hbar+\mathcal D_0\), the induced
Hilbert--Schmidt norm in \cref{def:r_state_dissipation} gives
\begin{align}
 \mathcal E(B)
 &=-\frac{\im}{\hbar}[H-H_0,B]+(\mathcal D-\mathcal D_0)(B),
 \label{eq:general_physical_perturbation}\\
 \|\mathcal E\|_{2\to2}
 &\leq\frac{2\|H-H_0\|_{\mathrm{op}}}{\hbar}
              +\|\mathcal D-\mathcal D_0\|_{2\to2}.
 \label{eq:general_physical_coupling_bound}
\end{align}
Here the commutator bound follows from
\(\|VB-BV\|_{\mathrm{HS}}\leq
2\|V\|_{\mathrm{op}}\|B\|_{\mathrm{HS}}\), with \(V=H-H_0\).
Only the change \(\mathcal D-\mathcal D_0\) enters the perturbation
\(\mathcal E\); the total dissipation strength
\(\gamma_{\mathcal D}=\|\mathcal D\|_{2\to2}\) may remain finite.

For specified orthogonal sector projections \(P_L+P_R=I\), choose
\(H_0=P_LHP_L+P_RHP_R\). The  {total intersector coupling frequency}
\(\Omega_{\mathrm{tun}}^{\mathrm{op}}\) is
\begin{align}
 \Omega_{\mathrm{tun}}^{\mathrm{op}}
 &=\frac{2\|P_LHP_R\|_{\mathrm{op}}}{\hbar}
 =\frac{2\|H-H_0\|_{\mathrm{op}}}{\hbar}.
 \label{eq:general_intersector_tunneling_frequency}
\end{align}
The equality follows from the off-diagonal block structure, as in
\cref{eq:qld_sector_coupling_norm}. For the QLD sectors
\(P_L^{(r)},P_R^{(r)}\), this gives
the frequency \(\Omega_{\mathrm{tun}}^{\mathrm{op}}\) in
\cref{eq:qld_total_intersector_frequency}: it includes all direct coherent
paths between the sectors, whereas the selected-pair frequency
\(\Omega_{\mathrm{tun}}^{(r)}\) in \cref{eq:r_state_coupling_frequency}
measures one such pair.

The weak-perturbation construction of
\cite{macieszczak2016towards,macieszczak2021classical} separates the
eigenvalues emerging from the stationary eigenvalue \(0\) from the
remaining spectrum. The following theorem expresses this separation
using \(\chi_M\) from \cref{def:chi_separation}. We call a
Lindbladian \(\mathcal L\) relaxing when its stationary eigenvalue
\(0\) is simple and all other decay rates \(\Gamma_i\) are positive.

\begin{theorem_}[Spectral separation \(\chi_M\) from weak perturbations \(\mathcal E\)]
\label{thm:weak_sector_certificate}
For the reference Lindbladian \(\mathcal L_0\) above and every
\(0<\chi_c<1\), there exists \(\varepsilon>0\), depending only on
\(\mathcal L_0,\chi_c\), such that every relaxing Lindbladian
\(\mathcal L=\mathcal L_0+\mathcal E\) satisfies
\begin{align}
 0<\frac{\|\mathcal E\|_{2\to2}}{\Delta_{\min}}<\varepsilon
 \quad\Longrightarrow\quad
 \chi_{\max}(\mathcal L)\geq\chi_M(\mathcal L)>\chi_c.
 \label{eq:general_weak_transfer_epsilon_window}
\end{align}
For the two-sector decomposition \(P_L+P_R=I\) above, a sufficient
condition is
\begin{align}
 0<\frac{\Omega_{\mathrm{tun}}^{\mathrm{op}}
       +\|\mathcal D-\mathcal D_0\|_{2\to2}}{\Delta_{\min}}
 <\varepsilon.
 \label{eq:general_physical_transfer_threshold}
\end{align}
\end{theorem_}

The condition in \cref{eq:general_physical_transfer_threshold} compares
the  {total intersector coupling frequency} \(\Omega_{\mathrm{tun}}^{\mathrm{op}}\)
and dissipative coupling \(\mathcal D-\mathcal D_0\) with the internal
relaxation gap \(\Delta_{\min}\). The proof in
\cref{app:proof_weak_sector_certificate} uses spectral continuity
to separate the decay rates emerging from zero from the remaining
rates \(\Gamma_i\), \(i\geq M+1\), yielding the threshold
condition. If \(\mathcal D=\mathcal D_0\),
the dissipation strength \(\gamma_{\mathcal D}\) need not be
small. Conversely, neither \(\gamma_{\mathcal D}\) nor the
selected-pair frequency \(\Omega_{\mathrm{tun}}^{(r)}\) alone controls
this condition. The threshold \(\varepsilon\) depends on the full
reference generator \(\mathcal L_0\), rather than only its gap
\(\Delta_{\min}\).

The quadratic slow decay in \cref{thm:qld_sector_origin} is a more
specific instance of stationary-space perturbation theory
\cite{macieszczak2016towards,macieszczak2021classical}.
For an analytic Lindbladian family
\(\mathcal L(\vartheta)=\mathcal L_0+\vartheta\mathcal A
+O(\vartheta^2)\), let
\(\mathscr P_0=\lim_{t\to\infty}e^{t\mathcal L_0}\).
If the projected first-order perturbation satisfies
\(\mathscr P_0\mathcal A\mathscr P_0=0\), the effective generator
on \(\operatorname{Ran}\mathscr P_0\) is \(O(\vartheta^2)\).
For a relaxing family \(\mathcal L(\vartheta)\), this gives
\(\Gamma_M(\vartheta)=O(\vartheta^2)\), while the fast rates
\(\Gamma_i(\vartheta)\), \(i\geq M+1\), remain bounded below
by \(\Delta_{\min}/2\) for sufficiently small \(\vartheta>0\).
 {For QLD, the cancellation in
\cref{eq:qld_no_first_order_population_transfer} gives the direct
bound \(\Gamma_2\leq C(\Omega_{\mathrm{tun}}^{\mathrm{op}})^2/
\Delta_{\min}^{(r)}\) in \cref{eq:qld_quadratic_spectral_rates},
without introducing an interpolation parameter.}
In the two-sector QLD case \(M=2\), the slow nonstationary direction
describes redistribution between the two sectors. For general \(M\), the
\(M-1\) slow directions need not all describe classical populations.

We next increase the dissipation strength \(\gamma_{\mathcal D}\)
at fixed Hamiltonian \(H\) and fixed normalized dissipator
\(\widehat{\mathcal D}=\mathcal D/\gamma_{\mathcal D}\), with
\(\|\widehat{\mathcal D}\|_{2\to2}=1\). Set
\(\mathcal K=-\im[H,\cdot]/\hbar\) and define the Hamiltonian
bandwidth frequency
\(\omega_{\mathrm{bw}}=[\lambda_{\max}(H)-\lambda_{\min}(H)]/\hbar
=\|\mathcal K\|_{2\to2}\), as in
\cref{eq:r_state_hamiltonian_matrix_units,eq:r_state_hamiltonian_bandwidth_norm}.
The rescaled family is
\begin{align}
 \mathcal L_{\gamma_{\mathcal D}}
 &=\mathcal K+\gamma_{\mathcal D}\widehat{\mathcal D}.
 \label{eq:general_strong_dissipation_family}
\end{align}
Assume the full-system dissipator \(\widehat{\mathcal D}\) has a
simple stationary eigenvalue \(0\) and positive nonstationary decay
rates \(d_i=-\operatorname{Re}\lambda_i(\widehat{\mathcal D})\),
ordered as \(0<d_2\leq\cdots\leq d_{d_H^2}\).

\begin{theorem_}[Spectral exclusion by strong dissipation \(\gamma_{\mathcal D}\)]
\label{thm:general_strong_dissipation_exclusion}
Fix \(H,\widehat{\mathcal D}\) as above, with
\(\omega_{\mathrm{bw}}>0\). If
\(\chi_{\max}(\widehat{\mathcal D})<\chi_c<1\), there exists
\(\varepsilon>0\), independent of \(\gamma_{\mathcal D}\), such that
\begin{align}
 0<\frac{\omega_{\mathrm{bw}}}{\gamma_{\mathcal D}}<\varepsilon
 \quad\Longrightarrow\quad
 \chi_{\max}(\mathcal L_{\gamma_{\mathcal D}})<\chi_c.
 \label{eq:general_strong_dissipation_epsilon_exit}
\end{align}
\end{theorem_}

The exclusion guarantee relies on the stated spectral assumptions on
\(\widehat{\mathcal D}\) and does not cover a degenerate stationary
space \(\ker\widehat{\mathcal D}\). The complete proof, based on a
spectral-continuity estimate, is presented in
\cref{app:proof_general_strong_dissipation}.
Under the same relaxation assumptions,
\(\chi_{\max}(\widehat{\mathcal D})>\chi_c\) instead makes the
window persist for sufficiently large \(\gamma_{\mathcal D}\);
equality requires separate analysis. If \(\omega_{\mathrm{bw}}=0\),
then \(\mathcal K=0\) and
\(\chi_{\max}(\mathcal L_{\gamma_{\mathcal D}})
=\chi_{\max}(\widehat{\mathcal D})\) for every
\(\gamma_{\mathcal D}>0\).

Together, \cref{thm:weak_sector_certificate,thm:general_strong_dissipation_exclusion}
show how weak intersector coupling can create the spectral window
\(\chi_{\max}>\chi_c\), and how its exclusion by increasing the
dissipation strength \(\gamma_{\mathcal D}\) depends on the normalized
dissipator \(\widehat{\mathcal D}\). To use these mechanisms for
optimization, we must also determine how an initial state \(\rho_0\)
populates the slow modes and how long its trajectory \(\rho(t)\)
takes to approach the target. \Cref{sec:metastability_control} therefore
develops finite-time bounds to guide dissipative control at fixed
target accuracy \(\varepsilon\).

\section{Mechanism-guided dissipative control for optimization}
\label{sec:metastability_control}

\Cref{sec:origin_metastability} motivates controlling a Lindbladian
\(\mathcal L\) by tuning both the balance and overall strength of
dissipation. At fixed Hamiltonian \(H\), the anisotropy
\(\delta_{\mathcal D}\) and dissipation strength
\(\gamma_{\mathcal D}^{(2)}\) determine the spectral window in the
symmetric two-state model (\cref{thm:two_state_lower_edge}). For
general \(\mathcal L\), the corresponding controls act on the
normalized dissipator \(\widehat{\mathcal D}\) and its strength
\(\gamma_{\mathcal D}\); whether increasing
\(\gamma_{\mathcal D}\) excludes the window depends on
\(\widehat{\mathcal D}\)
(\cref{thm:general_strong_dissipation_exclusion}).

We now study how to leverage metastability in quantum dynamics for optimization. We choose QLD as a concrete example, with tunable coefficients
\(\mu(t),\nu(t)\) in the projected jump operator
\(L^{(r)}=\mu x_r+\im\nu p_r\) \cite{chen2023quantum}.
Throughout this section, the potential \(V(x)\) and energy projector
\(P_r\) are fixed, so these coefficients modify the dissipator
\(\mathcal D_r\) without changing the retained operators
\(H_r,x_r,p_r\) or coupling frequency \(\Omega_{\mathrm{tun}}^{(r)}\).
For a baseline pair \((\mu_{\mathrm b},\nu_{\mathrm b})\) with
nonzero dissipator \(\mathcal D_{r,\mathrm b}\), common rescaling
by \(c>0\) gives
\begin{align}
 (\mu,\nu)&=c(\mu_{\mathrm b},\nu_{\mathrm b}),
 &\mathcal D_r&=c^2\mathcal D_{r,\mathrm b},
 &\gamma_{\mathcal D}^{(r)}&=c^2\gamma_{\mathcal D,\mathrm b}^{(r)}.
 \label{eq:control_qld_common_scaling}
\end{align}

Tuning \(\mu,\nu\) therefore implements two control routes:
\begin{itemize}[leftmargin=*,itemsep=4pt,topsep=4pt]
 \item \textbf{Tuning the anisotropy \(\delta_{\mathcal D}\).}
 Change \(\nu/\mu\), for \(\mu>0\), during a finite stage of duration \(T_p\)
 to suppress slow-mode overlap, then restore the baseline generator
 \(\mathcal L_b\)
 (\cref{thm:two_state_dissipative_pulse,thm:qld_dissipative_pulse}).
 The anisotropy \(\delta_{\mathcal D}\) describes the symmetric
 two-state case; for general \(r\), this route changes the normalized
 dissipator \(\widehat{\mathcal D}_r\). Ratio tuning can also change
 \(\gamma_{\mathcal D}^{(r)}\), so keeping it fixed requires
 compensating common rescaling.
 \item \textbf{Tuning the dissipation strength
 \(\gamma_{\mathcal D}^{(r)}\).}
 Commonly rescale \(\mu,\nu\) at fixed \(\widehat{\mathcal D}_r\),
 using a switch or a ramp of duration \(T_r\), then hold a value
 with verified relaxation and target accuracy
 (\cref{thm:two_state_strength_control,thm:qld_strength_control}).
 This route need not reduce \(\chi_{\max}\). No common direction
 of strength adjustment works for every potential \(V(x)\);
 strong-dissipation exclusion requires the hypotheses of
 \cref{thm:qld_strong_dissipation_exclusion}.
\end{itemize}

The finite-stage route adapts established slow-mode preparation and
temporary-reset strategies
\cite{carollo2021mpemba,kochsiek2022accelerating,bao2025temporary}
to the available QLD coefficients \(\mu(t),\nu(t)\), rather than
assuming access to arbitrary unitary or reset operations.

The spectral window \(\chi_{\max}>\chi_c\) identifies a candidate
bottleneck, but its effect on convergence depends on the initial
slow-mode overlap, the fast transient, and the chosen target
\cite{macieszczak2016towards,szehr2015spectral}. We therefore compare
complete trajectories \(\rho(t)\) at a fixed accuracy rather than
infer a time gain from spectral separation alone. For a target
\(\rho_{\mathrm{tar}}\), define the convergence time
\(T_\varepsilon[\rho;\rho_{\mathrm{tar}}]\) by
\begin{align}
 T_\varepsilon[\rho;\rho_{\mathrm{tar}}]
 =\inf\{T\geq0:
   D_{\mathrm{tr}}(\rho(t),\rho_{\mathrm{tar}})\leq\varepsilon
   \text{ for every }t\geq T\}.
 \label{eq:control_settling_time}
\end{align}
The infimum is \(+\infty\) if the set is empty. All comparisons below
use the same initial state \(\rho_0\), target \(\rho_{\mathrm{tar}}\),
and tolerance \(\varepsilon\); time spent applying control is included.
The admissible controls and their finite maximum strengths are fixed
before comparison. Our certificates compare with a specified baseline,
not the optimal constant control, and do not impose equal
time-integrated dissipative cost unless stated separately.
If \(\rho_{\mathrm{tar}}=\rho_{\mathrm{ss}}\), trace-distance
contractivity makes the distance nonincreasing after the final generator
\(\mathcal L\) has been fixed. For a ground-state target
\(\rho_{\mathrm{tar}}=\ket{\phi_0}\bra{\phi_0}\), the stationary
bias \(D_{\mathrm{tr}}(\rho_{\mathrm{ss}},\rho_{\mathrm{tar}})\)
must additionally be controlled.

With this comparison fixed, \cref{sec:two_state_control} gives explicit
coefficient and duration choices, and \cref{sec:qld_general_control}
extends the tests to general potentials \(V(x)\), including finite
switching errors in \cref{rem:control_robustness}. In each certified time gain,
a controlled upper bound must beat a baseline lower bound, including
the pulse duration \(T_p\) or ramp duration \(T_r\).

\subsection{Exact tuning in the two-state model}
\label{sec:two_state_control}

The symmetric two-state generator \(\mathcal L_2^{\mathrm{TP}}\)
provides explicit versions of both routes when the coefficients
\(\mu,\nu\) can be varied independently. By
\cref{thm:two_state_lower_edge}, increasing their common magnitude
at fixed anisotropy \(\delta_{\mathcal D}\) does not remove an
existing spectral window \(\chi_{\max}>\chi_c>1/3\). Nevertheless,
one can either change their ratio to remove slow-mode overlap or
increase their common magnitude to improve a finite-time contraction
bound. The latter need not close the spectral window. We first
determine the attainable stationary
accuracy, since faster relaxation to \(\rho_{\mathrm{ss}}\) need not
mean faster preparation of the ground state \(\ket{\phi_0}\).
Choose the energy-basis phases so that
\(x_{01}>0\), and take \(a_x,b\geq0\). With
\(S=a_x^2+b^2>0\) and \(\delta_{\mathcal D}=(a_x^2-b^2)/S\), the
stationary population from \cref{eq:symmetric_stationary_bloch} obeys
\begin{align}
 \langle\phi_0|\rho_{\mathrm{ss}}|\phi_0\rangle
 &=\frac{1+2a_xb/S}{2}
 =\frac{1+\sqrt{1-\delta_{\mathcal D}^2}}{2}.
 \label{eq:control_two_state_ground_population}
\end{align}
In the spectral window \(\chi_{\max}>\chi_c>1/3\),
\cref{eq:two_state_lower_edge_separations,eq:two_state_splitting_squared}
give \(\chi_{\max}=\chi_2\leq|\delta_{\mathcal D}|\). Hence
\begin{align}
 D_{\mathrm{tr}}(\rho_{\mathrm{ss}},\ket{\phi_0}\bra{\phi_0})
 &=1-\langle\phi_0|\rho_{\mathrm{ss}}|\phi_0\rangle
 \geq\frac{1-\sqrt{1-\chi_2^2}}{2}
 >\frac{1-\sqrt{1-\chi_c^2}}{2}.
 \label{eq:control_two_state_accuracy_tradeoff}
\end{align}
The first equality holds because \(\rho_{\mathrm{ss}}\) is diagonal
in the energy basis \(\ket{\phi_0},\ket{\phi_1}\). Thus a highly
separated symmetric two-state generator \(\mathcal L_2^{\mathrm{TP}}\)
does not simultaneously have an arbitrarily accurate ground-state
stationary state \(\rho_{\mathrm{ss}}\). This exact restriction must
not be hidden by calling every relaxation experiment ground-state
preparation.

We next keep the stationary target \(\rho_{\mathrm{ss}}\) of an
overdamped baseline generator \(\mathcal L_b\), rather than replacing
it with a more convenient target. Following the slow-overlap
criterion of \cite{carollo2021mpemba}, we construct a finite QLD
stage that realizes this criterion through anisotropy tuning.
Hold the Hamiltonian \(H_2\) and
quadratic jump budget \(\operatorname{Tr}(L^{(2)\dagger}L^{(2)})=2S\)
fixed. Define its slow and fast transverse rates \(\Gamma_s,\Gamma_f\)
by
\begin{align}
 \Gamma_s&=2S-\sqrt{\Delta_B},
 &\Gamma_f&=2S+\sqrt{\Delta_B},
 &0<\Gamma_s&<\Gamma_f<4S,
 \label{eq:control_two_state_rates}
\end{align}
where \(\Delta_B>0\) and \(\omega_{01}>0\). During a finite control
interval, choose balanced coefficients
\(a_x^{\mathrm c}=b^{\mathrm c}=\sqrt{S/2}\). They are implemented
without changing \(H_2\) by
\begin{align}
 \mu_{\mathrm c}&=\frac{\sqrt{S/2}}{x_{01}},
 &\nu_{\mathrm c}&=\frac{\sqrt{S/2}}{m\omega_{01}x_{01}}.
 \label{eq:control_balanced_jump_coefficients}
\end{align}
By \cref{eq:symmetric_adjacent_separation_summary}, the control generator
\(\mathcal L_c\) has \(\chi_{\max}(\mathcal L_c)=1/3\).
Write the transverse Bloch vector as
\(w(t)=(u_x(t),u_y(t))^{\mathsf T}\), with initial value \(w_0\),
and choose a real left eigenvector \(\ell_s\) of the baseline block
\(A_\perp\) such that
\(\ell_s^{\mathsf T}A_\perp=-\Gamma_s\ell_s^{\mathsf T}\)
and \(\|\ell_s\|_2=1\). Its overlap \(\ell_s^{\mathsf T}w\)
measures the slow component of the coherence
\(\rho_{2,01}=(u_x-\im u_y)/2\).
The schedule reduces the larger of \(a_x,b\) and increases the
smaller until \(a_x=b\), keeping \(S\) fixed; it holds these values
for a duration \(T_p\), then restores the original coefficients
\(\mu,\nu\). Equivalently, the anisotropy follows
\(\delta_{\mathcal D}\to0\to\delta_{\mathcal D}\).
The dissipator norm \(\gamma_{\mathcal D}^{(2)}
=4S\sqrt{2-\delta_{\mathcal D}^2}\) increases during balancing,
but the essential change is its structure, not merely its magnitude.
The balanced pulse rotates \(w\) until
\(\ell_s^{\mathsf T}w=0\), after which the original generator
\(\mathcal L_b\) has no slow transverse component to relax.
The fixed budget \(2S\) is a projected jump budget, not a claim of
equal laboratory power. If bath parameters \(\eta,T,\Omega\) are
constrained as in \cref{eq:qld_thermal_mu,eq:qld_thermal_nu}, reachability of
\cref{eq:control_balanced_jump_coefficients} must be checked separately.

For a nonzero initial overlap \(\ell_s^{\mathsf T}w_0\), write
\(\ell_s=(\ell_x,\ell_y)^{\mathsf T}\) and define the pulse
duration \(T_p\) by
\begin{align}
 T_p&=\frac{1}{\omega_{01}}
 \left[\operatorname{atan2}\!\left(
 -\ell_s^{\mathsf T}w_0,\,
 \ell_xu_{y,0}-\ell_yu_{x,0}\right)\bmod\pi\right],
 \qquad 0<T_p<\frac{\pi}{\omega_{01}}.
 \label{eq:control_explicit_pulse_time}
\end{align}
Here \(\operatorname{atan2}(y,x)\) denotes the polar angle of
\((x,y)\), and the representative modulo \(\pi\) lies in
\((0,\pi)\).

\begin{theorem_}[Optimization by balancing the dissipative anisotropy]
\label{thm:two_state_dissipative_pulse}
For the symmetric baseline generator \(\mathcal L_b\) satisfying
\cref{eq:control_two_state_rates}, assume
\(\ell_s^{\mathsf T}w_0\ne0\) and that the balanced coefficients
\(\mu_{\mathrm c},\nu_{\mathrm c}\) in
\cref{eq:control_balanced_jump_coefficients} are admissible.
At fixed \(H_2,S\), apply them for the duration \(T_p\) in
\cref{eq:control_explicit_pulse_time}, then restore \(\mathcal L_b\).
This removes the slow transverse component exactly. If
\(\rho_p=e^{T_p\mathcal L_c}\rho_0\) and
\(D_p=D_{\mathrm{tr}}(\rho_p,\rho_{\mathrm{ss}})\), then, for
every tolerance \(\varepsilon>0\),
\begin{align}
 D_{\mathrm{tr}}(e^{t\mathcal L_b}\rho_p,\rho_{\mathrm{ss}})
 &\leq D_p e^{-\Gamma_f t},\qquad t\geq0,
 \label{eq:control_two_state_fast_relaxation}\\
 T_\varepsilon^{\mathrm{ctrl}}
 &\leq T_p+\frac1{\Gamma_f}\log\max\{1,D_p/\varepsilon\},
 \label{eq:control_two_state_time_upper}\\
 T_\varepsilon^{\mathrm{base}}
 &\geq\frac1{\Gamma_s}
 \log\max\{1,|\ell_s^{\mathsf T}w_0|/(2\varepsilon)\}.
 \label{eq:control_two_state_time_lower}
\end{align}
Whenever the upper bound in \cref{eq:control_two_state_time_upper}
is smaller than the lower bound in \cref{eq:control_two_state_time_lower},
the protocol is strictly faster at the same accuracy \(\varepsilon\).
For each fixed initial state \(\rho_0\) satisfying the overlap
condition, this holds for all sufficiently small \(\varepsilon>0\).
\end{theorem_}

Balancing \(\delta_{\mathcal D}=0\) removes the damping contrast
identified in \cref{thm:two_state_lower_edge}, so coherent rotation
can bring the initial component \(w_0\) into the fast direction.
Restoring \(\mathcal L_b\) preserves its spectrum and the target
\(\rho_{\mathrm{ss}}\), but removes the trajectory's
slow overlap; the time bound includes \(T_p\).
The proof in \cref{app:proof_two_state_dissipative_pulse} also
discusses timing errors and the required access to
\(\mu_{\mathrm c},\nu_{\mathrm c}\). Exact cancellation uses the
symmetric block \(A_\perp\), ideal switching, and knowledge of
\(w_0,\ell_s\); their estimation cost is not included in the
evolution time. At finite accuracy \(\varepsilon\), the residual-error
test in \cref{rem:control_robustness} relaxes exact cancellation.

The second route keeps \(-1<\delta_{\mathcal D}<1\) fixed and
increases the quadratic strength \(S(t)\) continuously and
nondecreasingly from \(S_{\mathrm b}>0\) to a finite admissible
\(S_{\mathrm h}>S_{\mathrm b}\) over \(T_r>0\), then holds it
constant. Implement this through
\((\mu(t),\nu(t))=\sqrt{S(t)/S_{\mathrm b}}
(\mu_{\mathrm b},\nu_{\mathrm b})\). All generators share
\(\rho_{\mathrm{ss}}\), since
\(u_{z,\mathrm{ss}}=\sqrt{1-\delta_{\mathcal D}^2}\) is independent
of \(S(t)\). Increasing \(S\) need not increase the exact gap
\(\Gamma_2\) monotonically, so we instead use the contraction bound
\(\alpha_2(t)\) and initial distance \(D_0\):
\begin{align}
 \alpha_2(t)&=2(1-|\delta_{\mathcal D}|)S(t)
 =\frac{1-|\delta_{\mathcal D}|}
        {2\sqrt{2-\delta_{\mathcal D}^2}}
        \gamma_{\mathcal D}^{(2)}(t),
 &D_0&=D_{\mathrm{tr}}(\rho_0,\rho_{\mathrm{ss}}).
 \label{eq:two_state_control_contraction_rate}
\end{align}
Here \(\alpha_2(t)\) bounds damping in the least-damped Bloch
direction, rather than equalling \(\Gamma_2(t)\); write
\(\alpha_{2,\mathrm h}=2(1-|\delta_{\mathcal D}|)S_{\mathrm h}>0\).

\begin{theorem_}[Optimization by continuous strength modulation]
\label{thm:two_state_strength_control}
For the symmetric two-state model with fixed
\(-1<\delta_{\mathcal D}<1\), the preceding continuous
increase-and-hold schedule satisfies
\begin{align}
 D_{\mathrm{tr}}(\rho(t),\rho_{\mathrm{ss}})
 &\leq D_0\exp\!\left[-\int_0^t\alpha_2(u)\,\mathrm du\right],
 \label{eq:two_state_control_integrated_contraction}\\
 T_\varepsilon^{\mathrm{ctrl}}
 &\leq T_r+\frac{1}{\alpha_{2,\mathrm h}}
 \log\max\left\{1,\frac{D_0}{\varepsilon}
          e^{-\int_0^{T_r}\alpha_2(u)\,\mathrm du}\right\}.
 \label{eq:two_state_strength_time_upper}
\end{align}
If the baseline is overdamped with
\(\ell_s^{\mathsf T}w_0\ne0\), this upper bound certifies a
strict speedup whenever it is smaller than
\cref{eq:control_two_state_time_lower}. In particular,
\(\alpha_{2,\mathrm h}>\Gamma_s\) guarantees a strict speedup
for all sufficiently small \(\varepsilon>0\), with fixed finite
\(T_r,S_{\mathrm h}\). No bound on \(\dot S(t)\) is required
within this time-local model.
\end{theorem_}

The proof in \cref{app:proof_two_state_strength_control} shows how
integrated damping replaces overlap cancellation. The sufficient
condition \(S_{\mathrm h}>\Gamma_s/[2(1-|\delta_{\mathcal D}|)]\)
must fit within the available control ceiling; it excludes
\(|\delta_{\mathcal D}|=1\). This route need not close the spectral
window or improve the stationary accuracy in
\cref{eq:control_two_state_accuracy_tradeoff}. If immediate use of
\(S_{\mathrm h}\) is admissible, it maximizes the contraction
integral within this family; the ramp is not claimed to outperform
that switch.

\subsection{Tuning QLD for general potentials}
\label{sec:qld_general_control}

For a general potential \(V(x)\), additional states \(\ket{\phi_n}\)
need not preserve the planar rotation of
\cref{thm:two_state_dissipative_pulse} or the common stationary
state of \cref{thm:two_state_strength_control}. With \(H_r,P_r\)
fixed, ratio tuning instead changes the intersector transfer and
energy selectivity of \(L^{(r)}=\mu x_r+\im\nu p_r\).
The matrix-element calculations in
\cref{app:proof_qld_dissipative_pulse} connect these changes to
the weak-transfer mechanism of \cref{thm:qld_sector_origin}.
To determine whether they bypass an occupied slow component,
we adapt temporary dissipative preparation and restoration
\cite{bao2025temporary}, restricting the auxiliary generator
\(\mathcal L_c\) to accessible QLD coefficients
\(\mu_{\mathrm c},\nu_{\mathrm c}\).

Let the baseline generator \(\mathcal L_b=\mathcal L_r^{\mathrm{TP}}\)
relax to \(\rho_{\mathrm{ss}}^{\mathrm b}\), with a simple real
slow eigenvalue \(-\Gamma_s\), \(\Gamma_s>0\), strictly separated
from the other nonstationary decay rates. Choose its Hermitian left
eigenmatrix \(F_s\), satisfying
\(\mathcal L_b^\dagger F_s=-\Gamma_sF_s\), so that
\(\|F_s\|_{\mathrm{op}}=1\) and
\(w_s=\operatorname{Tr}(F_s\rho_0)/2>0\).
Stationarity gives
\(\operatorname{Tr}(F_s\rho_{\mathrm{ss}}^{\mathrm b})=0\).
Thus \(F_s\) measures the long-lived displacement, which may
represent population redistribution between wells but need not equal
a bare well population. Suppose an admissible pair
\(\mu_{\mathrm c},\nu_{\mathrm c}\) defines
\(L_{\mathrm c}^{(r)}=\mu_{\mathrm c}x_r+\im\nu_{\mathrm c}p_r\)
and a generator \(\mathcal L_c\) relaxing to \(\sigma_c\), with
\(D_{\mathrm{tr}}(e^{t\mathcal L_c}\rho_0,\sigma_c)
\leq C_c e^{-f_ct}\) for \(t\geq0\) and \(C_c,f_c>0\).
The stationary expectation \(m_c=\operatorname{Tr}(F_s\sigma_c)\)
tests whether this control eventually reverses the sign of the
occupied slow component.

\begin{theorem_}[A finite structure-tuning stage that removes the slow component]
\label{thm:qld_dissipative_pulse}
Under the preceding baseline and admissibility conditions, assume
\(m_c<0\). There exists a duration \(T_p\) such that
\begin{align}
 0<T_p&\leq\frac1{f_c}\log\frac{2C_c}{|m_c|},
 &\operatorname{Tr}(F_s\rho_p)&=0,
 &\rho_p&=e^{T_p\mathcal L_c}\rho_0.
 \label{eq:qld_pulse_cancellation_time}
\end{align}
Restore \(\mathcal L_b\) at \(T_p\). For any \(f\) strictly
between \(\Gamma_s\) and all remaining nonstationary decay
rates, there is a finite \(C_p\geq0\) such that
\begin{align}
 D_{\mathrm{tr}}(e^{t\mathcal L_b}\rho_p,
                    \rho_{\mathrm{ss}}^{\mathrm b})
 &\leq C_p e^{-ft},\qquad t\geq0,
 \label{eq:qld_pulse_fast_relaxation}\\
 T_\varepsilon^{\mathrm{ctrl}}
 &\leq T_p+\frac1f\log\max\{1,C_p/\varepsilon\},
 &T_\varepsilon^{\mathrm{base}}
 &\geq\frac1{\Gamma_s}\log\max\{1,w_s/\varepsilon\}.
 \label{eq:qld_pulse_time_comparison}
\end{align}
The protocol is strictly faster whenever these bounds are strictly
ordered, in particular for all sufficiently small
\(\varepsilon>0\) at fixed control parameters.
\end{theorem_}

The condition \(m_c<0\) concerns reachable coefficients
\(\mu_{\mathrm c},\nu_{\mathrm c}\), not an arbitrarily prescribed
reset state \(\sigma_c\). Once it holds, a zero of
\(\operatorname{Tr}(F_s e^{t\mathcal L_c}\rho_0)\) determines
\(T_p\) within the stated finite bracket; the dissipation strength
\(\gamma_{\mathcal D}^{(r)}\) need not increase. The proof in
\cref{app:proof_qld_dissipative_pulse} establishes the bracket and
time comparison. The condition is sufficient, not necessary:
the two-state construction reaches zero by transient rotation.

\begin{remark_}[Robustness at finite accuracy]
\label{rem:control_robustness}
Exact cancellation is unnecessary for a prescribed tolerance
\(\varepsilon\). For the baseline \(\mathcal L_b\) above, let an
admissible stage prepare \(\rho_p\) at \(T_p\), and set
\(A_p=\|\mathcal P_s(\rho_p-\rho_{\mathrm{ss}}^{\mathrm b})\|_1/2\),
where \(\mathcal P_s\) projects onto \(-\Gamma_s\).
Let \(\mathcal Q_f\) project onto the remaining nonstationary
spectrum and choose \(C_p\geq0\), \(f>\Gamma_s\) such that
\(\|e^{t\mathcal L_b}\mathcal Q_f
(\rho_p-\rho_{\mathrm{ss}}^{\mathrm b})\|_1/2\leq C_pe^{-ft}\)
for \(t\geq0\). Suppose the ideal and implemented CPTP schedules
start from the same \(\rho_0\), keep \(H_r\) fixed, and restore
\(\mathcal L_b\) at \(T_p\) and \(T_f\geq T_p\), respectively.
If their endpoint discrepancy satisfies
\(D_{\mathrm{tr}}(\widetilde\rho(T_f),\rho(T_f))
\leq\epsilon_{\mathrm{sw}}\), then
\begin{align}
 D_{\mathrm{tr}}(\widetilde\rho(T_f+t),\rho_{\mathrm{ss}}^{\mathrm b})
 &\leq A_pe^{-\Gamma_st}+C_pe^{-ft}+\epsilon_{\mathrm{sw}},
 \qquad t\geq0.
 \label{eq:control_robust_postpulse_distance}
\end{align}
Consequently, a finite-accuracy improvement is certified if this
bound is at most \(\varepsilon\) for some \(t\geq0\) with
\(T_f+t<\Gamma_s^{-1}\log\max\{1,w_s/\varepsilon\}\).
The derivation and coefficient-based bounds on
\(\epsilon_{\mathrm{sw}}\) are given in
\cref{app:proof_qld_dissipative_pulse}. A nonzero residual amplitude
\(A_p\), or switching error that repopulates the slow mode, can
restore the asymptotic rate \(\Gamma_s\) without precluding this
finite-accuracy improvement; \(\epsilon_{\mathrm{sw}}\) is an error
bound, not a physical asymptotic floor.
\end{remark_}

Common-strength tuning provides a complementary route, but now the
stationary state generally varies with
\(\gamma_{\mathcal D}^{(r)}\). The spectral screening criterion
\cref{thm:qld_strong_dissipation_exclusion} must therefore be
supplemented by contraction and target-accuracy conditions.
Fix \(\widehat{\mathcal D}_r\), a baseline strength
\(\gamma_{\mathrm b}>0\), and a finite admissible
\(\gamma_{\mathrm h}>\gamma_{\mathrm b}\). For \(T_r>0\) and a
continuous nondecreasing profile \(h:[0,1]\to[0,1]\), with
\(h(0)=0\), \(h(1)=1\), use
\begin{align}
 \gamma_{\mathcal D}^{(r)}(t)
 &=\begin{cases}
 \gamma_{\mathrm b}+(\gamma_{\mathrm h}-\gamma_{\mathrm b})h(t/T_r),
       &0\leq t\leq T_r,\\
 \gamma_{\mathrm h},&t>T_r,
 \end{cases}
 \label{eq:qld_control_ramp_schedule}\\
 (\mu(t),\nu(t))
 &=\sqrt{\gamma_{\mathcal D}^{(r)}(t)/\gamma_{\mathrm b}}
       (\mu_{\mathrm b},\nu_{\mathrm b}).
 \label{eq:qld_control_ramp_coefficients}
\end{align}
Here \(\mu_{\mathrm b},\nu_{\mathrm b}\) generate
\(\gamma_{\mathrm b}\widehat{\mathcal D}_r\), all intermediate
coefficients must be admissible, and \(H_r,P_r\), hence
\(\Omega_{\mathrm{tun}}^{(r)}\), remain fixed. A direct switch is
the limiting choice \(T_r=0\).

Let \(\widehat{\mathcal D}_r\) have a simple stationary eigenvalue
\(0\) with density operator \(\sigma_D\). On
\(\mathfrak T_r=\{B=P_rBP_r:\operatorname{Tr}B=0\}\), assume
the damping bound below and define the inverse norm \(R_D\):
\begin{align}
 -\operatorname{Re}\langle B,\widehat{\mathcal D}_r(B)\rangle
 &\geq a_D\|B\|_{\mathrm{HS}}^2
 \quad(B\in\mathfrak T_r),\qquad a_D>0,
 \label{eq:qld_control_coercivity}\\
 R_D&=\|(\widehat{\mathcal D}_r|_{\mathfrak T_r})^{-1}\|_{2\to2}.
 \label{eq:qld_control_inverse_norm}
\end{align}
Positivity of \(a_D\) controls every traceless direction \(B\)
and is stronger than a positive spectral gap; \(R_D\) measures
the sensitivity of \(\sigma_D\) to \(\mathcal K_r\).
For a target density operator \(\rho_{\mathrm{tar}}\) supported in
\(\operatorname{Ran}P_r\), define its bias bound \(b_r\) by
\begin{align}
 b_r&=D_{\mathrm{tr}}(\sigma_D,\rho_{\mathrm{tar}})
       +\frac{\sqrt r R_D\omega_{\mathrm{bw}}^{(r)}}
                    {2\gamma_{\mathrm h}}.
 \label{eq:qld_control_target_bias_bound}
\end{align}
Increasing \(\gamma_{\mathrm h}\) reduces only the second term.
For comparison, assume
\(\mathcal L_b=\mathcal K_r+\gamma_{\mathrm b}\widehat{\mathcal D}_r\)
relaxes to \(\rho_{\mathrm{ss}}^{\mathrm b}\) and has a Hermitian
left eigenmatrix \(F_s\) with
\(\mathcal L_b^\dagger F_s=-\Gamma_sF_s\), \(\Gamma_s>0\),
\(\|F_s\|_{\mathrm{op}}=1\). Set
\(w_s=|\operatorname{Tr}(F_s\rho_0)|/2>0\) and
\(b_{\mathrm b}=D_{\mathrm{tr}}(\rho_{\mathrm{ss}}^{\mathrm b},
\rho_{\mathrm{tar}})\). The observable \(F_s\) bounds the baseline
time from below, while \(a_D,b_r\) bound the controlled time
from above.

\begin{theorem_}[Optimization by increasing and holding the dissipation]
\label{thm:qld_strength_control}
\label{prop:qld_target_accuracy_control}
\label{cor:qld_fixed_target_speedup}
Under \cref{eq:qld_control_coercivity}, apply the schedule
\cref{eq:qld_control_ramp_schedule,eq:qld_control_ramp_coefficients}
at fixed \(H_r\). The final generator
\(\mathcal L_{\mathrm h}=\mathcal K_r+
\gamma_{\mathrm h}\widehat{\mathcal D}_r\) has a unique
stationary state \(\rho_{\mathrm{ss}}^{\mathrm h}\), and, for
every initial state \(\rho_0\) supported in \(\operatorname{Ran}P_r\)
and \(t\geq0\),
\begin{align}
 D_{\mathrm{tr}}(\rho(T_r+t),\rho_{\mathrm{tar}})
 &\leq b_r+\sqrt{r/2}\,e^{-a_D\gamma_{\mathrm h}t}.
 \label{eq:qld_control_convergence_bound}
\end{align}
For \(b_r<\varepsilon\), a strict speedup over the preceding
baseline is certified by
\begin{align}
 T_r+\frac{1}{a_D\gamma_{\mathrm h}}
   \log\max\left\{1,\frac{\sqrt{r/2}}{\varepsilon-b_r}\right\}
 &<\frac{1}{\Gamma_s}
   \log\max\left\{1,\frac{w_s}{\varepsilon+b_{\mathrm b}}\right\}.
 \label{eq:qld_speedup_schedule_condition}
\end{align}
If \(\rho_{\mathrm{tar}}=\rho_{\mathrm{ss}}^{\mathrm b}\), the
baseline coefficients may be restored after this accuracy is attained.
No adiabatic bound on \(\dot\gamma_{\mathcal D}^{(r)}\) is required.
\end{theorem_}

The proof in \cref{app:proof_qld_strength_control} turns the
damping and bias bounds \(a_D,b_r\) into the time comparison
\cref{eq:qld_speedup_schedule_condition}. Its strict margin is the
available budget for \(T_r\), not an adiabatic condition.
If \cref{thm:qld_strong_dissipation_exclusion} also applies at
\(\gamma_{\mathrm h}\), the held generator \(\mathcal L_{\mathrm h}\)
additionally lies outside the spectral window. Neither conclusion
is automatic for an arbitrary \(V(x)\) or larger \(r\): the
admissible strength must satisfy the contraction, bias, and time
tests. Restoration preserves the attained accuracy only for the
stationary target \(\rho_{\mathrm{tar}}=\rho_{\mathrm{ss}}^{\mathrm b}\);
finite return errors can be budgeted as in
\cref{rem:control_robustness}.

Together, the two routes convert the origin analysis into
conditional parameter-selection rules: ratio tuning suppresses
the occupied slow component, whereas common rescaling improves
absolute damping subject to target accuracy. For several slow
directions, the entire occupied spectral sector must be controlled,
as detailed in \cref{app:proof_qld_dissipative_pulse}.
The numerical comparisons in \cref{sec:numerical_control} test
these rules at fixed \(\rho_0,\rho_{\mathrm{tar}},\varepsilon\),
including schedules outside the sufficient contraction hypothesis.
They also benchmark direct switches and distinguish elapsed time from
peak and integrated dissipation. Their conclusions apply to the stated
retained dimension \(r\) and grid resolution, without asserting global
optimality or uniform validity beyond the tested finite model.

\begingroup
\color{black}
\section{Numerical experiments}
\label{sec:numerical_simulations}
The experiments address two questions: how do tunneling and dissipation
move a system into or out of the spectral window
\(\chi_{\max}>\chi_c\), and how can this information guide
dissipative control for optimization? In \cref{sec:numerical_origin},
we compare the boundaries of this window with the physical scales
identified in \cref{sec:origin_metastability}, first in the two-state
model and then with additional states and more complex potentials.
\Cref{sec:numerical_control} subsequently tests whether coefficient
tuning improves convergence at the same initial state, stationary
target, and tolerance. The spectral maps diagnose timescale separation;
the trajectory comparisons assess its consequences for optimization.

The common numerical settings are \(\hbar=m=\omega=1\), homogeneous
Dirichlet boundaries, a centered second-order discretization of
\(H\), and momentum matrix elements
\(p_{nm}=\im(E_n-E_m)x_{nm}\) obtained from the same discrete
Hamiltonian. We use the dissipator normalization
\(2L\rho L^\dagger-\{L^\dagger L,\rho\}\), distinguishing the
spatial grid dimension \(N\) from the retained dimension \(r\).
The spectral maps use \(\chi_c=19/21\), corresponding to an
adjacent-rate ratio of \(20\), with a common color range \([0,1]\).
Orange contours mark this chosen threshold, not a thermodynamic
phase transition; physical-scale panels have separately labelled
colorbars. Parameters specific to each scan are collected in the
figure captions.

\subsection{Origin of metastability}
\label{sec:numerical_origin}
\label{sec:numerical_projection}

\paragraph{Two-state double-well benchmark.}
We first ask whether changing dissipation alone can move the two-state
model across the spectral threshold \(\chi_c\), without changing
the potential \(V(x)\) or coupling frequency
\(\Omega_{\mathrm{tun}}^{(2)}\). To separate magnitude from
structure, define
\(x_*=\sqrt{\hbar/(m\omega)}\), \(p_*=\hbar/x_*\),
\(\widetilde\mu=\mu x_*/\sqrt{\omega}\), and
\(\widetilde\nu=\nu p_*/\sqrt{\omega}\), with scan coordinates
\begin{align}
 s:=\widetilde\mu+\widetilde\nu>0,\qquad
 q:=\frac{\widetilde\mu-\widetilde\nu}{s}\in(-1,1).
 \label{eq:numerical_dimensionless_couplings}
\end{align}
At fixed \(q\), the dissipation strength
\(\gamma_{\mathcal D}^{(r)}\) scales as \(s^2\); changing \(q\)
changes the dissipative structure. The inverse relations
\(\widetilde\mu=s(1+q)/2\) and
\(\widetilde\nu=s(1-q)/2\) specify the implemented coefficients
at each scan point.

For comparison across retained dimensions \(r\), the vertical axis is
\(\widehat s=s|x_{01}^{\mathrm{ref}}|/
[x_*\sqrt{\omega_{01}^{\mathrm{ref}}/\omega}]\),
where the reference frequency \(\omega_{01}^{\mathrm{ref}}\) and
position matrix element \(x_{01}^{\mathrm{ref}}\) are computed from
the two lowest energy eigenstates for the fixed potential and grid
specified in \cref{fig:numerical_two_state_maps}, then held fixed
as \(r\) varies. The figure places the spectral maximum
\(\chi_{\max}\) next to the
relative dissipation strength \(\gamma_{\mathcal D}^{(2)}/\omega_{01}\)
to test whether their contours track the same competition of scales.

\begin{figure}[htbp]
 \centering
 \includegraphics[width=\textwidth]{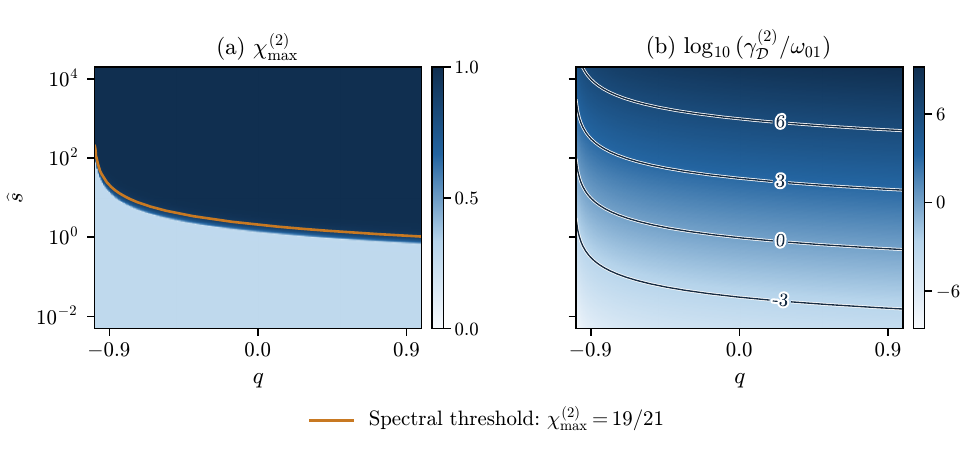}
 \caption{Dissipation-driven entry into and exit from the two-state
 spectral window \(\chi_{\max}>19/21\), at fixed
 \(E_b=4,\ell=6,E_\delta=10^{-4}\) on an \(N=128\) Dirichlet grid
 on \([-10,10]\). The axes show coefficient imbalance \(q\)
 horizontally and amplitude \(\widehat s\) logarithmically
 vertically. Colors show (a) \(\chi_{\max}\) from the
 Bloch matrix \(A\) and (b)
 \(\log_{10}(\gamma_{\mathcal D}^{(2)}/\omega_{01})\).
 The orange line is \(\chi_{\max}=19/21\);
 labelled contours in (b) give strength ratios.}
 \label{fig:numerical_two_state_maps}
\end{figure}

The logarithmic \(\widehat s\) axis in
\cref{fig:numerical_two_state_maps} reveals a sharp crossover between
weakly and strongly separated decay rates. At fixed \(q\), lowering
\(\widehat s\) across the orange contour takes \(\chi_{\max}\)
from above to below \(\chi_c\), removing the spectral window;
increasing \(\widehat s\) reverses this crossing, with no upper exit visible
in this scan. The threshold contour approximately follows
\(\gamma_{\mathcal D}^{(2)}/\omega_{01}\simeq4.5\) over most
of the \(q\) range. This near-alignment has a physical explanation:
at fixed anisotropy \(\delta_{\mathcal D}\),
\cref{thm:two_state_lower_edge} sets the symmetric boundary by the
ratio \(\gamma_{\mathcal D}^{(2)}/\Omega_{\mathrm{tun}}^{(2)}\).

We next ask whether weakening coherent transfer produces the
complementary change in \(\chi_{\max}\). In
\cref{fig:numerical_potential_controls}, we hold \(\mu,\nu\) fixed
and vary the dimensionless barrier width
\(\widetilde d_b=d_b/x_*\) and height
\(\widetilde E_b=E_b/(\hbar\omega)\). Comparing
\(\chi_{\max}\) with
\(\Omega_{\mathrm{tun}}^{(2)}/\omega_{01}\) identifies how the
geometry of \(V(x)\) changes the ability to transfer population
between the wells.

\begin{figure}[htbp]
 \centering
 \includegraphics[width=\textwidth]{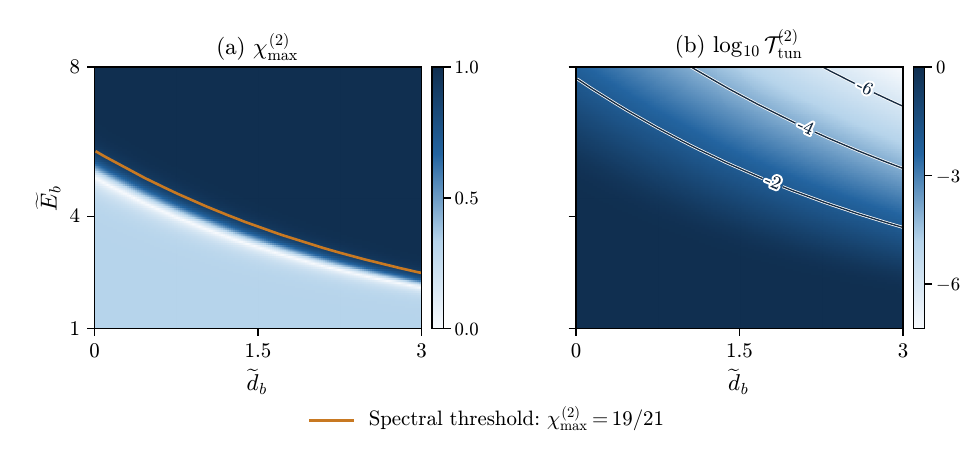}
 \caption{Barrier-controlled spectral separation in the two-state
 model on the same \(N=128\) grid.
 The \(121\times161\) points vary
 \(\widetilde d_b\in[0.01,3]\) and
 \(\widetilde E_b\in[1,8]\) at
 \(E_\delta=10^{-4}\) and
 \(\mu=\nu=1.8246\times10^{-3}\).
 Colors show (a) \(\chi_{\max}\) and (b)
 \(\log_{10}(\Omega_{\mathrm{tun}}^{(2)}/\omega_{01})\);
 the orange contour is \(\chi_{\max}=19/21\).
 Increasing the width \(\widetilde d_b\) or height \(\widetilde E_b\)
 suppresses coherent transfer and leads into the above-threshold region.
 The projected dissipator \(\mathcal D_2\) also varies across this scan.}
 \label{fig:numerical_potential_controls}
\end{figure}

The high-separation region in \cref{fig:numerical_potential_controls}
lies on the wider and higher barrier side of the orange contour.
The contour roughly tracks
\(\Omega_{\mathrm{tun}}^{(2)}/\omega_{01}\simeq0.4\); the same
computed points have
\(\gamma_{\mathcal D}^{(2)}/\Omega_{\mathrm{tun}}^{(2)}\simeq4\).
Their interpretation complements \cref{fig:numerical_two_state_maps}:
spectral separation grows when coherent transfer becomes weaker
relative to dissipation. In addition, the finite bias \(E_\delta\)
limits the transfer probability
\(P_{\max}^{(2)}=(\Omega_{\mathrm{tun}}^{(2)}/\omega_{01})^2\).
The dependence on this ratio agrees with the mechanism in
\cref{thm:biased_two_state_window}, although that theorem's limiting
criteria do not locate the present finite-strength boundary.
Changing \(V(x)\) also changes the projection \(P_2\) and dissipator
\(\mathcal D_2\); the observed contour is consequently a joint
tunneling--dissipation boundary, not a tunneling-only threshold.

These scans do not isolate the role of anisotropy
\(\delta_{\mathcal D}\), which remains close to one in
\cref{fig:numerical_two_state_maps}. We therefore vary the ratio
\(\nu/\mu\) while compensating its effect on
\(\gamma_{\mathcal D}^{(2)}\) by common rescaling of \(\mu,\nu\).
Panel (a) of \cref{fig:numerical_anisotropy_validation} independently controls
\(\delta_{\mathcal D}\) and
\(\gamma_{\mathcal D}^{(2)}/\Omega_{\mathrm{tun}}^{(2)}\)
in a fixed symmetric potential. Panel (b) tests the different
weak-dissipation mechanism at finite bias, comparing
\(\delta_{\mathcal D}\) with
\(\Omega_{\mathrm{tun}}^{(2)}/\omega_{01}\).
In both panels, colors and orange contours are computed from the
\(4\times4\) generator \(\mathcal L_2^{\mathrm{TP}}\);
dashed curves are independent theoretical predictions.

\begin{figure}[htbp]
 \centering
 \includegraphics[width=\textwidth]{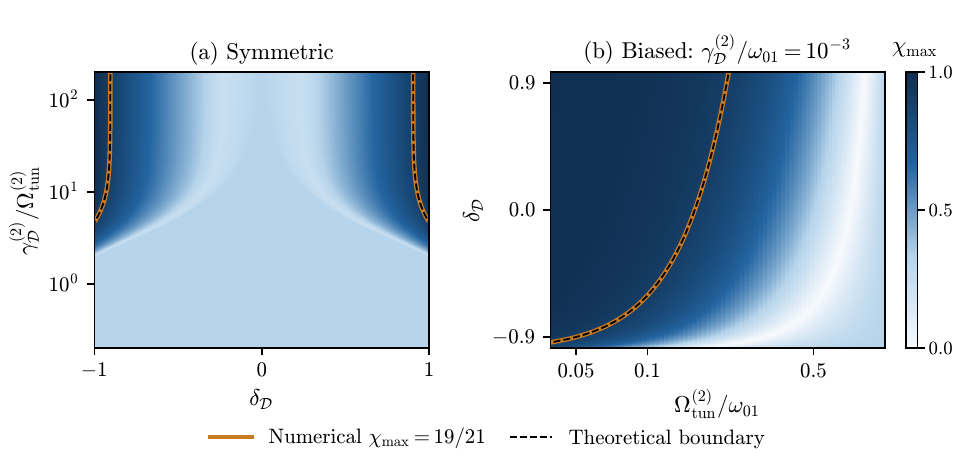}
 \caption{Dissipative anisotropy \(\delta_{\mathcal D}\) and
 spectral separation in two-state QLD, with \(E_b=2,d_b=1\)
 and an \(N=128\) grid on \([-9,9]\).
 (a) Symmetric potential \(E_\delta=0\): independently varied
 \(\delta_{\mathcal D}\) and
 \(\gamma_{\mathcal D}^{(2)}/\Omega_{\mathrm{tun}}^{(2)}\).
 (b) Biased potential \(E_\delta\in[10^{-4},0.3]\), at fixed
 \(\gamma_{\mathcal D}^{(2)}/\omega_{01}=10^{-3}\);
 the horizontal coordinate is \(\Omega_{\mathrm{tun}}^{(2)}/\omega_{01}\).
 Colors show \(\chi_{\max}\), and orange contours mark
 \(\chi_{\max}=19/21\). Dashed curves give the exact symmetric
 criterion of \cref{thm:two_state_lower_edge} in (a) and the
 weak-dissipation criterion of \cref{thm:biased_two_state_window} in (b).}
 \label{fig:numerical_anisotropy_validation}
\end{figure}

The two high-separation regions in
\cref{fig:numerical_anisotropy_validation}(a) require both large
\(|\delta_{\mathcal D}|\) and sufficiently strong relative
dissipation, as predicted by \cref{thm:two_state_lower_edge}.
The window closes without weakening the total dissipative action:
balancing the damping removes the contrast that sustained the slow
transverse mode. Across the computed grid, the numerical classification
agrees with the necessary and sufficient criterion, including the
absence of a window when \(|\delta_{\mathcal D}|\leq\chi_c\). Panel (b) shows why this balancing rule cannot be transferred
unchanged to biased models. Small
\(\Omega_{\mathrm{tun}}^{(2)}/\omega_{01}\) can sustain separation
even at \(\delta_{\mathcal D}=0\), because bias introduces an
additional contrast between population and coherence relaxation.
Increasing \(\delta_{\mathcal D}\) expands this weak-dissipation
window toward larger relative coupling frequencies, consistently
with \cref{thm:biased_two_state_window}. 

\paragraph{Multi-state models with complex potentials.}
Recall that \(r\) denotes the number of lowest-energy eigenstates of
the Hamiltonian \(H\) retained in the projected QLD model.
We now ask which mechanisms survive additional relaxation channels:
first by retaining more states in the double well, then by adding
wells or a spatial direction. Alongside the tunneling--dissipation
scans, we vary dissipative structure at fixed strength to test whether
uneven damping remains important beyond the two-state anisotropy
\(\delta_{\mathcal D}\).

We consider three potential families: the one-dimensional double well
\(V(x)\) in \cref{eq:piecewise_double_well_potential_0710},
a multiwell potential \(V_{\mathrm{mw}}\), and a two-dimensional
double well \(V_{\mathrm{2D}}\). Additional wells can introduce
several redistribution modes, while a second spatial direction
adds dissipative channels. The latter two families are
\begin{align}
 V_{\mathrm{mw}}(x)
 &=0.03(x+1.5)^2+E_{\mathrm{mw}}
       \left[1-\cos\!\left(\frac{2\pi(x+1.5)}{3}\right)\right],
 \label{eq:numerical_multiwell_potential}\\
 V_{\mathrm{2D}}(x,y)
 &=V(x)+\frac{\omega_y^2}{2}(y-c_{xy}x)^2,\qquad
 H_{\mathrm{2D}}=\frac{p_x^2+p_y^2}{2}+V_{\mathrm{2D}}(x,y).
 \label{eq:numerical_two_dimensional_potential}
\end{align}
In \(V_{\mathrm{2D}}\), \(V(x)\) is the symmetric \(E_b=2\) double well,
\(\omega_y=1.3\), and \(c_{xy}=0.35\).
The sector cut \(x=0\) partitions groups of wells in
\(V_{\mathrm{mw}}\); in two dimensions the dissipators of
\(L_x=\mu x+\im\nu p_x\) and
\(L_y=\mu y+\im\nu p_y\) are added before their norm is taken.
Representative members of these potential families are shown in
\cref{fig:numerical_potentials}.

\begin{figure}[htbp]
 \centering
 \includegraphics[width=\textwidth]{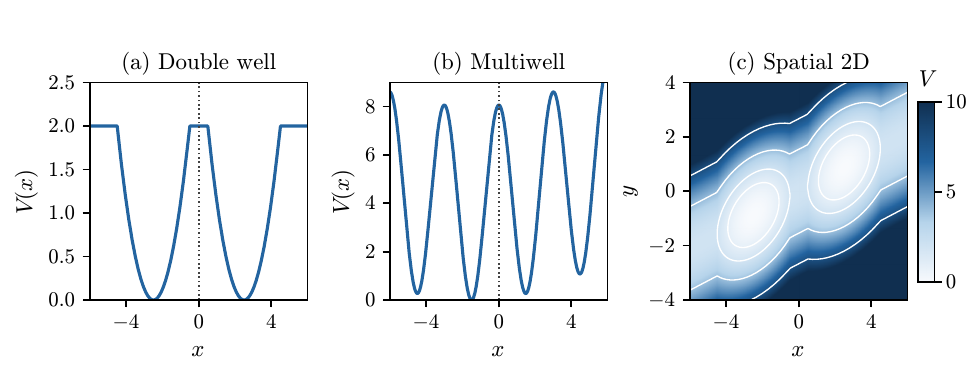}
 \caption{Representative landscapes for the spectral scans;
 the control experiments in \cref{sec:numerical_control} use separately
 specified parameters from the same families.
 (a) A symmetric double well with \(E_b=2,d_b=1\);
 (b) \(V_{\mathrm{mw}}\) from
 \cref{eq:numerical_multiwell_potential} at
 \(E_{\mathrm{mw}}=4\); and (c) \(V_{\mathrm{2D}}\) from
 \cref{eq:numerical_two_dimensional_potential}.
 Dotted lines mark the sector cut \(x=0\).
 In (c), color is potential energy clipped at \(10\), with
 contours at \(1,2,4,8\); this display grid is finer
 than the dynamical grid.}
 \label{fig:numerical_potentials}
\end{figure}

The landscapes in \cref{fig:numerical_potentials} distinguish
several competing minima in \(V_{\mathrm{mw}}\) from two wells
in the coupled geometry of \(V_{\mathrm{2D}}\).
We begin with the one-dimensional double well, repeating the
original scans in \cref{fig:numerical_hierarchy}
with \(r=4,8,16\) and the unprojected \(N=128\) generator
\(\mathcal L_N\). 

\begin{figure}[!htbp]
 \centering
 \includegraphics[width=\textwidth]{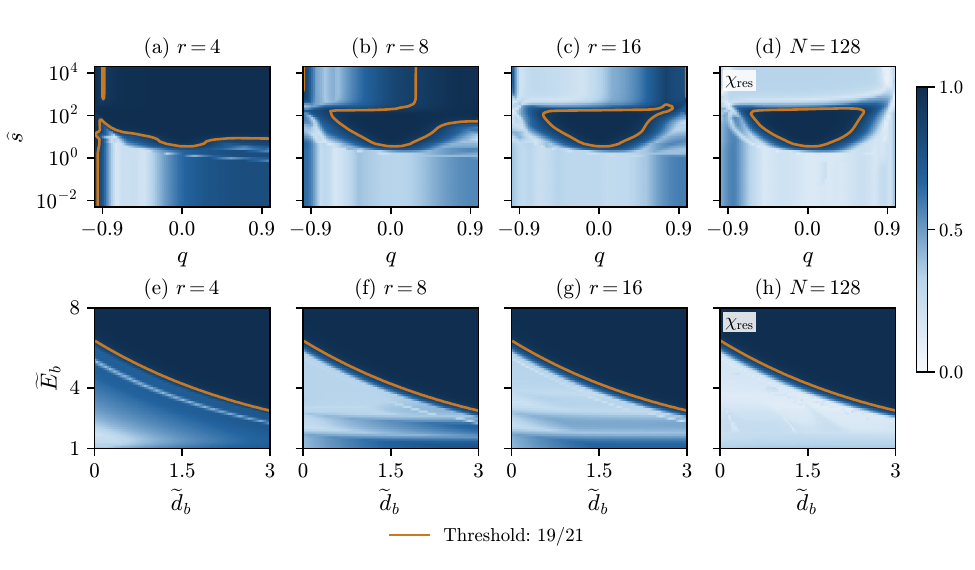}
 \caption{The scans of \cref{fig:numerical_two_state_maps,fig:numerical_potential_controls}
 repeated for \(r=4,8,16\) and the unprojected \(N=128\)
 grid. Top: \(q,\widehat s\)
 at fixed potential; bottom: \(\widetilde d_b,\widetilde E_b\)
 at fixed \(\mu,\nu\). Reduced columns show complete
 reduced-spectrum \(\chi_{\max}\). The full-grid column uses
 \(32\) resolved low-frequency eigenvalues rather than the full
 spectrum. Orange denotes the \(19/21\) contour.
 Near \(q=0\), the higher-dimensional panels show an upper exit
 from the spectral window that is absent in the two-state scan.}
 \label{fig:numerical_hierarchy}
\end{figure}

Near \(q=0\), increasing \(s\) in
\cref{fig:numerical_hierarchy} first opens and then closes the
spectral window for \(r=8,16\) and the resolved full-grid spectrum,
whereas the \(r=2,4\) scans remain strongly separated at large
\(s\). The barrier-induced boundary also persists as \(r\)
increases, consistently with the transfer bottleneck in
\cref{thm:qld_sector_origin}. Agreement between the \(r=16\) and full-grid maps
concerns these principal features.

To identify why strong dissipation closes some windows but not others,
we next separate dissipative structure from magnitude. In
\cref{fig:numerical_structure_hierarchy}, the horizontal coordinate
\(\widetilde\nu/\widetilde\mu\) varies the relative contributions
of momentum \(p_r\) and position \(x_r\) to \(L^{(r)}\);
at each ratio, common
rescaling of \(\mu,\nu\) sets the prescribed norm
\(\gamma_{\mathcal D}^{(r)}\). Each horizontal slice therefore
holds \(H_r\), \(\Omega_{\mathrm{tun}}^{(r)}\), and
\(\gamma_{\mathcal D}^{(r)}\) fixed.
For the reduced models, we compare the numerical \(\chi_{\max}\)
with \(\chi_{\max}(\widehat{\mathcal D}_r)\), the limiting diagnostic
in \cref{thm:qld_strong_dissipation_exclusion}.

\begin{figure}[!htbp]
 \centering
 \includegraphics[width=\textwidth]{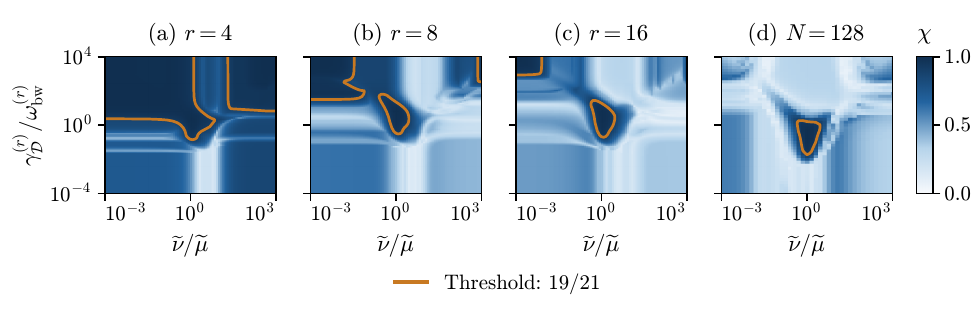}
 \caption{Dissipative structure and strength controlled independently
 in a symmetric double well: (a)--(c) the \(r=4,8,16\)
 reductions and (d) the full-grid Lindbladian \(\mathcal L_N\)
 with \(N=128\).
 The potential \(V(x)\) has \(E_b=2,d_b=1,E_\delta=0\),
 discretized on \(N=128\) points in \([-9,9]\).
 All panels share the coefficient-ratio and relative-strength ranges,
 using each model's bandwidth
 \(\omega_{\mathrm{bw}}^{(r)}=(E_{r-1}-E_0)/\hbar\),
 with \(r=N\) in (d).
 Colors show complete reduced-spectrum \(\chi_{\max}\) in (a)--(c).}
 \label{fig:numerical_structure_hierarchy}
\end{figure}

The windows in \cref{fig:numerical_structure_hierarchy} can close
along a horizontal slice: changing the channel weights can remove
spectral separation without changing the overall dissipative action.
For example, at \(r=16\) and
\(\gamma_{\mathcal D}^{(16)}/\omega_{\mathrm{bw}}^{(16)}=1\),
changing \(\widetilde\nu/\widetilde\mu\) from \(1\) to \(0.3\)
reduces \(\chi_{\max}\) from approximately \(0.987\) to
\(0.480\). Independently computed dissipator spectra explain
the different vertical responses:
at ratio \(1\), \(\chi_{\max}(\widehat{\mathcal D}_r)\)
is approximately \(0.941,0.696,0.268\) for \(r=4,8,16\),
respectively. The first exceeds \(\chi_c\), whereas the latter
two satisfy the strict spectral condition of
\cref{thm:qld_strong_dissipation_exclusion} and exhibit an upper
exit. This comparison directly tests the theorem's structural
condition: additional levels can change which damping patterns
remain separated, but neither larger \(r\) nor larger
\(\gamma_{\mathcal D}^{(r)}\) alone guarantees exclusion.

We next test whether the distinction between transfer, damping
strength, and damping structure remains useful for the multiwell
potential \(V_{\mathrm{mw}}\) and the two-dimensional double well
\(V_{\mathrm{2D}}\).
To compare their spectral response on common physical axes,
\cref{fig:numerical_landscape_maps} plots
\(\chi_{\max}^{(16)}\) against the measured coupling frequency
\(\Omega_{\mathrm{tun}}^{(16)}/\omega\) and dissipation strength
\(\gamma_{\mathcal D}^{(16)}/\omega\). The former varies
monotonically along each sampled potential family, while the
latter is independently set by common rescaling of \(\mu,\nu\).

\begin{figure}[htbp]
 \centering
 \includegraphics[width=\textwidth]{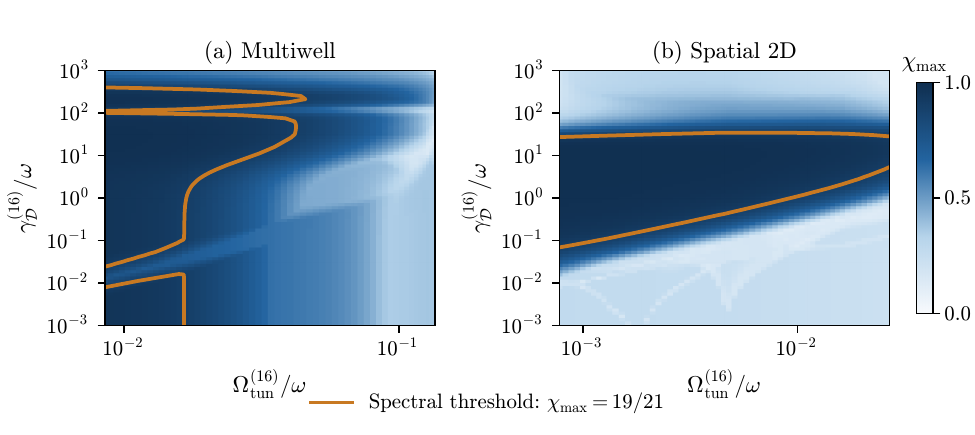}
 \caption{Full reduced-spectrum \(\chi_{\max}\) for two
 \(r=16\) landscape families, each on a \(61\times81\) grid.
 The horizontal coordinate is measured
 \(\Omega_{\mathrm{tun}}^{(16)}/\omega\);
 the vertical range is
 \(\gamma_{\mathcal D}^{(16)}/\omega\in[10^{-3},10^3]\).
 (a) Multiwell:
 \(E_{\mathrm{mw}}\in[0.25,4]\), \(\nu/\mu=0.3\),
 and \(N=128\) on \([-7,7]\).
 (b) Two-dimensional double well:
 \(d_b\in[0.05,2]\), \(\nu/\mu=1\), on a
 \(40\times28\) grid. Orange denotes \(\chi_{\max}=19/21\).}
 \label{fig:numerical_landscape_maps}
\end{figure}

The two-dimensional map in \cref{fig:numerical_landscape_maps}(b)
has a bounded window whose lower boundary rises with
\(\Omega_{\mathrm{tun}}^{(16)}\): stronger coherent transfer
requires stronger dissipation before the decay rates separate.
At larger \(\gamma_{\mathcal D}^{(16)}\), the window closes again.
The multiwell map in panel (a) instead contains disjoint windows
with maximizing cuts \(i^\ast=2,3,4\), indicating different
numbers of separated slow modes. This is consistent with the
multi-sector mechanism of \cref{thm:weak_sector_certificate},
although the map alone does not establish its stationary-sector
assumptions. 

Because the preceding scans keep \(\nu/\mu\) fixed, they cannot
determine whether the windows are consequences of dissipation
strength alone. We therefore repeat the fixed-norm structure scan
at one fixed potential from each family in
\cref{fig:numerical_structure_landscapes}. As in
\cref{fig:numerical_structure_hierarchy},
horizontal comparisons change only the channel weights. Separate
calculations of \(\chi_{\max}(\widehat{\mathcal D}_{16})\)
test the strong-dissipation condition.
For the two-dimensional generator, this comparison uses the sum
of both dissipators, as required by the general criterion in
\cref{thm:general_strong_dissipation_exclusion}.

\begin{figure}[!htbp]
 \centering
 \includegraphics[width=\textwidth]{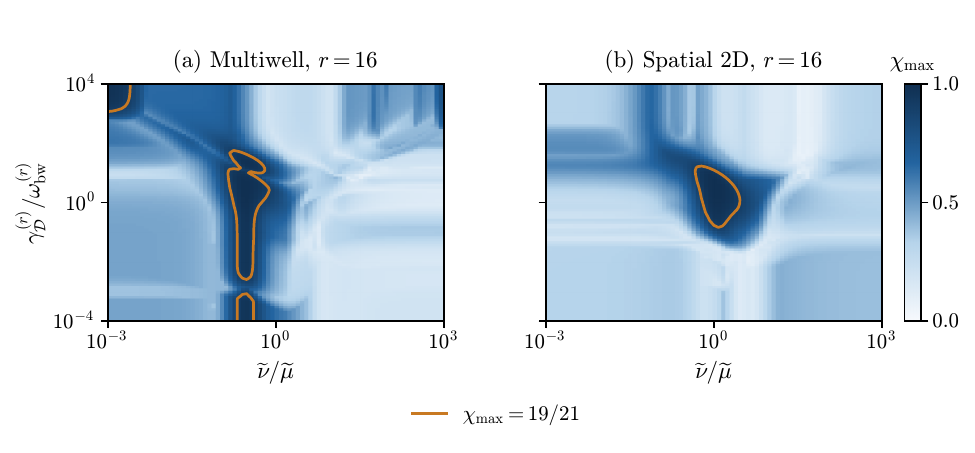}
 \caption{Fixed-norm tests of dissipative structure in complex
 landscapes, with \(r=16\).
 (a) The multiwell potential \(V_{\mathrm{mw}}\) at
 \(E_{\mathrm{mw}}=4\), on \(N=128\) points in \([-7,7]\).
 (b) The two-dimensional potential \(V_{\mathrm{2D}}\) at
 \(E_b=2,d_b=1,E_\delta=0,\omega_y=1.3,c_{xy}=0.35\),
 on a \(40\times28\) grid in \([-7,7]\times[-5,5]\), with
 common coefficients \(\mu,\nu\) for \(L_x,L_y\).
 Each \(82\times97\) scan varies
 \(\widetilde\nu/\widetilde\mu\in[10^{-3},10^3]\) horizontally
 and \(\gamma_{\mathcal D}^{(16)}/\omega_{\mathrm{bw}}^{(16)}
 \in[10^{-4},10^4]\) vertically, with
 \(\omega_{\mathrm{bw}}^{(16)}=(E_{15}-E_0)/\hbar\).
 Colors give complete reduced-spectrum \(\chi_{\max}\),
 and orange marks \(\chi_{\max}=19/21\).
 }
 \label{fig:numerical_structure_landscapes}
\end{figure}

Both maps in \cref{fig:numerical_structure_landscapes} contain
horizontal crossings, confirming that dissipative structure remains
an independent control of spectral separation. At
\(\gamma_{\mathcal D}^{(16)}/\omega_{\mathrm{bw}}^{(16)}=1\),
increasing \(\widetilde\nu/\widetilde\mu\) from \(0.3\) to
\(1\) closes the multiwell window but opens the two-dimensional
window. Thus the two-state instruction to balance
\(\delta_{\mathcal D}\) does not become a universal coefficient
ratio for larger models. The strong-dissipation ends instead
approach the independently computed limits
\(\chi_{\max}(\widehat{\mathcal D}_{16})\):
where this limit lies strictly below \(\chi_c\), sufficiently
strong dissipation excludes the window, consistently with
\cref{thm:qld_strong_dissipation_exclusion,thm:general_strong_dissipation_exclusion}.
The smallest sampled ratios in the multiwell model retain an
above-threshold dissipator limit, illustrating why increasing
\(\gamma_{\mathcal D}^{(16)}\) alone need not ensure exclusion.
\FloatBarrier

\subsection{Dissipative control of QLD for optimization}
\label{sec:numerical_control}

We first identify a dynamical plateau when immplementing QLD for optimization, then test whether the two
control routes in \cref{sec:qld_general_control} shorten the time
needed to approach the same stationary target.
We fix one potential from each family in
\cref{eq:piecewise_double_well_potential_0710,eq:numerical_multiwell_potential,eq:numerical_two_dimensional_potential}
and evolve all spatial-grid states: \(N=128\) in one dimension and
\(N=12\times6=72\) in two dimensions. For the double well, the initial state \(\rho_0\) is a
normalized real Gaussian wave packet with center \(-\ell/2\) and
position standard deviation \(0.30\). For the multiwell potential,
we dephase a Gaussian state \(\rho_G\), with center \(-3.8\) and
standard deviation \(0.25\), in the full grid-energy basis:
\(\rho_0=\sum_{n=0}^{127}P_n\rho_G P_n\), where
\(P_n=|\phi_n\rangle\langle\phi_n|\) projects onto an eigenstate
\(\phi_n\) of the grid Hamiltonian \(H\). This preparation removes
initial energy coherences without truncating the evolved state space.
In two dimensions,
\(\rho_0\) is the ground state of the grid Hamiltonian restricted
to \(x<0\), embedded in the complete grid. Within each example,
all protocols use this same \(\rho_0\) and the same baseline
stationary target \(\rho_{\mathrm{ss}}^{\mathrm b}\).

\begin{figure}[!htbp]
 \centering
 \includegraphics[width=\textwidth]{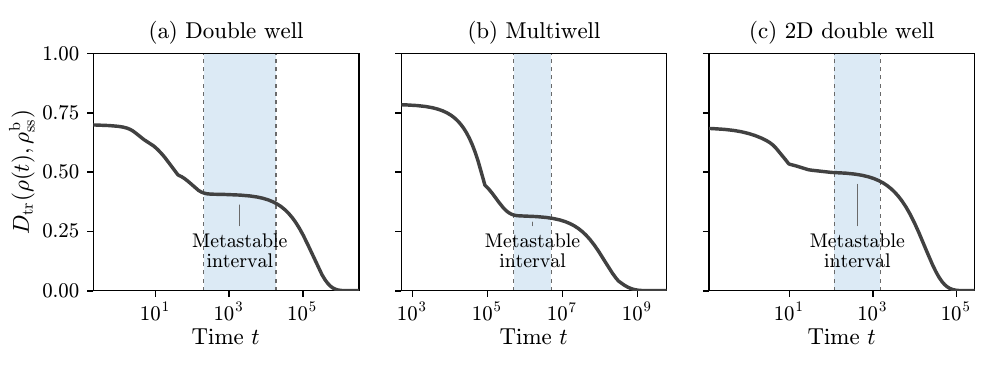}
 \caption{Baseline relaxation without dissipative control: (a) the
 symmetric double well, (b) the multiwell potential, and (c) the
 two-dimensional double well. The trace distance
 \(D_{\mathrm{tr}}(\rho(t),\rho_{\mathrm{ss}}^{\mathrm b})\) is plotted
 on a linear vertical axis against logarithmic time, resolving
 the initial transient, intermediate plateau, and final relaxation.
 The potential and dissipative parameters match
 \cref{fig:numerical_full_control}. The intervals are approximately
 \([203,18535]\), \([5\times10^5,5\times10^6]\), and
 \([120,1500]\).}
 \label{fig:numerical_full_baseline_plateaus}
\end{figure}

\Cref{fig:numerical_full_baseline_plateaus} shows the metastable plateau without dissipative control. To distinguish these plateaus from flat scalar curves, we bound the full-state
displacement throughout each blue band. All three
trajectories eventually leave these intervals and approach
\(\rho_{\mathrm{ss}}^{\mathrm b}\).

Ratio tuning changes \(\nu/\mu\) in one or two preparation stages
at fixed full dissipator norm
\(\gamma_{\mathcal D}=\|\mathcal D\|_{2\to2}\); strength tuning
keeps \(\nu/\mu\) fixed, raises \(\gamma_{\mathcal D}(t)\)
linearly over \(T_r\), and holds it for \(T_h\).
Both restore the baseline coefficients after preparation. We select one protocol per route and potential
from a finite search. All preparation time is included. The reported
controlled completion time is the first tolerance crossing after
restoration, which is a conservative upper bound on \(T_\varepsilon\) in
\cref{eq:control_settling_time}. We compare the same protocols at
\(\varepsilon=10^{-3},10^{-4},10^{-6}\), testing whether the gain
persists beyond an early crossing at loose accuracy.

\begin{figure}[p]
 \centering
 \includegraphics[width=0.90\textwidth]{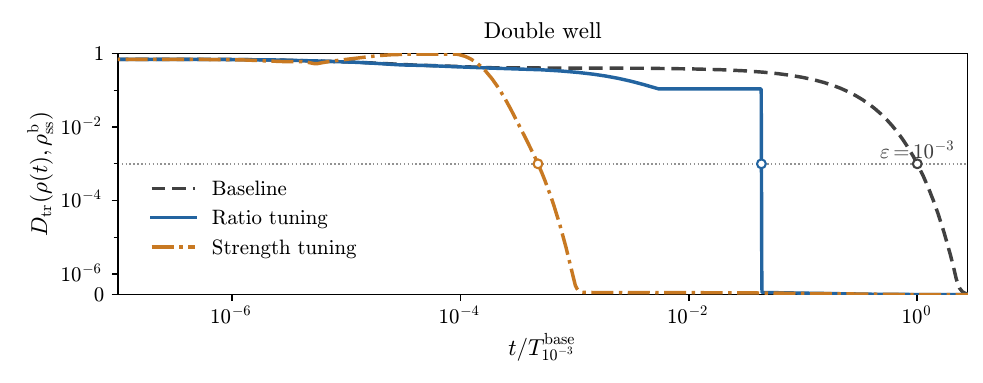}\par
 \vspace{0.5\baselineskip}
 \includegraphics[width=0.90\textwidth]{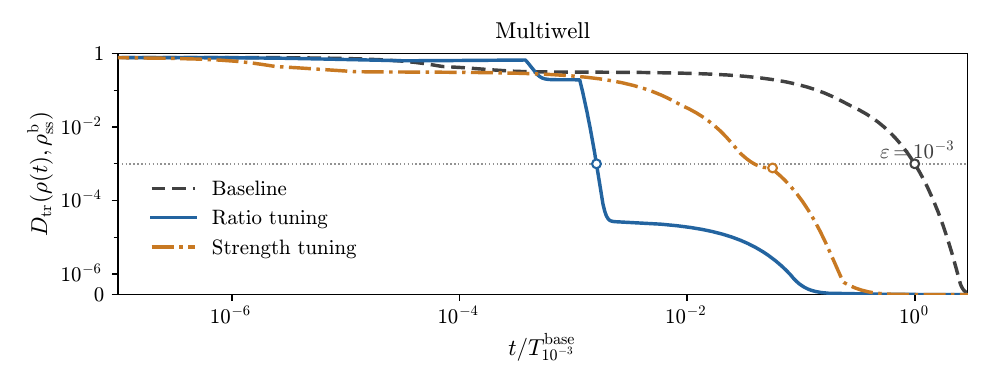}\par
 \vspace{0.5\baselineskip}
 \includegraphics[width=0.90\textwidth]{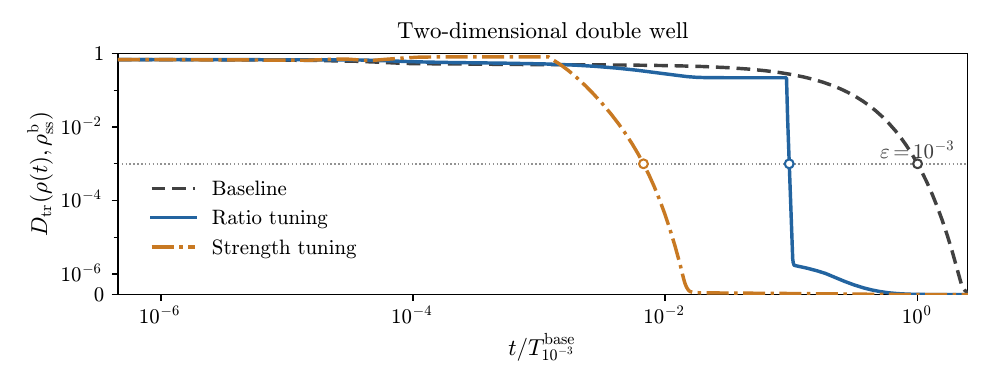}
 \caption{Full-grid dissipative control for the double well (top),
 multiwell potential (middle), and two-dimensional double well
 (bottom), using \(N=128,128,72\), respectively. Baseline
 (gray dashed), ratio tuning (blue), and strength tuning (orange)
 share the initial state \(\rho_0\) and target
 \(\rho_{\mathrm{ss}}^{\mathrm b}\) within each panel; both controls
 restore baseline. Logarithmic time includes preparation and is
 normalized by the respective baseline completion times
 \(T_{10^{-3}}^{\mathrm{base}}=1.1140\times10^6\),
 \(1.19824\times10^9\), and \(109427.40\).
 Circles mark post-restoration completion at the dotted tolerance
 \(10^{-3}\). Trace distance is logarithmic above \(10^{-6}\)
 and linear below it.}
 \label{fig:numerical_full_control}
\end{figure}

\paragraph{Double well.}
The symmetric double well provides the closest full-model test of
the slow-transfer mechanism. At \(E_b=3,d_b=1,E_\delta=0\), the
baseline \(\mu=\nu=0.08\) on \(N=128\) interior points in
\([-10,10]\) has a resolved separation
\(\chi_{\mathrm{res}}\simeq0.9996\).
After an initial decrease from approximately \(0.70\) to \(0.41\),
the trace distance changes little throughout
\(t\in[203,18535]\).
The top panel of \cref{fig:numerical_full_control} tests temporary
changes in ratio or strength: ratio tuning uses \(\nu/\mu=2\)
at fixed \(\gamma_{\mathcal D}=3.20502\) for \(T_p=47865.68\),
whereas strength tuning raises the norm one hundredfold over
\(T_r=40\) and holds it for \(T_h=62.6752\).

At \(\varepsilon=10^{-3}\), the baseline completion time
\(1.1140\times10^6\) falls to \(48061\) with ratio tuning and
\(531.25\) with strength tuning. Both protocols reduce the
occupied slow-component weight from \(0.407\) to approximately
\(10^{-7}\) before restoration, realizing the finite-accuracy
suppression mechanism in \cref{rem:control_robustness}.
The gain survives a tighter tolerance at \(10^{-6}\). The same
protocols improve the completion time by approximately \(49\)
and \(2249\). 

\paragraph{Multiwell potential.}
Additional wells test whether the same strategy can address several
slow redistribution directions. We use the potential family
\(V_{\mathrm{mw}}\) in \cref{eq:numerical_multiwell_potential}, with
barrier amplitude \(E_{\mathrm{mw}}=8\), \(N=128\) points in
\([-7,7]\), baseline ratio
\(\nu/\mu=0.17\), and norm \(\gamma_{\mathcal D}=0.003\).
The distance decreases from \(0.787\) to \(0.319\) before the
interval \([5\times10^5,5\times10^6]\), throughout which the
full-state displacement remains below \(0.030\).
To suppress the occupied slow sector, fixed-norm ratio tuning uses
\(\nu/\mu=2\) for \(453418.89\) time units, then \(0.5\) for
\(911873.47\). Alternatively, strength tuning raises the norm
thirtyfold over \(T_r=10\) and holds it for
\(T_h=6.756503525\times10^7\).

In the middle panel of \cref{fig:numerical_full_control}, ratio tuning reduces
the completion time at \(\varepsilon=10^{-3}\) from
\(1.19824\times10^9\) to \(1.91830\times10^6\). This example supports the need to control the
occupied slow sector in \cref{rem:control_robustness}; increasing
\(\gamma_{\mathcal D}\) can also improve absolute relaxation times
without necessarily eliminating relative spectral separation.

\paragraph{Two-dimensional double well.}
We retain both spatial coordinates and both jump operators
\(L_x,L_y\) of \cref{eq:numerical_two_dimensional_potential}.
We fix \(E_b=4,d_b=1,E_\delta=0\), retaining
\(\omega_y=1.3,c_{xy}=0.35\). The full \(12\times6\) grid on
\([-7,7]\times[-5,5]\) evolves all \(72^2\) density-matrix
entries. Its baseline \(\chi_{\mathrm{res}}\simeq0.987\) remains
approximately \(0.987\) on a \(14\times6\) grid when the ratio
and full dissipator norm are held fixed. This example tests whether
the control mechanism survives an additional spatial direction.
From baseline coefficients
\((\mu_{\mathrm b},\nu_{\mathrm b})=(0.228465,0.114233)\),
ratio tuning applies \(\mu=\nu=0.228047\) to both spatial channels
at fixed \(\gamma_{\mathcal D}=10\) for \(T_p=10000\).
Strength tuning instead raises the norm thirtyfold over \(T_r=10\)
and holds it for \(T_h=120\). The bottom panel of \cref{fig:numerical_full_control} shows gains of
\(10.4\) and \(149\) at \(\varepsilon=10^{-3}\), for ratio
and strength tuning, respectively. The same schedules retain gains
of \(13.9\) and \(151\) at \(10^{-4}\), and approximately
\(11.0\) and \(154\) at \(10^{-6}\). 
\FloatBarrier
\endgroup

\section{Conclusions and future work}
\label{sec:conclusion_future_work}

Metastability separates rapid local relaxation from slow global
redistribution. We have examined when this separation becomes a
spectral bottleneck and how an accessible dissipative control can
bypass it. We use QLD as a concrete example to study how coherent dynamics
and dissipation jointly influence the formation and suppression
of metastability. The symmetric
two-state model yields an exact spectral-window criterion. For the asymmetric two-state model, we identify conditions for
well-separated relaxation timescales under weak or strong dissipation.
The multi-state and general Lindbladian analyses extend these results:
weak coupling between sectors can produce well-separated relaxation
timescales, while sufficiently strong dissipation can remove this
separation when the dissipator itself has no spectral bottleneck. 

This analysis also shows that a slow mode in the generator need not
remain populated along a controlled trajectory.
We establish conditions for reaching the same target faster by
either suppressing slowly decaying components or adjusting the dissipation strength. The comparison includes the time
spent on control and requires the final stationary state to remain
within the prescribed accuracy of the target.
Double-well, multiwell, and
spatially two-dimensional numerical experiments exhibit shorter fixed-target
convergence times under both routes in QLD models. Changing the dissipative structure can accelerate convergence even
when the dissipator norm is held fixed. The improvement can also
persist after a temporary control ends and the original dynamics
is restored. In the protocols tested, gradual parameter changes
offer no advantage over direct switches. These results show how
identifying the slowly decaying components can guide the design
of dissipative control.

Future work could be:
\begin{itemize}
    \item Investigate whether quantum replica exchange
\cite{chen2025quantum,leng2025operator} can remove the metastable region and thereby reduces mixing times to the optimal state in optimization. Quantum transport bounds~\cite{rakovszky2024bottlenecks} could help
determine how much these mixing times can be reduced.
\item Structural results for thermal metastability
\cite{bergamaschi2025structural} also motivate examining information
retained in metastable states for optimization problems. 
\item An optimization advantage from a shorter metastable plateau also
requires an accurate target, numerical convergence, and accounting
for the total computational cost. 
\end{itemize}

\section*{AI statement}
The authors used GPT-5.5, GPT-5.6, and GPT-6 (OpenAI) to assist with
intermediate proof steps and algebraic checks, numerical implementation
and debugging, result analysis and visualization, literature searches,
and manuscript editing. All theorem statements and main research ideas
were formulated by the authors, who take full responsibility for the
correctness of the mathematical arguments and numerical results,
the accuracy of citations, and the overall content of this work.

\bibliography{ref.bib}

@article{catli2026efficient,
  title={Efficient Quantum Optimization via Dynamical Simulation},
  author={Catli, Ahmet Burak and Simon, Sophia and Wiebe, Nathan},
  journal={PRX Quantum},
  volume={7},
  number={3},
  pages={033026},
  year={2026},
  publisher={APS},
  doi = {10.1103/t59v-8flk},
  url = {https://doi.org/10.1103/t59v-8flk},
}

@article{olmos2012facilitated,
  author  = {Olmos, Beatriz and Lesanovsky, Igor and Garrahan, Juan P.},
  title   = {Facilitated Spin Models of Dissipative Quantum Glasses},
  journal = {Physical Review Letters},
  year    = {2012},
  volume  = {109},
  number  = {2},
  pages   = {020403},
  doi     = {10.1103/PhysRevLett.109.020403},
  url = {https://doi.org/10.1103/PhysRevLett.109.020403},
}

@article{fink2018signatures,
  author  = {Fink, Thomas and Schade, Anne and H{\"o}fling, Sven
             and Schneider, Christian and {\.I}mamo{\u{g}}lu, Ata{\c{c}}},
  title   = {Signatures of a dissipative phase transition in photon
             correlation measurements},
  journal = {Nature Physics},
  year    = {2018},
  volume  = {14},
  number  = {4},
  pages   = {365--369},
  doi     = {10.1038/s41567-017-0020-9},
  url = {https://doi.org/10.1038/s41567-017-0020-9},
}

@article{minganti2018spectral,
  author  = {Minganti, Fabrizio and Biella, Alberto and Bartolo, Nicola
             and Ciuti, Cristiano},
  title   = {Spectral theory of {Liouvillians} for dissipative phase
             transitions},
  journal = {Physical Review A},
  year    = {2018},
  volume  = {98},
  number  = {4},
  pages   = {042118},
  doi     = {10.1103/PhysRevA.98.042118},
  url = {https://doi.org/10.1103/PhysRevA.98.042118},
}

@article{denchev2016computational,
  author  = {Denchev, Vasil S. and Boixo, Sergio and Isakov, Sergei V.
             and Ding, Nan and Babbush, Ryan and Smelyanskiy, Vadim
             and Martinis, John and Neven, Hartmut},
  title   = {What is the Computational Value of Finite-Range Tunneling?},
  journal = {Physical Review X},
  year    = {2016},
  volume  = {6},
  number  = {3},
  pages   = {031015},
  doi     = {10.1103/PhysRevX.6.031015},
  url = {https://doi.org/10.1103/PhysRevX.6.031015},
}

@article{muthukrishnan2016tunneling,
  author  = {Muthukrishnan, Siddharth and Albash, Tameem
             and Lidar, Daniel A.},
  title   = {Tunneling and Speedup in Quantum Optimization for
             Permutation-Symmetric Problems},
  journal = {Physical Review X},
  year    = {2016},
  volume  = {6},
  number  = {3},
  pages   = {031010},
  doi     = {10.1103/PhysRevX.6.031010},
  url = {https://doi.org/10.1103/PhysRevX.6.031010},
}

@article{plenio2008dephasing,
  author  = {Plenio, M. B. and Huelga, S. F.},
  title   = {Dephasing-assisted transport: quantum networks and biomolecules},
  journal = {New Journal of Physics},
  year    = {2008},
  volume  = {10},
  number  = {11},
  pages   = {113019},
  doi     = {10.1088/1367-2630/10/11/113019},
  url = {https://doi.org/10.1088/1367-2630/10/11/113019},
}

@article{syassen2008strong,
  author  = {Syassen, N. and Bauer, D. M. and Lettner, M. and Volz, T.
             and Dietze, D. and Garc{\'i}a-Ripoll, J. J. and Cirac, J. I.
             and Rempe, G. and D{\"u}rr, S.},
  title   = {Strong Dissipation Inhibits Losses and Induces Correlations
             in Cold Molecular Gases},
  journal = {Science},
  year    = {2008},
  volume  = {320},
  number  = {5881},
  pages   = {1329--1331},
  doi     = {10.1126/science.1155309},
  url = {https://doi.org/10.1126/science.1155309},
}

@article{chebotarev1998sufficient,
  author  = {Chebotarev, Alexander M. and Fagnola, Franco},
  title   = {Sufficient Conditions for Conservativity of Minimal
             Quantum Dynamical Semigroups},
  journal = {Journal of Functional Analysis},
  year    = {1998},
  volume  = {153},
  number  = {2},
  pages   = {382--404},
  doi     = {10.1006/jfan.1997.3189},
  url = {https://doi.org/10.1006/jfan.1997.3189},
}

@article{geman1986diffusions,
  title   = {Diffusions for Global Optimization},
  author  = {Geman, Stuart and Hwang, Chii-Ruey},
  journal = {SIAM Journal on Control and Optimization},
  volume  = {24},
  number  = {5},
  pages   = {1031--1043},
  year    = {1986},
  doi     = {10.1137/0324060},
  url = {https://doi.org/10.1137/0324060},
}

@article{chiang1987diffusion,
  title   = {Diffusion for Global Optimization in {$\mathbb{R}^n$}},
  author  = {Chiang, Tzuu-Shuh and Hwang, Chii-Ruey and Sheu, Shuenn-Jyi},
  journal = {SIAM Journal on Control and Optimization},
  volume  = {25},
  number  = {3},
  pages   = {737--753},
  year    = {1987},
  doi     = {10.1137/0325042},
  url = {https://doi.org/10.1137/0325042},
}

@inproceedings{zhang2017hitting,
  title     = {A Hitting Time Analysis of Stochastic Gradient {Langevin} Dynamics},
  author    = {Zhang, Yuchen and Liang, Percy and Charikar, Moses},
  booktitle = {Proceedings of the 2017 Conference on Learning Theory},
  series    = {Proceedings of Machine Learning Research},
  volume    = {65},
  pages     = {1980--2022},
  year      = {2017},
  publisher = {PMLR},
  url       = {https://proceedings.mlr.press/v65/zhang17b.html}
}

@article{gao2022global,
  title   = {Global Convergence of Stochastic Gradient {Hamiltonian Monte Carlo} for Nonconvex Stochastic Optimization: Nonasymptotic Performance Bounds and Momentum-Based Acceleration},
  author  = {Gao, Xuefeng and G{\"u}rb{\"u}zbalaban, Mert and Zhu, Lingjiong},
  journal = {Operations Research},
  volume  = {70},
  number  = {5},
  pages   = {2931--2947},
  year    = {2022},
  doi     = {10.1287/opre.2021.2162},
  url = {https://doi.org/10.1287/opre.2021.2162},
}

@article{chak2023generalized,
  title   = {On the Generalized {Langevin} Equation for Simulated Annealing},
  author  = {Chak, Martin and Kantas, Nikolas and Pavliotis, Grigorios A.},
  journal = {SIAM/ASA Journal on Uncertainty Quantification},
  volume  = {11},
  number  = {1},
  pages   = {139--167},
  year    = {2023},
  doi     = {10.1137/21M1462970},
  url = {https://doi.org/10.1137/21M1462970},
}

@article{herman2023constrained,
  title={Constrained optimization via quantum {Zeno} dynamics},
  author={Herman, Dylan and Shaydulin, Ruslan and Sun, Yue and Chakrabarti, Shouvanik and Hu, Shaohan and Minssen, Pierre and Rattew, Arthur and Yalovetzky, Romina and Pistoia, Marco},
  journal={Communications Physics},
  volume={6},
  number={1},
  pages={219},
  year={2023},
  publisher={Nature Publishing Group UK London},
  doi = {10.1038/s42005-023-01331-9},
  url = {https://doi.org/10.1038/s42005-023-01331-9},
}

@inproceedings{chakrabarti2026provable,
  title={Provable Speedups in Convex Optimization via Quantum Dynamics},
  author={Chakrabarti, Shouvanik and Herman, Dylan and Watkins, Jacob and Fontana, Enrico and Augustino, Brandon and Kim, Junhyung Lyle and Pistoia, Marco},
  booktitle={21st Conference on the Theory of Quantum Computation, Communication and Cryptography (TQC 2026)},
  pages={4--1},
  year={2026},
  organization={Schloss Dagstuhl--Leibniz-Zentrum f{\"u}r Informatik},
  doi = {10.4230/LIPIcs.TQC.2026.4},
  url = {https://doi.org/10.4230/LIPIcs.TQC.2026.4},
}

@article{abbas2024challenges,
  title={Challenges and opportunities in quantum optimization},
  author={Abbas, Amira and Ambainis, Andris and Augustino, Brandon and B{\"a}rtschi, Andreas and Buhrman, Harry and Coffrin, Carleton and Cortiana, Giorgio and Dunjko, Vedran and Egger, Daniel J and Elmegreen, Bruce G and others},
  journal={Nature Reviews Physics},
  volume={6},
  number={12},
  pages={718--735},
  year={2024},
  publisher={Nature Publishing Group},
  doi = {10.1038/s42254-024-00770-9},
  url = {https://doi.org/10.1038/s42254-024-00770-9},
}

@article{wang2026efficient,
  title   = {Efficient learning of quantum interactions from thermal metastable states},
  author  = {Wang, Bingrun and Ye, Qi and Chen, Chi-Fang},
  journal = {arXiv preprint arXiv:2610.01538},
  year    = {2026},
  url     = {https://arxiv.org/abs/2610.01538}
}

@article{li2026quantum,
  title   = {Quantum {Markov} State Models for Metastable Dynamics},
  author  = {Li, Hao-En and Lin, Lin and Ragone, Michael},
  journal = {arXiv preprint arXiv:2609.40214},
  year    = {2026},
  url     = {https://arxiv.org/abs/2609.40214}
}

@article{escobar2026metastability,
  title   = {On the Metastability of the Mean-Field Interchange Model for Local Dimension {$\geq 3$}},
  author  = {Escobar, Sergio and Lin, Lin and Ragone, Michael and Stubbs, Kevin D.},
  journal = {arXiv preprint arXiv:2610.01987},
  year    = {2026},
  url     = {https://arxiv.org/abs/2610.01987}
}

@article{lindblad1976generators,
  title={On the generators of quantum dynamical semigroups},
  author={Lindblad, Goran},
  journal={Communications in Mathematical Physics},
  volume={48},
  number={2},
  pages={119--130},
  year={1976},
  publisher={Springer},
  doi = {10.1007/BF01608499},
  url = {https://doi.org/10.1007/BF01608499},
}

@article{bergamaschi2025structural,
  title={A Structural Theory of Quantum Metastability: {M}arkov Properties and Area Laws},
  author={Bergamaschi, Thiago and Chen, Chi-Fang and Vazirani, Umesh},
  journal={arXiv preprint arXiv:2510.08538},
  year={2025},
  url = {https://arxiv.org/abs/2510.08538},
}

@article{chen2025quantum,
  title={Quantum Replica Exchange},
  author={Chen, Zherui and Basso, Joao and Ding, Zhiyan and Lin, Lin},
  journal={arXiv preprint arXiv:2510.07291},
  year={2025},
  doi = {10.48550/arXiv.2510.07291},
  url = {https://doi.org/10.48550/arXiv.2510.07291},
}

@article{rakovszky2024bottlenecks,
  title={Bottlenecks in Quantum Channels and Finite Temperature Phases of Matter},
  author={Rakovszky, Tibor and Placke, Benedikt and Breuckmann, Nikolas P. and Khemani, Vedika},
  journal={Physical Review Letters},
  volume={137},
  number={5},
  pages={050402},
  year={2026},
  doi={10.1103/y9z2-52rl},
  url={https://doi.org/10.1103/y9z2-52rl},
  eprint={2412.09598},
  archivePrefix={arXiv}
}

@article{leng2025operator,
  title={Operator-level quantum acceleration of non-logconcave sampling},
  author={Leng, Jiaqi and Ding, Zhiyan and Chen, Zherui and Lin, Lin},
  journal={Proceedings of the National Academy of Sciences},
  volume={123},
  number={8},
  pages={e2512789123},
  year={2026},
  doi={10.1073/pnas.2512789123},
  url={https://doi.org/10.1073/pnas.2512789123},
  eprint={2505.05301},
  archivePrefix={arXiv}
}

@article{chen2023quantum,
  title={Quantum {Langevin} Dynamics for Optimization},
  author={Chen, Zherui and Lu, Yuchen and Wang, Hao and Liu, Yizhou and Li, Tongyang},
  journal={Communications in Mathematical Physics},
  volume={406},
  number={3},
  pages={52},
  year={2025},
  doi={10.1007/s00220-025-05234-4},
  url={https://doi.org/10.1007/s00220-025-05234-4},
  eprint={2311.15587},
  archivePrefix={arXiv}
}

@article{leng2023quantum,
  title={Quantum {H}amiltonian descent},
  author={Leng, Jiaqi and Hickman, Ethan and Li, Joseph and Wu, Xiaodi},
  journal={arXiv preprint arXiv:2303.01471},
  year={2023},
  doi = {10.48550/arXiv.2303.01471},
  url = {https://doi.org/10.48550/arXiv.2303.01471},
}

@inproceedings{chen2024local,
title={{Local minima in quantum systems}},
author={Chen, Chi-Fang and Huang, Hsin-Yuan and Preskill, John and Zhou, Leo},
booktitle={Proceedings of the 56th Annual ACM Symposium on Theory of Computing},
pages={1323--1330},
year={2024},
eprint="2309.16596",
  doi = {10.1145/3618260.3649675},
  url = {https://doi.org/10.1145/3618260.3649675},
}

@article{leng2025quantum,
  title={Quantum {H}amiltonian Descent for Non-smooth Optimization},
  author={Leng, Jiaqi and Zheng, Yufan and Jia, Zhiyuan and Fan, Lei and Zhao, Chaoyue and Peng, Yuxiang and Wu, Xiaodi},
  journal={arXiv preprint arXiv:2503.15878},
  year={2025},
  doi = {10.48550/arXiv.2503.15878},
  url = {https://doi.org/10.48550/arXiv.2503.15878},
}

@article{macieszczak2016towards,
  title={Towards a theory of metastability in open quantum dynamics},
  author={Macieszczak, Katarzyna and Gu{\c{t}}{\u{a}}, M{\u{a}}d{\u{a}}lin and Lesanovsky, Igor and Garrahan, Juan P},
  journal={Physical Review Letters},
  volume={116},
  number={24},
  pages={240404},
  year={2016},
  doi={10.1103/PhysRevLett.116.240404},
  eprint={1512.05801},
  archivePrefix={arXiv},
  url = {https://doi.org/10.1103/PhysRevLett.116.240404},
}

@article{brown2024unravelling,
  author={Brown, Calum A. and Macieszczak, Katarzyna and Jack, Robert L.},
  title={Unraveling metastable {Markovian} open quantum systems},
  journal={Physical Review A},
  volume={109}, pages={022244}, year={2024},
  doi={10.1103/PhysRevA.109.022244},
  eprint={2308.14107}, archivePrefix={arXiv},
  url = {https://doi.org/10.1103/PhysRevA.109.022244},
}

@article{macieszczak2021classical,
  author={Macieszczak, Katarzyna and Rose, Dominic C. and Lesanovsky, Igor and Garrahan, Juan P.},
  title={Theory of classical metastability in open quantum systems},
  journal={Physical Review Research},
  volume={3}, pages={033047}, year={2021},
  doi={10.1103/PhysRevResearch.3.033047},
  eprint={2006.01227}, archivePrefix={arXiv},
  url = {https://doi.org/10.1103/PhysRevResearch.3.033047},
}

@inproceedings{azouit2016adiabatic,
  author={Azouit, Remi and Sarlette, Alain and Rouchon, Pierre},
  title={Adiabatic elimination for open quantum systems with effective {Lindblad} master equations},
  booktitle={2016 IEEE 55th Conference on Decision and Control (CDC)},
  pages={4559--4565},
  year={2016}, eprint={1603.04630}, archivePrefix={arXiv},
  doi={10.1109/CDC.2016.7798963},
  url={https://doi.org/10.1109/CDC.2016.7798963}
}

@article{carollo2021mpemba,
  author={Carollo, Federico and Lasanta, Antonio and Lesanovsky, Igor},
  title={Exponentially Accelerated Approach to Stationarity in {Markovian} Open Quantum Systems through the {Mpemba} Effect},
  journal={Physical Review Letters}, volume={127}, pages={060401},
  year={2021}, doi={10.1103/PhysRevLett.127.060401},
  eprint={2103.05020}, archivePrefix={arXiv},
  url = {https://doi.org/10.1103/PhysRevLett.127.060401},
}

@article{kochsiek2022accelerating,
  author={Kochsiek, Simon and Carollo, Federico and Lesanovsky, Igor},
  title={Accelerating the approach of dissipative quantum spin systems towards stationarity through global spin rotations},
  journal={Physical Review A},
  volume={106},
  pages={012207},
  year={2022},
  doi={10.1103/PhysRevA.106.012207},
  eprint={2204.05339},
  archivePrefix={arXiv},
  url = {https://doi.org/10.1103/PhysRevA.106.012207},
}

@article{szehr2015spectral,
  author={Szehr, Oleg and Reeb, David and Wolf, Michael M.},
  title={Spectral convergence bounds for classical and quantum {Markov} processes},
  journal={Communications in Mathematical Physics},
  volume={333}, pages={565--595}, year={2015},
  doi={10.1007/s00220-014-2188-5},
  eprint={1301.4827}, archivePrefix={arXiv},
  url = {https://doi.org/10.1007/s00220-014-2188-5},
}

@article{zanardi2014coherent,
  author={Zanardi, Paolo and Campos Venuti, Lorenzo},
  title={Coherent Quantum Dynamics in Steady-State Manifolds of Strongly Dissipative Systems},
  journal={Physical Review Letters}, volume={113}, pages={240406}, year={2014},
  doi={10.1103/PhysRevLett.113.240406},
  eprint={1404.4673}, archivePrefix={arXiv},
  url = {https://doi.org/10.1103/PhysRevLett.113.240406},
}

@article{bao2025temporary,
  author={Bao, Ruicheng and Hou, Zhonghuai},
  title={Accelerating Quantum Relaxation via Temporary Reset: A {Mpemba}-Inspired Approach},
  journal={Physical Review Letters},
  volume={135},
  pages={150403},
  year={2025},
  doi={10.1103/g94p-7421},
  eprint={2212.11170},
  archivePrefix={arXiv},
  url = {https://doi.org/10.1103/g94p-7421},
}

@article{liu2023tunneling,
  author={Liu, Yizhou and Su, Weijie J. and Li, Tongyang},
  title={On Quantum Speedups for Nonconvex Optimization via Quantum Tunneling Walks},
  journal={Quantum},
  volume={7},
  pages={1030},
  year={2023},
  doi={10.22331/q-2023-06-02-1030},
  eprint={2209.14501},
  archivePrefix={arXiv},
  url = {https://doi.org/10.22331/q-2023-06-02-1030},
}

@article{rose2016metastability,
  author={Rose, Dominic C. and Macieszczak, Katarzyna and Lesanovsky, Igor and Garrahan, Juan P.},
  title={Metastability in an open quantum {Ising} model},
  journal={Physical Review E},
  volume={94},
  number={5},
  pages={052132},
  year={2016},
  doi={10.1103/PhysRevE.94.052132},
  url = {https://doi.org/10.1103/PhysRevE.94.052132},
}

@article{casteels2017critical,
  author={Casteels, W. and Fazio, R. and Ciuti, C.},
  title={Critical dynamical properties of a first-order dissipative phase transition},
  journal={Physical Review A},
  volume={95},
  number={1},
  pages={012128},
  year={2017},
  doi={10.1103/PhysRevA.95.012128},
  url = {https://doi.org/10.1103/PhysRevA.95.012128},
}

@article{rodriguez2017probing,
  author={Rodriguez, S. R. K. and Casteels, W. and Storme, F. and Carlon Zambon, N. and Sagnes, I. and Le Gratiet, L. and Galopin, E. and Lema{\^i}tre, A. and Amo, A. and Ciuti, C. and Bloch, J.},
  title={Probing a Dissipative Phase Transition via Dynamical Optical Hysteresis},
  journal={Physical Review Letters},
  volume={118},
  number={24},
  pages={247402},
  year={2017},
  doi={10.1103/PhysRevLett.118.247402},
  url = {https://doi.org/10.1103/PhysRevLett.118.247402},
}

@inproceedings{welling2011bayesian,
  title = {{Bayesian} Learning via Stochastic Gradient {Langevin} Dynamics},
  author = {Welling, Max and Teh, Yee Whye},
  booktitle = {Proceedings of the 28th International Conference on Machine Learning},
  pages = {681--688},
  year = {2011},
  url = {https://icml.cc/2011/papers/398_icmlpaper.pdf}
}

@inproceedings{raginsky2017nonconvex,
  title = {Non-convex learning via Stochastic Gradient {Langevin} Dynamics: a nonasymptotic analysis},
  author = {Raginsky, Maxim and Rakhlin, Alexander and Telgarsky, Matus},
  booktitle = {Proceedings of the 2017 Conference on Learning Theory},
  series = {Proceedings of Machine Learning Research},
  volume = {65},
  pages = {1674--1703},
  publisher = {PMLR},
  year = {2017},
  url = {https://proceedings.mlr.press/v65/raginsky17a.html}
}

@inproceedings{leng2025gradient,
  title={Quantum Optimization via Gradient-Based {H}amiltonian Descent},
  author={Leng, Jiaqi and Shi, Bin},
  booktitle={Proceedings of the 42nd International Conference on Machine Learning},
  pages={33941--33959},
  year={2025},
  volume={267},
  series={Proceedings of Machine Learning Research},
  publisher={PMLR},
  url={https://proceedings.mlr.press/v267/leng25a.html},
  eprint={2505.14670},
  archivePrefix={arXiv}
}
\bibliographystyle{alpha-href}

\appendix
\newpage

\section{Dynamics of a non-diagonalizable Lindbladian \texorpdfstring{\(\mathcal L\)}{L}}
\label{app:non_diagonalizable_dynamics}

The decay rates \(\Gamma_j\) in \cref{eq:decay_rates_Gamma} are
defined even when the finite-dimensional generator \(\mathcal L\)
is not diagonalizable. Its propagator \(e^{t\mathcal L}\) must then
be expanded in generalized eigenmatrices. We give that expansion to
identify the polynomial factors affecting finite-time decay. Under
column-stacking vectorization, the generator matrix
\(\widehat{\mathcal L}\) and initial state \(\rho_0\) satisfy
\begin{align}
 \frac{\mathrm d}{\mathrm dt}|\rho(t)\rangle\!\rangle
 &=\widehat{\mathcal L}|\rho(t)\rangle\!\rangle,
 \qquad |\rho(t)\rangle\!\rangle
 =e^{t\widehat{\mathcal L}}|\rho_0\rangle\!\rangle.
 \label{eq:jordan_vectorized_dynamics}
\end{align}

Choose a complete Jordan basis and let \(\alpha\) label its blocks.
For the block eigenvalue \(\lambda_\alpha\) and block size
\(q_\alpha\geq1\), write the chain as
\begin{align}
 |\mathbf R_{\alpha,0}\rangle\!\rangle,\ldots,
 |\mathbf R_{\alpha,q_\alpha-1}\rangle\!\rangle,
 \qquad |\mathbf R_{\alpha,j}\rangle\!\rangle
 =\operatorname{vec}(\mathbf R_{\alpha,j}).
 \label{eq:jordan_chain_basis}
\end{align}
The ordinary eigenmatrix \(\mathbf R_{\alpha,0}\) and generalized
eigenmatrices \(\mathbf R_{\alpha,j}\) obey
\begin{align}
 (\mathcal L-\lambda_\alpha\mathcal I)(\mathbf R_{\alpha,0})&=0,
 \label{eq:jordan_chain_eigenmatrix}\\
 (\mathcal L-\lambda_\alpha\mathcal I)(\mathbf R_{\alpha,j})
 &=\mathbf R_{\alpha,j-1},\qquad1\leq j<q_\alpha.
 \label{eq:jordan_chain_recursion}
\end{align}
Here \(\mathcal I\) is the identity superoperator. The subscript
\(j\) indexes a chain, not the
projection size \(r\) used in the main text.

Let \(\mathcal P_\alpha\) be the projection onto this Jordan block
along the other chosen blocks, and let \(\mathcal I_\alpha\) be
the identity on \(\operatorname{Ran}\mathcal P_\alpha\). The
restricted generator and its nilpotent part \(\mathcal N_\alpha\)
satisfy
\begin{align}
 \mathcal L|_{\operatorname{Ran}\mathcal P_\alpha}
 &=\lambda_\alpha\mathcal I_\alpha+\mathcal N_\alpha,
 \qquad \mathcal N_\alpha^{q_\alpha}=0,
 \label{eq:jordan_block_form}\\
 \mathcal N_\alpha\mathbf R_{\alpha,0}&=0,\qquad
 \mathcal N_\alpha\mathbf R_{\alpha,j}=\mathbf R_{\alpha,j-1}
 \quad(1\leq j<q_\alpha).
 \label{eq:jordan_nilpotent_action}
\end{align}
Since \(\lambda_\alpha\mathcal I_\alpha\) commutes with
\(\mathcal N_\alpha\), the exponential series terminates:
\begin{align}
 e^{t\mathcal L}|_{\operatorname{Ran}\mathcal P_\alpha}
 &=e^{t\lambda_\alpha}e^{t\mathcal N_\alpha}
 =e^{t\lambda_\alpha}\left(\mathcal I_\alpha+t\mathcal N_\alpha
       +\frac{t^2}{2!}\mathcal N_\alpha^2+\cdots
       +\frac{t^{q_\alpha-1}}{(q_\alpha-1)!}
                        \mathcal N_\alpha^{q_\alpha-1}\right).
 \label{eq:jordan_block_propagator}
\end{align}
Applying this propagator to \(\mathbf R_{\alpha,j}\) gives
\begin{align}
 e^{t\mathcal L}\mathbf R_{\alpha,j}
 &=e^{t\lambda_\alpha}\sum_{p=0}^{j}\frac{t^p}{p!}
                              \mathbf R_{\alpha,j-p}.
 \label{eq:jordan_chain_evolution}
\end{align}
For the first three chain entries, where present, this reads
\begin{align}
 e^{t\mathcal L}\mathbf R_{\alpha,0}
 &=e^{t\lambda_\alpha}\mathbf R_{\alpha,0},
 \label{eq:jordan_chain_evolution_r0}\\
 e^{t\mathcal L}\mathbf R_{\alpha,1}
 &=e^{t\lambda_\alpha}(\mathbf R_{\alpha,1}
                              +t\mathbf R_{\alpha,0}),
 \label{eq:jordan_chain_evolution_r1}\\
 e^{t\mathcal L}\mathbf R_{\alpha,2}
 &=e^{t\lambda_\alpha}(\mathbf R_{\alpha,2}
             +t\mathbf R_{\alpha,1}+\tfrac{t^2}{2}\mathbf R_{\alpha,0}).
 \label{eq:jordan_chain_evolution_r2}
\end{align}

Completeness of the Jordan basis allows expansion of any initial
operator \(\rho_0\). Crucially, a single decaying block contains
only a component of \(\rho_0\), not generally a density operator.
For coefficients \(c_{\alpha,j}\), write
\begin{align}
 \mathcal P_\alpha\rho_0
 &=\sum_{j=0}^{q_\alpha-1}c_{\alpha,j}\mathbf R_{\alpha,j}.
 \label{eq:jordan_initial_expansion}
\end{align}
Indeed, for \(\lambda_\alpha\ne0\), trace preservation and
\cref{eq:jordan_chain_eigenmatrix,eq:jordan_chain_recursion} give
\begin{align}
 \lambda_\alpha\operatorname{Tr}\mathbf R_{\alpha,0}&=0,
 \qquad
 \lambda_\alpha\operatorname{Tr}\mathbf R_{\alpha,j}
   =-\operatorname{Tr}\mathbf R_{\alpha,j-1}=0.
 \label{eq:jordan_nonstationary_traces}
\end{align}
Induction therefore makes every such component traceless. It cannot
equal a normalized initial state \(\rho_0\). Applying
\cref{eq:jordan_chain_evolution} to
\cref{eq:jordan_initial_expansion} and collecting the coefficient of
each \(\mathbf R_{\alpha,k}\) yields
\begin{align}
 e^{t\mathcal L}\mathcal P_\alpha\rho_0
 &=e^{t\lambda_\alpha}
   \sum_{j=0}^{q_\alpha-1}c_{\alpha,j}
   \sum_{p=0}^{j}\frac{t^p}{p!}\mathbf R_{\alpha,j-p},
 \nonumber\\
 &=e^{t\lambda_\alpha}
   \sum_{k=0}^{q_\alpha-1}
   \left(\sum_{j=k}^{q_\alpha-1}
      c_{\alpha,j}\frac{t^{j-k}}{(j-k)!}\right)\mathbf R_{\alpha,k}.
 \label{eq:jordan_collected_block_expansion}
\end{align}
The index substitution \(k=j-p\) gives the second expression. A
physical trajectory \(\rho(t)\) requires the sum over all blocks:
\begin{align}
 |\rho(t)\rangle\!\rangle
 &=\sum_\alpha e^{t\lambda_\alpha}
   \sum_{k=0}^{q_\alpha-1}
   \left(\sum_{j=k}^{q_\alpha-1}
    c_{\alpha,j}\frac{t^{j-k}}{(j-k)!}\right)
                         |\mathbf R_{\alpha,k}\rangle\!\rangle,
 \label{eq:jordan_full_vectorized_expansion}\\
 \rho(t)
 &=\sum_\alpha e^{t\lambda_\alpha}
   \sum_{k=0}^{q_\alpha-1}
   \left(\sum_{j=k}^{q_\alpha-1}
    c_{\alpha,j}\frac{t^{j-k}}{(j-k)!}\right)\mathbf R_{\alpha,k}.
 \label{eq:jordan_full_density_expansion}
\end{align}
When all \(q_\alpha=1\), this reduces to
\(\rho(t)=\sum_k e^{t\lambda_k}c_k\mathbf R_k\), as in
\cref{eq:diagonalizable_state_expansion}. For larger \(q_\alpha\),
the possible polynomial factors are \(1,t,\ldots,t^{q_\alpha-1}\).

Nontrivial Jordan blocks at peripheral eigenvalues
\(\operatorname{Re}\lambda_\alpha=0\) are impossible for a
finite-dimensional Lindbladian \(\mathcal L\). To see this, write
an arbitrary operator as \(B=A+\im C\) with Hermitian \(A,C\).
Trace-norm contraction on Hermitian operators gives the uniform bound
\begin{align}
 \|e^{t\mathcal L}B\|_1
 &\leq\|A\|_1+\|C\|_1\leq2\|B\|_1,
 \qquad t\geq0.
 \label{eq:jordan_semigroup_boundedness}
\end{align}
If a chain had \(q_\alpha\geq2\), then
\cref{eq:jordan_chain_evolution_r1} and
\(|e^{t\lambda_\alpha}|=1\) would imply
\begin{align}
 \|e^{t\mathcal L}\mathbf R_{\alpha,1}\|_1
 &\geq t\|\mathbf R_{\alpha,0}\|_1
                 -\|\mathbf R_{\alpha,1}\|_1\longrightarrow\infty,
 \label{eq:jordan_peripheral_contradiction}
\end{align}
contradicting \cref{eq:jordan_semigroup_boundedness}. Thus every
peripheral eigenvalue \(\lambda_\alpha\), including zero, is
semisimple. This is a finite-dimensional consequence of the channel
property, not a genericity assumption.

For a decaying block, set \(g_\alpha=-\operatorname{Re}\lambda_\alpha>0\).
For each integer \(p\geq0\), the polynomial factor does not change
the asymptotic exponential rate:
\begin{align}
 t^p|e^{t\lambda_\alpha}|&=t^pe^{-g_\alpha t}\longrightarrow0,
 \label{eq:jordan_polynomial_decay}\\
 -\lim_{t\to\infty}\frac1t\log(t^pe^{-g_\alpha t})&=g_\alpha.
 \label{eq:jordan_asymptotic_exponent}
\end{align}
It does change finite-time bounds. For \(0<f<g_\alpha\),
maximizing \(t^pe^{-(g_\alpha-f)t}\) gives
\begin{align}
 t^pe^{-g_\alpha t}
 &\leq
 \begin{cases}
  e^{-ft},&p=0,\\
  \left(\dfrac{p}{e(g_\alpha-f)}\right)^p e^{-ft},&p\geq1.
 \end{cases}
 \label{eq:jordan_polynomial_envelope}
\end{align}
For \(p\geq1\), the maximum occurs at
\(t=p/(g_\alpha-f)\). This explicitly explains how fast Jordan
contributions can be absorbed into a larger prefactor \(C_f\),
provided the chosen rate \(f\) is strictly smaller than their
asymptotic rates \(g_\alpha\).

Finally, for a relaxing generator \(\mathcal L\), collect the
coefficient matrices \(A_{\alpha,p}\) and fast remainder
\(R_f(t)\) as
\begin{align}
 A_{\alpha,p}
 &=\frac1{p!}\sum_{k=0}^{q_\alpha-1-p}
                  c_{\alpha,k+p}\mathbf R_{\alpha,k},
 \qquad0\leq p<q_\alpha,
 \label{eq:jordan_polynomial_coefficients}\\
 R_f(t)&=\sum_{\alpha\in\mathrm{fast}}e^{t\lambda_\alpha}
                       \sum_{p=0}^{q_\alpha-1}t^pA_{\alpha,p}.
 \label{eq:jordan_fast_remainder}\\
 \rho(t)&=\rho_{\mathrm{ss}}
       +\sum_{\alpha\in\mathrm{slow}}e^{t\lambda_\alpha}
                      \sum_{p=0}^{q_\alpha-1}t^pA_{\alpha,p}
       +R_f(t).
 \label{eq:jordan_slow_mode_expansion}
\end{align}
The sets of slow and fast blocks exclude the stationary eigenvalue
\(0\) and include complete conjugate sectors. If the slow blocks
have \(q_\alpha=1\), their sum reduces to
\(\sum_{k=2}^{M}e^{t\lambda_k}c_k\mathbf R_k\).
Neither this simplification nor \(\Gamma_M\ll\Gamma_{M+1}\)
alone guarantees a nearly constant trajectory \(\rho(t)\).
The modal amplitudes and the fast remainder \(R_f(t)\) must also
be controlled over the proposed time interval, while the slow
component must change little there. This finite-time qualification
is required in metastability theory \cite{macieszczak2021classical}
and motivates the fixed-target comparison in
\cref{sec:metastability_control}.

\section{Proofs for the origin of metastability}
\label{app:proof_of_theorems}

This appendix follows the order of the formal theorems in
\cref{sec:origin_metastability}.
Each subsection proves one theorem, using the assumptions and
notation stated with that result. Shared arguments are cited rather
than repeated; in particular, the QLD strong-dissipation criterion
is a specialization of the general Lindbladian criterion.

\addtocontents{toc}{\protect\begingroup\protect\value{tocdepth}=1\relax}

\subsection{Proof of \texorpdfstring{\cref{thm:two_state_lower_edge}}{Theorem 3.1}}
\label{app:two_state_lower_edge}

For the proof of \cref{thm:two_state_lower_edge}, use the
auxiliary transverse damping scale \(S\) and ratio \(g\):
\begin{align}
 S&=a_x^2+b^2>0,
 &\delta_{\mathcal D}&=\frac{a_x^2-b^2}{S},
 &g&=\frac{2S}{\Omega_{\mathrm{tun}}^{(2)}}.
 \label{eq:two_state_competing_scales}
\end{align}
The anisotropy \(\delta_{\mathcal D}\) is the quantity defined
in \cref{eq:two_state_dissipative_anisotropy}.
Here \(g>0\), \(|\delta_{\mathcal D}|\leq1\), and
\(\Omega_{\mathrm{tun}}^{(2)}=\omega_{01}>0\).

\begin{proof}
At \(E_\delta=0\), parity gives \(a_0=a_z=0\). The Bloch
matrix \(A\) in \cref{eq:exact_bloch_affine} has the longitudinal
eigenvalue \(\lambda_z=-4S\) and transverse block
\begin{align}
 A_\perp
 &=\begin{pmatrix}-4b^2&\omega_{01}\\-\omega_{01}&-4a_x^2\end{pmatrix}.
 \label{eq:symmetric_transverse_block}
\end{align}
Expanding its characteristic determinant gives
\begin{align}
 \det(\lambda I_2-A_\perp)
 &=(\lambda+4b^2)(\lambda+4a_x^2)+\omega_{01}^2
 \nonumber\\
 &=(\lambda+2S)^2-4(a_x^2-b^2)^2+\omega_{01}^2
 =(\lambda+2S)^2-\Delta_B.
 \label{eq:symmetric_transverse_characteristic_square}
\end{align}
Thus \(\lambda_\pm=-2S\pm\sqrt{\Delta_B}\), as stated in
\cref{eq:symmetric_transverse_eigenvalues}. If \(\Delta_B\leq0\),
their real parts coincide, including the repeated-root case, and
\begin{align}
 (\Gamma_2,\Gamma_3,\Gamma_4)&=(2S,2S,4S),
 \label{eq:symmetric_underdamped_rate_ordering}\\
 \chi_2&=0,\qquad \chi_3=\chi_{\max}=\frac13<\chi_c.
 \label{eq:symmetric_underdamped_chi}
\end{align}
For \(\Delta_B>0\), the identity
\(4S^2-\Delta_B=16a_x^2b^2+\omega_{01}^2>0\) gives
\(0<\xi=\sqrt{\Delta_B}/(2S)<1\). The ordered decay rates are
\begin{align}
 (\Gamma_2,\Gamma_3,\Gamma_4)
 &=\bigl(2S(1-\xi),2S(1+\xi),4S\bigr).
 \label{eq:symmetric_overdamped_rate_ordering}
\end{align}
Substitution into \cref{def:chi_separation} gives
\begin{align}
 \chi_2=\xi,\qquad
 \chi_3=\frac{1-\xi}{3+\xi}\leq\frac13,\qquad 0<\xi<1.
 \label{eq:two_state_lower_edge_separations}
\end{align}
Only the separation \(\chi_2\) can therefore exceed \(\chi_c\).
Using \cref{eq:two_state_competing_scales} and
\(\Omega_{\mathrm{tun}}^{(2)}=\omega_{01}\), we obtain
\begin{align}
 \xi^2
 &=\frac{4(a_x^2-b^2)^2-\omega_{01}^2}{4S^2}
 =\delta_{\mathcal D}^2
       -\frac{(\Omega_{\mathrm{tun}}^{(2)})^2}{4S^2}
 =\delta_{\mathcal D}^2-g^{-2}.
 \label{eq:two_state_splitting_squared}
\end{align}
If \(|\delta_{\mathcal D}|\leq\chi_c\), no positive \(g\)
can give \(\xi>\chi_c\). For
\(|\delta_{\mathcal D}|>\chi_c\), the equivalent condition is
\begin{align}
 \chi_{\max}>\chi_c
 \quad\Longleftrightarrow\quad
 g>\frac1{\sqrt{\delta_{\mathcal D}^2-\chi_c^2}}.
 \label{eq:two_state_origin_iff}
\end{align}
This strict inequality already implies \(\Delta_B>0\), so it adds
no underdamped branch.

To express the same condition in the dissipation strength
\(\gamma_{\mathcal D}^{(2)}\), {first evaluate its
induced Hilbert--Schmidt norm. In the symmetric model,
\(L^{(2)}=a_x\sigma_x+\im b\sigma_y\). Substituting into
\cref{eq:two_state_dissipator} gives
\(\mathcal D_2(I_2)=8a_xb\sigma_z\),
\(\mathcal D_2(\sigma_x)=-4b^2\sigma_x\),
\(\mathcal D_2(\sigma_y)=-4a_x^2\sigma_y\), and
\(\mathcal D_2(\sigma_z)=-4S\sigma_z\).
Hence the representation in the orthonormal operator basis
\((I_2,\sigma_x,\sigma_y,\sigma_z)/\sqrt2\) is
\begin{align}
 [\mathcal D_2]
 &=4\begin{pmatrix}
 0&0&0&0\\
 0&-b^2&0&0\\
 0&0&-a_x^2&0\\
 2a_xb&0&0&-S
 \end{pmatrix}.
 \label{eq:two_state_symmetric_dissipator_matrix}
\end{align}
Its nonzero rows are orthogonal, so its singular values are their
Euclidean norms, together with zero. Since \(S=a_x^2+b^2\),
\begin{align}
 \gamma_{\mathcal D}^{(2)}
 &=4\max\{b^2,a_x^2,\sqrt{S^2+4a_x^2b^2}\}
 =4\sqrt{S^2+4a_x^2b^2}.
 \label{eq:two_state_symmetric_dissipation_norm}
\end{align}
Combining this identity with the definition of
\(\delta_{\mathcal D}\) gives}
\begin{align}
 4a_x^2b^2
 &=S^2-(a_x^2-b^2)^2=S^2(1-\delta_{\mathcal D}^2),
 \label{eq:two_state_damping_product_identity}\\
 \gamma_{\mathcal D}^{(2)}
 &=4S\sqrt{2-\delta_{\mathcal D}^2},\qquad
 \frac{\gamma_{\mathcal D}^{(2)}}{\Omega_{\mathrm{tun}}^{(2)}}
 =2g\sqrt{2-\delta_{\mathcal D}^2}.
 \label{eq:two_state_norm_damping_relation}
\end{align}
Since \(2-\delta_{\mathcal D}^2\geq1\), inversion and substitution
into \cref{eq:two_state_splitting_squared} give
\begin{align}
 S&=\frac{\gamma_{\mathcal D}^{(2)}}
              {4\sqrt{2-\delta_{\mathcal D}^2}},
 \label{eq:two_state_norm_scale_inversion}\\
 \xi^2
 &=\delta_{\mathcal D}^2
 -4(2-\delta_{\mathcal D}^2)
       \left(\frac{\Omega_{\mathrm{tun}}^{(2)}}
                   {\gamma_{\mathcal D}^{(2)}}\right)^2.
 \label{eq:two_state_scaled_splitting}
\end{align}
Consequently, the threshold condition can also be rearranged as
\begin{align}
 &\begin{aligned}[b]
 \chi_{\max}>\chi_c
 &\Longleftrightarrow
 \delta_{\mathcal D}^2
 -4(2-\delta_{\mathcal D}^2)
       \left(\frac{\Omega_{\mathrm{tun}}^{(2)}}
                   {\gamma_{\mathcal D}^{(2)}}\right)^2
 >\chi_c^2\\
 &\Longleftrightarrow
 (\gamma_{\mathcal D}^{(2)})^2
       (\delta_{\mathcal D}^2-\chi_c^2)
 >4(2-\delta_{\mathcal D}^2)(\Omega_{\mathrm{tun}}^{(2)})^2>0.
 \end{aligned}
 \label{eq:two_state_lower_edge_rearranged}
\end{align}
{The first equivalence in
\cref{eq:two_state_lower_edge_rearranged} is precisely the
necessary and sufficient condition in
\cref{eq:two_state_spectral_iff}. At equality, the squared
splitting in \cref{eq:two_state_scaled_splitting} is
\(\xi^2=\chi_c^2>0\), so
\(\xi=\chi_{\max}=\chi_c\).} Within the overdamped branch,
differentiation at fixed
\(\delta_{\mathcal D},\Omega_{\mathrm{tun}}^{(2)}\) gives
\begin{align}
 \frac{\partial\xi}{\partial\gamma_{\mathcal D}^{(2)}}
 &=\frac{4(2-\delta_{\mathcal D}^2)
                  (\Omega_{\mathrm{tun}}^{(2)})^2}
         {(\gamma_{\mathcal D}^{(2)})^3\xi}>0,
 \label{eq:two_state_splitting_monotonicity}\\
 \lim_{\gamma_{\mathcal D}^{(2)}\to\infty}\xi
 &=|\delta_{\mathcal D}|.
 \label{eq:two_state_splitting_limit}
\end{align}
On the nonempty window, \(\chi_{\max}=\xi>\chi_c>1/3\);
these relations exclude a finite upper exit.
\end{proof}

\subsection{Proof of \texorpdfstring{\cref{thm:biased_two_state_window}}{Theorem 3.2}}
\label{app:two_state_biased_limits}

For the proof of \cref{thm:biased_two_state_window}, separate
the fixed dissipative structure from its magnitude.
At fixed potential \(V(x)\) and relative jump coefficients
\(\mu,\nu\), common rescaling leaves
\(\mathcal D_2/\gamma_{\mathcal D}^{(2)}\) fixed. To separate its
structure from its magnitude, use the squared frequency ratio
\(y\) and normalized Bloch matrix \(\mathsf B(y)\):

\begin{align}
 y&=\left(\frac{\omega_{01}}{\gamma_{\mathcal D}^{(2)}}\right)^2>0,
 \label{eq:biased_scaled_frequency}\\
 \mathsf B(y)&=\frac{A}{\gamma_{\mathcal D}^{(2)}}
 =\frac1{\gamma_{\mathcal D}^{(2)}}
 \begin{pmatrix}-\alpha_A&0&h_-\\0&-\beta_A&0\\h_+&0&-\gamma_A\end{pmatrix}
 +\sqrt y\begin{pmatrix}0&1&0\\-1&0&0\\0&0&0\end{pmatrix}.
 \label{eq:biased_scaled_bloch_matrix}
\end{align}

Every ratio
\(\alpha_A/\gamma_{\mathcal D}^{(2)},
\beta_A/\gamma_{\mathcal D}^{(2)},
\gamma_A/\gamma_{\mathcal D}^{(2)},h_\pm/\gamma_{\mathcal D}^{(2)}\)
is constant along this scan. Thus \(y\) is not a new independent
control parameter: weak and strong relative dissipation correspond
to \(y\to\infty\) and \(y\to0\), respectively.
The matrix \(\mathsf B(0)\) represents the action of
\(\mathcal D_2/\gamma_{\mathcal D}^{(2)}\) on traceless operators;
the affine source in \cref{eq:exact_bloch_affine} does not enter
its nonstationary spectrum.

Let \(z_i\), \(i=1,2,3\), be the eigenvalues of \(\mathsf B(y)\),
ordered by their negative real parts, and write
\begin{align}
 d_i&=-\operatorname{Re}z_i
 =\frac{\Gamma_{i+1}}{\gamma_{\mathcal D}^{(2)}},
 \qquad d_1\leq d_2\leq d_3,
 \label{eq:biased_normalized_decay_rates}\\
 \chi_{i+1}
 &=\frac{d_{i+1}-d_i}{d_{i+1}+d_i},\qquad i=1,2.
 \label{eq:biased_scale_invariant_separation}
\end{align}
The index shift omits the stationary rate \(\Gamma_1=0\).
The common scale \(\gamma_{\mathcal D}^{(2)}\) cancels, but
the competition with \(H_2\) remains through \(y\).

\begin{proof}
The symmetric part of \(\mathsf B(y)\) follows directly from
\cref{eq:biased_alpha_beta_coefficients,eq:biased_bloch_coefficients}:
\begin{align}
 \frac{\mathsf B(y)+\mathsf B(y)^{\mathsf T}}2
 &=-\frac4{\gamma_{\mathcal D}^{(2)}}
 \begin{pmatrix}
 b^2+a_z^2&0&-a_xa_z\\
 0&a_x^2+a_z^2&0\\
 -a_xa_z&0&b^2+a_x^2
 \end{pmatrix}.
 \label{eq:biased_symmetric_part}
\end{align}
Its \(xz\) block is
\(-4[b^2I_2+(a_z,-a_x)^{\mathsf T}(a_z,-a_x)]/
\gamma_{\mathcal D}^{(2)}\). Thus its three eigenvalues are
\begin{align}
 -\frac{4b^2}{\gamma_{\mathcal D}^{(2)}},\qquad
 -\frac{4(a_x^2+a_z^2)}{\gamma_{\mathcal D}^{(2)}},\qquad
 -\frac{4(a_x^2+a_z^2+b^2)}{\gamma_{\mathcal D}^{(2)}}.
 \label{eq:biased_symmetric_part_eigenvalues}
\end{align}
For a right eigenvector \(v\ne0\) with \(\mathsf B(y)v=zv\),
\begin{align}
 \operatorname{Re}z
 &=\frac{v^\dagger[\mathsf B(y)+\mathsf B(y)^{\mathsf T}]v}
           {2v^\dagger v},
 \label{eq:biased_real_part_rayleigh}\\
 0<\frac{4\min\{b^2,a_x^2+a_z^2\}}{\gamma_{\mathcal D}^{(2)}}
 &\leq d_i\leq
 \frac{4(a_x^2+a_z^2+b^2)}{\gamma_{\mathcal D}^{(2)}}
 =\frac{c_2}{2\gamma_{\mathcal D}^{(2)}}.
 \label{eq:biased_normalized_rate_bounds}
\end{align}
The lower bound is positive by the stated assumptions; its value
is independent of the overall dissipation strength
\(\gamma_{\mathcal D}^{(2)}\). Taking the trace gives
\begin{align}
 d_1+d_2+d_3
 &=\frac{c_2}{\gamma_{\mathcal D}^{(2)}},
 \label{eq:biased_normalized_rate_sum}\\
 d_3&\leq d_1+d_2\leq2d_2,\qquad
 \chi_3=\frac{d_3-d_2}{d_3+d_2}\leq\frac13.
 \label{eq:biased_fast_pair_separation_bound}
\end{align}
 {This proves the bound on \(\chi_3\) and the
selection \(i^\ast=2\) whenever \(\chi_{\max}>\chi_c\),
including repeated real parts and nonreal pairs.}

For weak relative dissipation, separate the Hamiltonian and
dissipative contributions to \(A\):
\begin{align}
 \frac A{\omega_{01}}
 &=\begin{pmatrix}0&1&0\\-1&0&0\\0&0&0\end{pmatrix}
 +\frac{\gamma_{\mathcal D}^{(2)}}{\omega_{01}}\mathsf B(0).
 \label{eq:biased_weak_coupling_decomposition}
\end{align}
The first term in \cref{eq:biased_weak_coupling_decomposition}
is normal, with simple eigenvalues \(0,\pm\im\).
The corresponding normalized eigenvectors are
\((0,0,1)^{\mathsf T}\) and
\((1,\pm\im,0)^{\mathsf T}/\sqrt2\).
Their first-order corrections are
\begin{align}
 (0,0,1)\mathsf B(0)(0,0,1)^{\mathsf T}
 &=-\frac{\gamma_A}{\gamma_{\mathcal D}^{(2)}},
 \label{eq:biased_weak_population_correction}\\
 \frac12(1,\mp\im,0)\mathsf B(0)(1,\pm\im,0)^{\mathsf T}
 &=-\frac{\alpha_A+\beta_A}{2\gamma_{\mathcal D}^{(2)}}.
 \label{eq:biased_weak_coupling_corrections}
\end{align}
Consequently the real and conjugate-pair decay rates
\(\Gamma_R,\Gamma_C\) obey
\begin{align}
 \Gamma_R&=\gamma_A+
 O\!\left(\frac{(\gamma_{\mathcal D}^{(2)})^2}{\omega_{01}}\right),
 \label{eq:biased_weak_coupling_rates}\\
 \Gamma_C&=\frac{\alpha_A+\beta_A}{2}+
 O\!\left(\frac{(\gamma_{\mathcal D}^{(2)})^2}{\omega_{01}}\right).
 \label{eq:biased_weak_coherence_rate}
\end{align}
The conjugate pair persists for sufficiently weak positive
dissipation.  {Define the weak-dissipation limit
\(\chi_0\coloneqq\lim_{\gamma_{\mathcal D}^{(2)}/\omega_{01}\to0^+}
\chi_{\max}\). Using \cref{eq:biased_pair_separation} gives}
\begin{align}
 &\begin{aligned}[b]
 \chi_0
 &=\frac{|\gamma_A-(\alpha_A+\beta_A)/2|}
          {\gamma_A+(\alpha_A+\beta_A)/2},\\
 &=\frac{|a_x^2+b^2-2a_z^2|}
          {3(a_x^2+b^2)+2a_z^2}.
 \end{aligned}
 \label{eq:biased_weak_separation_derivation}
\end{align}
Under the QLD identities in \cref{eq:bloch_coefficients}, \(b\ne0\)
implies \(x_{01}\ne0\), and \(a_x^2+a_z^2>0\) implies \(\mu\ne0\).
Hence \(a_x\ne0\). Diagonalizing \(x_2\), or using
\((\hbar\omega_{01})^2=(\epsilon^{(2)})^2+4(J^{(2)})^2\), yields

\begin{align}
 \left(\frac{\Omega_{\mathrm{tun}}^{(2)}}{\omega_{01}}\right)^2
 &=\frac{x_{01}^2}{x_{01}^2+(x_{00}-x_{11})^2/4}
 =\frac{a_x^2}{a_x^2+a_z^2}>0,
 \label{eq:biased_tunneling_position_contrast}\\
 \frac{2a_z^2}{a_x^2+b^2}
 &=(1+\delta_{\mathcal D})
       \bigl[(\omega_{01}/\Omega_{\mathrm{tun}}^{(2)})^2-1\bigr].
 \label{eq:biased_weak_physical_substitution}
\end{align}

Dividing the numerator and denominator in
\cref{eq:biased_weak_separation_derivation} by
\(a_x^2+b^2>0\) and substituting
 {\cref{eq:biased_weak_physical_substitution} gives}

\begin{align}
 \chi_0
 &=\frac{\left|1-(1+\delta_{\mathcal D})
          \bigl[(\omega_{01}/\Omega_{\mathrm{tun}}^{(2)})^2-1\bigr]\right|}
          {3+(1+\delta_{\mathcal D})
          \bigl[(\omega_{01}/\Omega_{\mathrm{tun}}^{(2)})^2-1\bigr]}.
 \label{eq:biased_weak_coupling_limit}
\end{align}

If \(a_x^2+b^2\geq2a_z^2\), then \(\chi_0\leq1/3<\chi_c\).
On the other branch, removing the absolute value gives
\begin{align}
 &\begin{aligned}[b]
 \chi_0>\chi_c
 &\Longleftrightarrow
 2(1-\chi_c)a_z^2>(1+3\chi_c)(a_x^2+b^2),\\
 &\Longleftrightarrow
 a_z^2>\frac{1+3\chi_c}{2(1-\chi_c)}(a_x^2+b^2).
 \end{aligned}
 \label{eq:biased_weak_metastability_condition}
\end{align}
 {Substituting \cref{eq:biased_weak_physical_substitution}
shows that \(\chi_0>\chi_c\) is equivalent to the parameter
inequality on the right of \cref{eq:biased_weak_physical_criterion};
the inequality also ensures the required branch
\(2a_z^2>a_x^2+b^2\). The same algebra gives equality precisely
when \(\chi_0=\chi_c\).}

For strong relative dissipation, \(y\to0\). The characteristic
determinant of \(\mathsf B(0)\) factors as
\begin{align}
 \det(zI_3-\mathsf B(0))
 &=\left(z+\frac{\beta_A}{\gamma_{\mathcal D}^{(2)}}\right)
 \left[
 \left(z+\frac{\alpha_A}{\gamma_{\mathcal D}^{(2)}}\right)
 \left(z+\frac{\gamma_A}{\gamma_{\mathcal D}^{(2)}}\right)
 -\frac{h_-h_+}{(\gamma_{\mathcal D}^{(2)})^2}\right].
 \label{eq:biased_strong_coupling_factorization}
\end{align}
The first root is \(-\beta_A/\gamma_{\mathcal D}^{(2)}\);
the other two roots are
\begin{align}
 z&=-\frac{\alpha_A+\gamma_A}{2\gamma_{\mathcal D}^{(2)}}
 \pm\frac{\sqrt{(\alpha_A-\gamma_A)^2/4+h_-h_+}}
               {\gamma_{\mathcal D}^{(2)}}.
 \label{eq:biased_strong_coupling_quadratic_roots}
\end{align}
Inserting \cref{eq:biased_alpha_beta_coefficients,eq:biased_bloch_coefficients}
gives all three roots explicitly:
\begin{align}
 z&=-\frac{4(a_x^2+a_z^2)}{\gamma_{\mathcal D}^{(2)}},
 \label{eq:biased_pure_dissipator_real_eigenvalue}\\
 z&=\frac{-2(a_x^2+a_z^2+2b^2)
        \pm2\sqrt{(a_x^2+a_z^2)^2-4b^2a_0^2}}
              {\gamma_{\mathcal D}^{(2)}}.
 \label{eq:biased_pure_dissipator_spectrum}
\end{align}

Define
\(\chi_\infty\coloneqq\chi_{\max}(\mathcal D_2/\gamma_{\mathcal D}^{(2)})\).
Ordering the negative real parts of the three roots \(z\) as
\(0<d_1\leq d_2\leq d_3\) gives
\(\chi_\infty=\max\{(d_2-d_1)/(d_2+d_1),
(d_3-d_2)/(d_3+d_2)\}\), by
\cref{eq:biased_scale_invariant_separation}; the stationary
eigenvalue \(0\) is excluded. This also applies when the square
root in \cref{eq:biased_pure_dissipator_spectrum} is imaginary.

The positive bound in \cref{eq:biased_normalized_rate_bounds}
and continuity of finite-dimensional spectra imply

\begin{align}
 \chi_{\max}(\mathcal L_2^{\mathrm{TP}})
 &\longrightarrow\chi_\infty
 \qquad\left(\frac{\omega_{01}}{\gamma_{\mathcal D}^{(2)}}\to0^+\right).
 \label{eq:biased_strong_separation_limit}
\end{align}

The dissipator \(\mathcal D_2\) has a unique stationary eigenvalue
zero because its traceless block \(\mathsf B(0)\) is strictly stable.
Repeated eigenvalues of \(\mathsf B(0)\) do not affect continuity
of this spectral ratio.

It remains to turn these limits into finite-dissipation criteria.
For assertion (i), the excluded parameter equality is equivalent to
\(\chi_0=\chi_c\). With \(\chi_0\ne\chi_c\), the weak limit
therefore gives \(\varepsilon_{\mathrm w}>0\) such that
\(\lvert\chi_{\max}-\chi_0\rvert<\lvert\chi_0-\chi_c\rvert/2\)
for every \(0<\gamma_{\mathcal D}^{(2)}/\omega_{01}<\varepsilon_{\mathrm w}\).
Similarly, for assertion (ii), \(\chi_\infty\ne\chi_c\) gives
\(\varepsilon_{\mathrm s}>0\) such that
\(\lvert\chi_{\max}-\chi_\infty\rvert<\lvert\chi_\infty-\chi_c\rvert/2\)
for every \(0<\omega_{01}/\gamma_{\mathcal D}^{(2)}<\varepsilon_{\mathrm s}\).
For either endpoint value \(\chi_\star\in\{\chi_0,\chi_\infty\}\),
the corresponding estimate gives
\(\chi_{\max}>\chi_\star-(\chi_\star-\chi_c)/2>\chi_c\)
when \(\chi_\star>\chi_c\), and
\(\chi_{\max}<\chi_\star+(\chi_c-\chi_\star)/2<\chi_c\)
when \(\chi_\star<\chi_c\). Thus
\begin{align}
 \operatorname{sgn}(\chi_{\max}-\chi_c)
 &=
 \begin{cases}
 \operatorname{sgn}(\chi_0-\chi_c),
   &0<\gamma_{\mathcal D}^{(2)}/\omega_{01}<\varepsilon_{\mathrm w},\\
 \operatorname{sgn}(\chi_\infty-\chi_c),
   &0<\omega_{01}/\gamma_{\mathcal D}^{(2)}<\varepsilon_{\mathrm s}.
 \end{cases}
 \label{eq:biased_endpoint_signs}
\end{align}
In \cref{eq:biased_endpoint_signs}, each line requires only its own
noncriticality condition. Combining the first line with
\cref{eq:biased_weak_metastability_condition,eq:biased_weak_physical_substitution}
proves the equivalence in \cref{eq:biased_weak_physical_criterion}.
The second line and the definition of \(\chi_\infty\) prove
\cref{eq:biased_strong_physical_criterion}. Both thresholds
\(\varepsilon_{\mathrm w},\varepsilon_{\mathrm s}\) depend only
on the fixed model and \(\chi_c\), not on the varying strength
\(\gamma_{\mathcal D}^{(2)}\).
\end{proof}

\subsection{Proof of \texorpdfstring{\cref{thm:qld_sector_origin}}{Theorem 3.3}}
\label{app:proof_qld_sector_origin}

\begin{proof}
We first relate the total frequency
\(\Omega_{\mathrm{tun}}^{\mathrm{op}}\) to both coherent and
dissipative transfer. Relative to the reference generator
\(\mathcal L_{0,r}\) in \cref{eq:qld_decoupled_reference}, define
the coupling operators \(V_{H,r},V_{L,r}\) and coupling norm
\(J_{\mathrm{op}}^{(r)}\) by
\begin{align}
 H_{0,r}&=\sum_{\alpha=L,R}P_\alpha^{(r)}H_rP_\alpha^{(r)},
 &V_{H,r}&=H_r-H_{0,r},\label{eq:qld_sector_hamiltonian_split}\\
 L_{0,r}&=\sum_{\alpha=L,R}P_\alpha^{(r)}L^{(r)}P_\alpha^{(r)},
 &V_{L,r}&=L^{(r)}-L_{0,r},\label{eq:qld_sector_jump_split}\\
 J_{\mathrm{op}}^{(r)}
 &=\|P_L^{(r)}H_rP_R^{(r)}\|_{\mathrm{op}}
 =\|V_{H,r}\|_{\mathrm{op}},
 &J^{(r)}&\leq J_{\mathrm{op}}^{(r)}.
 \label{eq:qld_sector_coupling_norm}
\end{align}
The norm identity follows because the square \(V_{H,r}^2\) has
diagonal blocks
\((P_L^{(r)}H_rP_R^{(r)})(P_R^{(r)}H_rP_L^{(r)})\) and
\((P_R^{(r)}H_rP_L^{(r)})(P_L^{(r)}H_rP_R^{(r)})\).
Evaluating the off-diagonal block between the normalized states
\(\ket{\psi_L^{(r)}},\ket{\psi_R^{(r)}}\) gives the inequality.
Consequently, \(\Omega_{\mathrm{tun}}^{\mathrm{op}}
=2J_{\mathrm{op}}^{(r)}/\hbar\), consistently with
\cref{eq:qld_total_intersector_frequency}.

Since \(P_L^{(r)},P_R^{(r)}\) are spectral projectors of
\(x_r\), the position contribution \(\mu x_r\) is sector
diagonal. Hence
\begin{align}
 V_{L,r}&=\im\nu\bigl(P_L^{(r)}p_rP_R^{(r)}
                         +P_R^{(r)}p_rP_L^{(r)}\bigr).
 \label{eq:qld_cross_jump_momentum}
\end{align}
Under the assumed identity \(p_r=\im m[H_r,x_r]/\hbar\),
the commutator \([H_{0,r},x_r]\) is sector diagonal, whereas
\([V_{H,r},x_r]\) is sector off-diagonal. Thus
\begin{align}
 V_{L,r}&=-\frac{\nu m}{\hbar}[V_{H,r},x_r],
 \label{eq:qld_cross_jump_commutator}\\
 \|V_{L,r}\|_{\mathrm{op}}
 &\leq\frac{|\nu|m\ell_r}{\hbar}J_{\mathrm{op}}^{(r)},
 \qquad
 \ell_r=\lambda_{\max}(x_r)-\lambda_{\min}(x_r).
 \label{eq:qld_cross_jump_bound}
\end{align}
For the bound, subtracting
\((\lambda_{\max}(x_r)+\lambda_{\min}(x_r))/2\) from \(x_r\)
leaves the commutator unchanged and reduces its operator norm to
\(\ell_r/2\). The product norm inequality then bounds the
commutator norm by \(\ell_r\|V_{H,r}\|_{\mathrm{op}}\).
At fixed \(H_{0,r},x_r,\mu,\nu\), the operator \(L_{0,r}\)
is fixed, and \cref{eq:qld_cross_jump_bound} controls the entire
operator \(V_{L,r}\) by \(\Omega_{\mathrm{tun}}^{\mathrm{op}}\).
No common rescaling of the individual entries of \(V_{H,r}\)
is required.

The stationary projection \(\mathscr P_{0,r}\) preserves the
sector populations and replaces each local state by its equilibrium
\(\rho_\alpha^{(r)}\):
\begin{align}
 \mathscr P_{0,r}(B)
 &=\sum_{\alpha=L,R}\rho_\alpha^{(r)}
                    \operatorname{Tr}(P_\alpha^{(r)}B),
 &\mathscr Q_{0,r}&=\mathcal I-\mathscr P_{0,r}.
 \label{eq:qld_local_equilibrium_projection}
\end{align}
The relaxation assumptions give
\(e^{t\mathcal L_{0,r}}\to\mathscr P_{0,r}\) and
\(\mathcal L_{0,r}\mathscr P_{0,r}
=\mathscr P_{0,r}\mathcal L_{0,r}=0\).
The stationary eigenvalue \(0\) is semisimple, since a nontrivial
Jordan chain there would contradict this bounded limit. Fast Jordan
blocks at \(z\ne0\) remain allowed.

Substituting the decompositions in
\cref{eq:qld_sector_hamiltonian_split,eq:qld_sector_jump_split}
into \(\mathcal L_r^{\mathrm{TP}}\) separates the terms linear
and quadratic in the coupling blocks \(V_{H,r},V_{L,r}\):
\begin{align}
 \mathcal L_r^{\mathrm{TP}}
 &=\mathcal L_{0,r}+\mathcal A_r+\mathcal B_r,
 \label{eq:qld_tunneling_generator_expansion}\\
 \mathcal A_r(B)
 &=\begin{aligned}[t]
   &-\frac{\im}{\hbar}[V_{H,r},B]
    +2L_{0,r}BV_{L,r}^\dagger+2V_{L,r}BL_{0,r}^\dagger\\
   &-\{L_{0,r}^\dagger V_{L,r}+V_{L,r}^\dagger L_{0,r},B\},
   \end{aligned}
 \label{eq:qld_tunneling_first_order}\\
 \mathcal B_r(B)&=\mathcal D[V_{L,r}](B).
 \label{eq:qld_tunneling_second_order}
\end{align}
For compatible operators \(U,B,W\), the product inequality
\(\|UBW\|_{\mathrm{HS}}\leq
\|U\|_{\mathrm{op}}\|W\|_{\mathrm{op}}\|B\|_{\mathrm{HS}}\)
and \cref{eq:qld_cross_jump_bound} give
\begin{align}
 &\begin{aligned}
   \|\mathcal A_r\|_{2\to2}
   &\leq\Omega_{\mathrm{tun}}^{\mathrm{op}}
        +8\|L_{0,r}\|_{\mathrm{op}}\|V_{L,r}\|_{\mathrm{op}}\\
   &\leq\bigl(1+4|\nu|m\ell_r\|L_{0,r}\|_{\mathrm{op}}\bigr)
              \Omega_{\mathrm{tun}}^{\mathrm{op}},
   \end{aligned}
 \label{eq:qld_linear_tunneling_norm_bound}\\
 \|\mathcal B_r\|_{2\to2}
 &\leq4\|V_{L,r}\|_{\mathrm{op}}^2
 \leq(|\nu|m\ell_r)^2(\Omega_{\mathrm{tun}}^{\mathrm{op}})^2.
 \label{eq:qld_quadratic_tunneling_norm_bound}
\end{align}
All coefficients in these bounds are fixed. Thus
\(\mathcal A_r=O(\Omega_{\mathrm{tun}}^{\mathrm{op}})\) and
\(\mathcal B_r=O((\Omega_{\mathrm{tun}}^{\mathrm{op}})^2)\)
uniformly over the allowed coupling blocks \(V_{H,r}\).
Here and below, operator estimates use the induced Hilbert--Schmidt
norm, with constants allowed to depend on the fixed local data.

Every term of \(\mathcal A_r(B)\) is sector off-diagonal when
\(B\) is sector diagonal. Since \(\mathscr P_{0,r}\) maps into
the diagonal sectors and annihilates the off-diagonal sectors,
\begin{align}
 \mathscr P_{0,r}\mathcal A_r\mathscr P_{0,r}=0,
 \qquad
 \mathscr P_{0,r}\mathcal L_r^{\mathrm{TP}}\mathscr P_{0,r}
 =\mathscr P_{0,r}\mathcal B_r\mathscr P_{0,r}.
 \label{eq:qld_no_first_order_population_transfer}
\end{align}
Thus the first-order perturbation \(\mathcal A_r\) cannot
directly redistribute the equilibrium populations.

Let \(F_r\) and \(F_{0,r}\) be the restrictions of
\(\mathscr Q_{0,r}\mathcal L_r^{\mathrm{TP}}\mathscr Q_{0,r}\)
and \(\mathscr Q_{0,r}\mathcal L_{0,r}\mathscr Q_{0,r}\)
to \(\operatorname{Ran}\mathscr Q_{0,r}\), respectively.
The gap \(\Delta_{\min}^{(r)}>0\) makes \(F_{0,r}\) invertible.
For sufficiently small \(\Omega_{\mathrm{tun}}^{\mathrm{op}}\)
and \(|z|\), the Neumann-series condition
\(\|F_{0,r}^{-1}(F_r-F_{0,r}-z\mathcal I)\|_{2\to2}<1/2\)
holds, and therefore
\begin{align}
 (F_r-z\mathcal I)^{-1}
 &=\bigl[\mathcal I+F_{0,r}^{-1}
                   (F_r-F_{0,r}-z\mathcal I)\bigr]^{-1}F_{0,r}^{-1},
 \label{eq:qld_fast_resolvent_identity}\\
 \|(F_r-z\mathcal I)^{-1}\|_{2\to2}
 &\leq2\|F_{0,r}^{-1}\|_{2\to2}.
 \label{eq:qld_fast_resolvent_bound}
\end{align}
Eigenvalue continuity at the fixed generator \(\mathcal L_{0,r}\)
gives exactly two eigenvalues \(z\) near zero, counted with algebraic
multiplicity; all remaining decay rates stay above
\(\Delta_{\min}^{(r)}/2\). For an eigenmatrix \(U\) of either
small eigenvalue \(z\), write \(X=\mathscr P_{0,r}U\) and
\(Y=\mathscr Q_{0,r}U\). Its projected equations yield
\begin{align}
 Y&=-(F_r-z\mathcal I)^{-1}
       \mathscr Q_{0,r}\mathcal L_r^{\mathrm{TP}}\mathscr P_{0,r}X,
 \label{eq:qld_tunneling_fast_elimination}\\
 zX&=\Bigl[\mathscr P_{0,r}\mathcal B_r\mathscr P_{0,r}
       -\mathscr P_{0,r}(\mathcal A_r+\mathcal B_r)\mathscr Q_{0,r}
        (F_r-z\mathcal I)^{-1}\mathscr Q_{0,r}
        (\mathcal A_r+\mathcal B_r)\mathscr P_{0,r}\Bigr]X.
 \label{eq:qld_tunneling_schur_equation}
\end{align}
Here \(X\ne0\), since otherwise the invertible fast equation
would give \(Y=0\) as well. The first term in brackets is
quadratic in \(\Omega_{\mathrm{tun}}^{\mathrm{op}}\), while each
factor adjoining the bounded inverse is at least linear.
Normalizing \(\|X\|_{\mathrm{HS}}=1\) gives
\begin{align}
 |z|=O\!\left((\Omega_{\mathrm{tun}}^{\mathrm{op}})^2\right).
 \label{eq:qld_quadratic_slow_eigenvalues}
\end{align}
Expanding the inverse in \cref{eq:qld_fast_resolvent_identity}
also identifies the leading population operator \(\mathscr M_r\):
\begin{align}
 \mathscr M_r
 &=\mathscr P_{0,r}\mathcal B_r\mathscr P_{0,r}
   -\mathscr P_{0,r}\mathcal A_r\mathscr Q_{0,r}F_{0,r}^{-1}
          \mathscr Q_{0,r}\mathcal A_r\mathscr P_{0,r},
 \label{eq:qld_second_order_population_operator}\\
 zX&=\Bigl[\mathscr M_r+
 O\!\left((\Omega_{\mathrm{tun}}^{\mathrm{op}})^3
          +|z|(\Omega_{\mathrm{tun}}^{\mathrm{op}})^2\right)\Bigr]X.
 \label{eq:qld_second_order_population_equation}
\end{align}
The two terms in \(\mathscr M_r\) describe direct dissipative
transfer and transfer mediated by rapidly relaxing coherences.
This realizes the stationary-space perturbation mechanism of
\cite{macieszczak2016towards}; see also \cite{azouit2016adiabatic}
for elimination of fast dissipative dynamics.

One small eigenvalue is the simple stationary eigenvalue \(0\).
The other, \(\lambda_2\), is real because Hermiticity preservation
requires nonreal eigenvalues to occur in conjugate pairs. It is
strictly negative: a Lindbladian has no eigenvalues with positive
real part, and a second zero would contradict simplicity.
For its eigenmatrix \(U\), trace preservation gives
\(\operatorname{Tr}U=0\); hence \(X\) is proportional to
\(\rho_L^{(r)}-\rho_R^{(r)}\).
\Cref{eq:qld_tunneling_fast_elimination} gives
\(\|Y\|_{\mathrm{HS}}=O(\Omega_{\mathrm{tun}}^{\mathrm{op}}/
\Delta_{\min}^{(r)})\), so the slow mode is predominantly a
redistribution of the sector populations.

By \cref{eq:qld_quadratic_slow_eigenvalues} and the fast spectral
bound, there exist dimensionless \(C,\varepsilon_0>0\) such that
\(0<\Omega_{\mathrm{tun}}^{\mathrm{op}}/\Delta_{\min}^{(r)}
<\varepsilon_0\) implies
\begin{align}
 0<\Gamma_2&\leq C\frac{(\Omega_{\mathrm{tun}}^{\mathrm{op}})^2}
                            {\Delta_{\min}^{(r)}},
 &\Gamma_3&\geq\tfrac12\Delta_{\min}^{(r)},
 \label{eq:qld_quadratic_spectral_rates}\\
 1-\chi_2
 &=\frac{2\Gamma_2}{\Gamma_3+\Gamma_2}
 \leq4C\left(\frac{\Omega_{\mathrm{tun}}^{\mathrm{op}}}
                    {\Delta_{\min}^{(r)}}\right)^2
 \longrightarrow0.
 \label{eq:qld_quadratic_spectral_window}
\end{align}
For the prescribed threshold \(\chi_c\), choose
\begin{align}
 \varepsilon
 &=\min\!\left\{\varepsilon_0,
   \sqrt{\frac{1-\chi_c}{2C(1+\chi_c)}}\right\}>0.
 \label{eq:qld_weak_tunneling_epsilon_choice}
\end{align}
Then \(0<\Omega_{\mathrm{tun}}^{\mathrm{op}}/
\Delta_{\min}^{(r)}<\varepsilon\) gives
\begin{align}
 \frac{\Gamma_2}{\Gamma_3}
 &\leq2C\left(\frac{\Omega_{\mathrm{tun}}^{\mathrm{op}}}
                   {\Delta_{\min}^{(r)}}\right)^2
 <\frac{1-\chi_c}{1+\chi_c}
 \quad\Longrightarrow\quad\chi_{\max}\geq\chi_2>\chi_c.
 \label{eq:qld_weak_tunneling_rate_ratio_threshold}
\end{align}
The dissipator norm \(\gamma_{\mathcal D}^{(r)}\) tends to the
positive reference norm \(\|\mathcal D[L_{0,r}]\|_{2\to2}\).
Positivity follows from \(\Delta_{\min}^{(r)}>0\): a purely
Hamiltonian reference generator could not damp its nonstationary
directions. Thus \cref{eq:qld_quadratic_spectral_rates} also gives
\begin{align}
 \frac{\Gamma_2}{\gamma_{\mathcal D}^{(r)}}
 &=O\!\left[
   \left(\frac{\Omega_{\mathrm{tun}}^{\mathrm{op}}}
               {\gamma_{\mathcal D}^{(r)}}\right)^2\right]
 \qquad(\Omega_{\mathrm{tun}}^{\mathrm{op}}\to0^+).
 \label{eq:qld_quadratic_tunneling_dissipation_ratio}
\end{align}
The constants \(\varepsilon,C\) depend on the fixed local data
and threshold \(\chi_c\), but not on the magnitude or direction
of \(V_{H,r}\). For a nonnormal generator \(\mathcal L_{0,r}\),
the inverse norm \(\|F_{0,r}^{-1}\|_{2\to2}\) in
\cref{eq:qld_fast_resolvent_bound} is not controlled by
\(\Delta_{\min}^{(r)}\) alone. Consequently, the threshold
\(\varepsilon\) is not universal across different local
generators \(\mathcal L_{0,r}\). The fixed-data assumptions exclude
a collapsing local gap \(\Delta_{\min}^{(r)}\) and an independent
dissipative transfer channel that remains finite as
\(\Omega_{\mathrm{tun}}^{\mathrm{op}}\to0\).
No diagonalization or auxiliary interpolation parameter was needed.
\end{proof}

\subsection{Proof of \texorpdfstring{\cref{thm:qld_strong_dissipation_exclusion}}{Theorem 3.4}}
\label{app:proof_qld_strong_dissipation}

\begin{proof}
Apply
\cref{thm:general_strong_dissipation_exclusion} to the retained
space \(\operatorname{Ran}P_r\): set \(d_H=r\),
\(H=H_r\), \(\widehat{\mathcal D}=\widehat{\mathcal D}_r\),
and \(\gamma_{\mathcal D}=\gamma_{\mathcal D}^{(r)}\).
The norm identity \cref{eq:r_state_hamiltonian_bandwidth_norm}
identifies the relative perturbation with
\(\omega_{\mathrm{bw}}^{(r)}/\gamma_{\mathcal D}^{(r)}\).
The general spectral limit
\cref{eq:general_strong_dissipation_spectral_limit} therefore gives
\begin{align}
 \chi_{\max}(\mathcal L_r^{\mathrm{TP}})
 &\longrightarrow\chi_{\max}(\widehat{\mathcal D}_r)
 \qquad\left(
 \frac{\gamma_{\mathcal D}^{(r)}}{\omega_{\mathrm{bw}}^{(r)}}
 \to\infty\right).
 \label{eq:projected_dissipative_limit}
\end{align}
The eigenvalue-continuity and strict-threshold argument is proved in
\cref{app:proof_general_strong_dissipation}. If
\(\omega_{\mathrm{bw}}^{(r)}=0\), then \(\mathcal K_r=0\)
and \(\chi_{\max}(\mathcal L_r^{\mathrm{TP}})
=\chi_{\max}(\widehat{\mathcal D}_r)\) for every
\(\gamma_{\mathcal D}^{(r)}>0\).
\end{proof}

\subsection{Proof of \texorpdfstring{\cref{thm:weak_sector_certificate}}{Theorem 3.5}}
\label{app:proof_weak_sector_certificate}

\begin{proof}
The stationary eigenvalue \(0\) of a finite-dimensional Lindbladian
\(\mathcal L_0\) is semisimple; this standard property also underlies
the stationary-space perturbation theory in
\cite{macieszczak2016towards,zanardi2014coherent}. Spectral continuity,
with algebraic multiplicity, therefore gives exactly \(M\) eigenvalues
near zero for sufficiently small \(\|\mathcal E\|_{2\to2}\).
The remaining eigenvalues have real parts at most
\(-\Delta_{\min}/2\). Since \(\mathcal L\) is relaxing, ordering
the decay rates \(\Gamma_i\) as in \cref{eq:decay_rates_Gamma} gives
\begin{align}
 \Gamma_M(\mathcal L)&\longrightarrow0,
 \qquad \Gamma_{M+1}(\mathcal L)\geq\frac{\Delta_{\min}}2
 \quad\text{as }\|\mathcal E\|_{2\to2}\to0^+.
 \label{eq:general_weak_transfer_rate_bounds}
\end{align}
Consequently, \(\Gamma_M/\Gamma_{M+1}\to0\). The threshold conversion
\begin{align}
 \chi_M(\mathcal L)>\chi_c
 &\quad\Longleftrightarrow\quad
 \frac{\Gamma_M(\mathcal L)}{\Gamma_{M+1}(\mathcal L)}
 <\frac{1-\chi_c}{1+\chi_c}
 \label{eq:general_weak_transfer_ratio_threshold}
\end{align}
and \(\chi_{\max}\geq\chi_M\) prove
\cref{eq:general_weak_transfer_epsilon_window} for some
\(\varepsilon>0\). The physical condition in
\cref{eq:general_physical_transfer_threshold} follows from
\cref{eq:general_physical_coupling_bound,eq:general_intersector_tunneling_frequency}.
No diagonalizability assumption is required.
\end{proof}

\subsection{Proof of \texorpdfstring{\cref{thm:general_strong_dissipation_exclusion}}{Theorem 3.6}}
\label{app:proof_general_strong_dissipation}

\begin{proof}
Fix a decay-rate tolerance \(0<\epsilon_d<d_2/2\).
Dividing \cref{eq:general_strong_dissipation_family} by
\(\gamma_{\mathcal D}>0\) gives
\begin{align}
 \left\|\frac{\mathcal L_{\gamma_{\mathcal D}}}{\gamma_{\mathcal D}}
                  -\widehat{\mathcal D}\right\|_{2\to2}
 &=\frac{\omega_{\mathrm{bw}}}{\gamma_{\mathcal D}}.
 \label{eq:general_strong_dissipation_norm_limit}
\end{align}
By finite-dimensional spectral continuity, counting algebraic
multiplicities, there exists \(\varepsilon_0>0\), depending only
on \(\widehat{\mathcal D},\epsilon_d\), such that
\(0<\omega_{\mathrm{bw}}/\gamma_{\mathcal D}<\varepsilon_0\)
allows the eigenvalues of
\(\mathcal L_{\gamma_{\mathcal D}}/\gamma_{\mathcal D}\) and
\(\widehat{\mathcal D}\) to be paired with errors smaller than
\(\epsilon_d\). Every eigenvalue paired with a nonzero eigenvalue
of \(\widehat{\mathcal D}\) has real part less than
\(-d_2+\epsilon_d<-d_2/2\). Trace preservation guarantees an
eigenvalue \(0\), so it must be the single eigenvalue paired with
the simple stationary eigenvalue \(0\) of \(\widehat{\mathcal D}\).
Sorting the negative real parts cannot increase their maximum
pairing error. Thus the ordered decay-rate errors \(e_i\) satisfy
\begin{align}
 e_i\coloneqq
 \frac{\Gamma_i(\mathcal L_{\gamma_{\mathcal D}})}{\gamma_{\mathcal D}}-d_i,
 \qquad |e_i|<\epsilon_d,
 \qquad d_i+e_i>d_2/2>0,
 \qquad 2\leq i\leq d_H^2.
 \label{eq:general_strong_dissipation_rate_limit}
\end{align}

The common scale \(\gamma_{\mathcal D}\) cancels in each
separation \(\chi_i\). Substituting the errors \(e_i\) and
combining the two fractions gives, for \(2\leq i<d_H^2\),
\begin{align}
 &\begin{aligned}[b]
 \left|\chi_i(\mathcal L_{\gamma_{\mathcal D}})
       -\frac{d_{i+1}-d_i}{d_{i+1}+d_i}\right|
 &=\frac{2|d_i e_{i+1}-d_{i+1}e_i|}
 {(d_i+d_{i+1})(d_i+d_{i+1}+e_i+e_{i+1})}\\
 &<\frac{2\epsilon_d}{d_i+d_{i+1}-2\epsilon_d}
 \leq\frac{\epsilon_d}{d_2-\epsilon_d}.
 \end{aligned}
 \label{eq:general_strong_dissipation_cut_error}
\end{align}
The denominators are positive because \(\epsilon_d<d_2/2\).
Using \(|\max_i a_i-\max_i b_i|\leq\max_i|a_i-b_i|\) gives
\begin{align}
 \left|\chi_{\max}(\mathcal L_{\gamma_{\mathcal D}})
       -\chi_{\max}(\widehat{\mathcal D})\right|
 &<\frac{\epsilon_d}{d_2-\epsilon_d}.
 \label{eq:general_strong_dissipation_max_error}
\end{align}
Now choose
\begin{align}
 \epsilon_d
 &=\frac{d_2[\chi_c-\chi_{\max}(\widehat{\mathcal D})]}
 {2+\chi_c-\chi_{\max}(\widehat{\mathcal D})},
 \qquad \varepsilon=\varepsilon_0.
 \label{eq:general_strong_dissipation_tolerance_choice}
\end{align}
The strict margin
\(0<\chi_c-\chi_{\max}(\widehat{\mathcal D})<1\)
ensures \(0<\epsilon_d<d_2/2\). Hence every
\(0<\omega_{\mathrm{bw}}/\gamma_{\mathcal D}<\varepsilon\)
satisfies
\begin{align}
 \chi_{\max}(\mathcal L_{\gamma_{\mathcal D}})
 &<\chi_{\max}(\widehat{\mathcal D})
   +\frac{\chi_c-\chi_{\max}(\widehat{\mathcal D})}{2}
 <\chi_c,
 \label{eq:general_strong_dissipation_strict_margin}
\end{align}
which proves \cref{eq:general_strong_dissipation_epsilon_exit}.
The threshold \(\varepsilon\) is independent of
\(\gamma_{\mathcal D}\). Since \(\epsilon_d\) in
\cref{eq:general_strong_dissipation_max_error} can be arbitrarily
small, the same estimate also establishes
\begin{align}
 \chi_{\max}(\mathcal L_{\gamma_{\mathcal D}})
 &\longrightarrow\chi_{\max}(\widehat{\mathcal D})
 \qquad(\gamma_{\mathcal D}\to\infty).
 \label{eq:general_strong_dissipation_spectral_limit}
\end{align}
This argument uses continuity of the eigenvalue multiset, not
diagonalizability or a linear eigenvalue-error bound in the
perturbation norm. It therefore also permits repeated nonstationary
eigenvalues and Jordan blocks of \(\widehat{\mathcal D}\).
\end{proof}

\section{Proofs for dissipative control}
\label{app:proof_dissipative_control}

This appendix gives the proofs of the theorems in
\cref{sec:metastability_control}. The first two subsections concern
the two-state model, and the final two treat QLD with general potentials.

\subsection{Proof of \texorpdfstring{\cref{thm:two_state_dissipative_pulse}}{Theorem 4.1}}
\label{app:proof_two_state_dissipative_pulse}

\begin{proof}
For balanced coefficients \(a_x^{\mathrm c}=b^{\mathrm c}\),
\cref{eq:symmetric_transverse_block,eq:symmetric_stationary_bloch} give
\begin{align}
 A_\perp^{\mathrm c}
 &=-2SI_2+\omega_{01}\begin{pmatrix}0&1\\-1&0\end{pmatrix},
 \label{eq:control_balanced_transverse_block}\\
 w(t)&=e^{-2St}R(-\omega_{01}t)w_0,
 &u_z(t)&=1+(u_{z,0}-1)e^{-4St},
 \label{eq:control_balanced_bloch_solution}\\
 R(-\theta)&=\begin{pmatrix}\cos\theta&\sin\theta\\
                              -\sin\theta&\cos\theta\end{pmatrix}.
 \label{eq:control_planar_rotation}
\end{align}
With \(\ell_s=(\ell_x,\ell_y)^{\mathsf T}\), the slow amplitude
after the rotation \(R(-\theta)\) is proportional to
\begin{align}
 \ell_s^{\mathsf T}R(-\theta)w_0
 &=(\ell_s^{\mathsf T}w_0)\cos\theta
   +(\ell_xu_{y,0}-\ell_yu_{x,0})\sin\theta.
 \label{eq:control_pulse_angle_equation}
\end{align}
Its values at \(\theta=0\) and \(\theta=\pi\) are nonzero with
opposite signs. Solving this sinusoidal equation gives the angle
\(\theta_p=\omega_{01}T_p\) in
\cref{eq:control_explicit_pulse_time}; the nonzero initial overlap
excludes both endpoints \(0,\pi\).
The vector \(w_p=e^{-2ST_p}R(-\theta_p)w_0\) is then in
\(\ker\ell_s^{\mathsf T}\). Since \(A_\perp\) has distinct
eigenvalues \(-\Gamma_s,-\Gamma_f\), this one-dimensional kernel
is exactly its fast right eigenspace. After restoring \(\mathcal L_b\),
\begin{align}
 w(T_p+t)&=e^{-\Gamma_f t}w_p,
 &u_z(T_p+t)-u_{z,\mathrm{ss}}
   &=e^{-4St}(u_{z,p}-u_{z,\mathrm{ss}}).
 \label{eq:control_postpulse_bloch_solution}
\end{align}
For qubit density operators \(\rho,\sigma\) with Bloch vectors
\(\boldsymbol u,\boldsymbol v\), the eigenvalues of
\(\rho-\sigma\) are \(\pm\|\boldsymbol u-\boldsymbol v\|_2/2\).
Thus their trace distance is half that vector distance, and
\begin{align}
 D_{\mathrm{tr}}(e^{t\mathcal L_b}\rho_p,\rho_{\mathrm{ss}})
 &=\frac12\sqrt{e^{-2\Gamma_ft}\|w_p\|_2^2
             +e^{-8St}(u_{z,p}-u_{z,\mathrm{ss}})^2}
 \leq D_pe^{-\Gamma_ft}.
 \label{eq:control_postpulse_distance_calculation}
\end{align}
Here \(\Gamma_f<4S\) follows from \(\omega_{01}>0\), which
implies \(\sqrt{\Delta_B}<2S\). This proves
\cref{eq:control_two_state_fast_relaxation,eq:control_two_state_time_upper}.

Under the baseline generator \(\mathcal L_b\), the left eigenvector
\(\ell_s\) gives
\begin{align}
 \ell_s^{\mathsf T}w(t) =e^{-\Gamma_st}\ell_s^{\mathsf T}w_0,
 \qquad
 D_{\mathrm{tr}}(e^{t\mathcal L_b}\rho_0,\rho_{\mathrm{ss}})
 \geq\frac12|\ell_s^{\mathsf T}w_0|e^{-\Gamma_st}.
 \label{eq:control_baseline_distance_lower}
\end{align}
Solving the necessary inequality in
\cref{eq:control_baseline_distance_lower} for distance at most
\(\varepsilon\) proves \cref{eq:control_two_state_time_lower}.
As \(\varepsilon\rightarrow 0^{+}\), the difference between the lower
baseline bound and upper control bound grows as
\((\Gamma_s^{-1}-\Gamma_f^{-1})\log(1/\varepsilon)\), up to
constants independent of \(\varepsilon\). Here \(D_p>0\): the
nonzero vector \(w_0\) remains nonzero under the invertible damped
rotation \(e^{-2ST_p}R(-\theta_p)\). This coefficient is positive,
so the finite preparation cost \(T_p\) is eventually outweighed.
\end{proof}

\paragraph{Timing, admissibility, and target accuracy.}
The duration \(T_p\) is a phase-matching time, not an adiabatic
timescale: making the switch slower does not by itself improve
cancellation. With otherwise exact coefficients and a duration error
\(\delta T\), where \(T_p+\delta T\geq0\),
\cref{eq:control_pulse_angle_equation} yields
\begin{align}
 \bigl|\ell_s^{\mathsf T}w(T_p+\delta T)\bigr|
 &=e^{-2S(T_p+\delta T)}\|w_0\|_2
       |\sin(\omega_{01}\delta T)|.
 \label{eq:control_pulse_timing_residual}
\end{align}
Indeed, the squared sum of the two sinusoidal coefficients in
\cref{eq:control_pulse_angle_equation} equals \(\|w_0\|_2^2\)
because \(\|\ell_s\|_2=1\). Thus finite timing error generally
restores a small slow tail. For finite ramps and imperfect coefficients,
\cref{eq:control_robust_time_comparison} gives an integrated
error criterion that preserves a speedup at a prescribed
\(\varepsilon\); it does not assert exact cancellation at fixed
nonzero error. The accessible thermal parameters also matter:
\cref{eq:qld_thermal_mu,eq:qld_thermal_nu} imply
\(\mu/\nu=m\Omega\coth[\hbar\Omega/(4k_BT)]\), whereas balancing
requires \(\mu_{\mathrm c}/\nu_{\mathrm c}=m\omega_{01}\).
At fixed \(\Omega\), changing only \(\eta\) cannot reach this ratio;
if \(\omega_{01}<\Omega\), no positive temperature \(T\) can reach
it either. The protocol is therefore conditional on reservoir access,
not a universal consequence of the QLD form.

If the desired target is instead the ground state
\(\ket{\phi_0}\bra{\phi_0}\), one can keep the balanced
coefficients \(\mu_{\mathrm c},\nu_{\mathrm c}\) rather than
restore the inaccurate baseline. The exact solution
\cref{eq:control_balanced_bloch_solution} gives
\begin{align}
 D_{\mathrm{tr}}(e^{t\mathcal L_c}\rho_0,
                   \ket{\phi_0}\bra{\phi_0})
 &\leq e^{-2St}D_{\mathrm{tr}}(\rho_0,\ket{\phi_0}\bra{\phi_0}).
 \label{eq:control_balanced_ground_target}
\end{align}
This is a ground-state preparation guarantee within the symmetric
two-state model under the same coefficient-access assumption. For
a tolerance below the baseline stationary bias in
\cref{eq:control_two_state_accuracy_tradeoff}, the baseline cannot
solve this task; that comparison is one of attainable accuracy,
not a finite speedup factor between two successful protocols.

\subsection{Proof of \texorpdfstring{\cref{thm:two_state_strength_control}}{Theorem 4.2}}
\label{app:proof_two_state_strength_control}

\begin{proof}
Let \(v(t)=(u_x(t),u_y(t),u_z(t)-u_{z,\mathrm{ss}})^{\mathsf T}\).
The symmetric Bloch equations in
\cref{eq:symmetric_transverse_block,eq:symmetric_stationary_bloch}
and \(a_x(t)^2=S(t)(1+\delta_{\mathcal D})/2\),
\(b(t)^2=S(t)(1-\delta_{\mathcal D})/2\) give
\begin{align}
 \dot v(t)&=\begin{pmatrix}
 -2S(t)(1-\delta_{\mathcal D})&\omega_{01}&0\\
 -\omega_{01}&-2S(t)(1+\delta_{\mathcal D})&0\\
 0&0&-4S(t)
 \end{pmatrix}v(t),
 \label{eq:two_state_control_error_dynamics}\\
 \frac{\mathrm d}{\mathrm dt}\|v(t)\|_2^2
 &=-4S(t)\bigl[(1-\delta_{\mathcal D})v_x^2
             +(1+\delta_{\mathcal D})v_y^2+2v_z^2\bigr]
 \nonumber\\
 &\leq-4S(t)(1-|\delta_{\mathcal D}|)\|v(t)\|_2^2
 =-2\alpha_2(t)\|v(t)\|_2^2.
 \label{eq:two_state_control_error_dissipation}
\end{align}
The terms containing \(\omega_{01}\) cancel in the quadratic
form. Integrating and using
\(D_{\mathrm{tr}}(\rho(t),\rho_{\mathrm{ss}})=\|v(t)\|_2/2\)
proves \cref{eq:two_state_control_integrated_contraction}.
For \(t=T_r+u\), the integrated damping is
\(\int_0^{T_r}\alpha_2(v)\,\mathrm dv+\alpha_{2,\mathrm h}u\).
Solving the decreasing bound for distance at most \(\varepsilon\)
gives \cref{eq:two_state_strength_time_upper}.

The baseline lower bound remains
\cref{eq:control_two_state_time_lower}. Since \(D_0>0\) and
the ramp has finite duration, the difference between that lower
bound and \cref{eq:two_state_strength_time_upper} is
\begin{align}
 \left(\frac1{\Gamma_s}-\frac1{\alpha_{2,\mathrm h}}\right)
       \log\frac1\varepsilon+O(1),
 \qquad\varepsilon \to 0^+.
 \label{eq:two_state_strength_speedup_limit}
\end{align}
It diverges to \(+\infty\) if
\(\alpha_{2,\mathrm h}>\Gamma_s\). The proof uses the values
\(S(t)\), not their derivatives, so it imposes no adiabatic
switching condition.
\end{proof}

\paragraph{The direction of strength tuning.}
Common-strength tuning requires a different diagnosis. At fixed
\(0<|\delta_{\mathcal D}|<1\), differentiating the exact slow
rate \(\Gamma_s=2S-\sqrt{4\delta_{\mathcal D}^2S^2-\omega_{01}^2}\)
in the overdamped region gives
\begin{align}
 \frac{\mathrm d\Gamma_s}{\mathrm dS}
 &=2-\frac{4\delta_{\mathcal D}^2S}
                  {\sqrt{4\delta_{\mathcal D}^2S^2-\omega_{01}^2}},
 \qquad
 \frac{\mathrm d\Gamma_s}{\mathrm dS}>0
 \ \Longleftrightarrow\ 
 2S|\delta_{\mathcal D}|\sqrt{1-\delta_{\mathcal D}^2}>
 \omega_{01}.
 \label{eq:control_two_state_gap_direction}
\end{align}
Both sides of the square-root comparison are positive, so squaring
is legitimate. Immediately above the overdamped boundary, increasing
\(S\) decreases \(\Gamma_s\); only beyond the threshold in
\cref{eq:control_two_state_gap_direction} does it increase
\(\Gamma_s\). At \(|\delta_{\mathcal D}|=1\), the derivative
is always negative. This supplies a local tuning direction, but
not a finite-time speedup certificate, because modal weights also
matter. The integrated contraction estimate in \cref{thm:two_state_strength_control} supplies the finite-time certificate.

\subsection{Proof of \texorpdfstring{\cref{thm:qld_dissipative_pulse}}{Theorem 4.3}}
\label{app:proof_qld_dissipative_pulse}

\begin{proof}
Trace duality and \(\|F_s\|_{\mathrm{op}}=1\) give
\begin{align}
 \operatorname{Tr}(F_s e^{t\mathcal L_c}\rho_0)
 &\leq m_c+
    |\operatorname{Tr}[F_s(e^{t\mathcal L_c}\rho_0-\sigma_c)]|
 \nonumber\\
 &\leq m_c+2C_c e^{-f_ct}.
 \label{eq:general_control_crossing_bracket}
\end{align}
At \(t=0\), this also implies
\(2C_c\geq\operatorname{Tr}(F_s\rho_0)-m_c>|m_c|\).
Thus \(f_c^{-1}\log(2C_c/|m_c|)>0\), and the right-hand side
of \cref{eq:general_control_crossing_bracket} is zero at that
time. Continuity and the positive initial expectation guarantee
the zero in \cref{eq:qld_pulse_cancellation_time}.

Let \(\mathbf R_s\) be the right eigenmatrix of
\(\mathcal L_b\) normalized by
\(\operatorname{Tr}(F_s\mathbf R_s)=1\). Simplicity of
\(-\Gamma_s\) makes its spectral projector
\(\mathcal P_s(B)=\mathbf R_s\operatorname{Tr}(F_sB)\).
With \(\mathcal P_0(B)=\rho_{\mathrm{ss}}^{\mathrm b}\operatorname{Tr}B\)
and \(\mathcal Q_f=\mathcal I-\mathcal P_0-\mathcal P_s\),
\begin{align}
 \mathcal P_s(\rho_p-\rho_{\mathrm{ss}}^{\mathrm b})
 &=\mathbf R_s\operatorname{Tr}(F_s\rho_p)=0,
 \label{eq:qld_pulse_slow_projection_zero}\\
 e^{t\mathcal L_b}\rho_p-\rho_{\mathrm{ss}}^{\mathrm b}
 &=e^{t\mathcal L_b}\mathcal Q_f
                  (\rho_p-\rho_{\mathrm{ss}}^{\mathrm b}).
 \label{eq:qld_pulse_fast_projection}
\end{align}
On the finite-dimensional fast subspace
\(\operatorname{Ran}\mathcal Q_f\), each Jordan contribution
has the form \(t^k e^{\lambda t}\). Since
\(-\operatorname{Re}\lambda>f\), the factor
\(t^k e^{-(-\operatorname{Re}\lambda-f)t}\) is bounded for
\(t\geq0\). Summing the finitely many contributions gives
\cref{eq:qld_pulse_fast_relaxation} and its convergence-time bound.
For the baseline, the adjoint equation instead gives
\begin{align}
 D_{\mathrm{tr}}(e^{t\mathcal L_b}\rho_0,
                    \rho_{\mathrm{ss}}^{\mathrm b})
 &\geq\tfrac12|\operatorname{Tr}(F_s e^{t\mathcal L_b}\rho_0)|
 =w_s e^{-\Gamma_st}.
 \label{eq:qld_pulse_baseline_lower}
\end{align}
This proves the second bound in \cref{eq:qld_pulse_time_comparison}.
For \(C_p>0\), the difference between the lower and upper time
bounds is \((\Gamma_s^{-1}-f^{-1})\log(1/\varepsilon)+O(1)\),
which tends to \(+\infty\). If \(C_p=0\), the pulse already
reaches \(\rho_{\mathrm{ss}}^{\mathrm b}\) at \(T_p\).
\end{proof}

\paragraph{QLD channel selection and reachability.}
The weak-transfer mechanism in \cref{thm:qld_sector_origin}
suggests first inspecting the intersector matrix elements of
the jump operator \(L^{(r)}\).
For a density operator \(\rho_L=P_L^{(r)}\rho_LP_L^{(r)}\),
the initial population-transfer rate into the right sector is
\begin{align}
 \left.\frac{\mathrm d}{\mathrm dt}
 \operatorname{Tr}(P_R^{(r)}\rho_r(t))\right|_{\rho_r(0)=\rho_L}
 &=\operatorname{Tr}[P_R^{(r)}\mathcal D_r(\rho_L)]
 =2\nu^2\|P_R^{(r)}p_rP_L^{(r)}\rho_L^{1/2}\|_{\mathrm{HS}}^2.
 \label{eq:control_qld_initial_sector_transfer}
\end{align}
The Hamiltonian commutator and both anticommutator terms have
zero trace against \(P_R^{(r)}\), because
\(P_R^{(r)}\rho_L=\rho_LP_R^{(r)}=0\). The gain term is
twice the squared Hilbert--Schmidt norm of
\(P_R^{(r)}L^{(r)}P_L^{(r)}\rho_L^{1/2}\), and
\(P_R^{(r)}x_rP_L^{(r)}=0\) removes its position contribution.
Thus increasing \(\nu\) at fixed \(\mu\) can strengthen an
initial escape channel without changing \(H_r\). This is a
candidate tuning direction, not a monotonicity theorem for
\(T_\varepsilon\): subsequent coherences, reverse transfer,
local relaxation, and the stationary state also change. If the
displayed momentum matrix element vanishes, this channel cannot
be opened by rescaling \(\nu\).

For several jumps \(L_j^{(r)}\), the gain is instead
\(2\sum_j\|P_R^{(r)}L_j^{(r)}P_L^{(r)}\rho_L^{1/2}\|_{\mathrm{HS}}^2\).
The simplification to a term proportional to \(\nu^2\) requires
\(P_R^{(r)}x_{j,r}P_L^{(r)}=0\) for the jump being considered.
Projected coordinates \(x_{j,r}\) need not commute, so this
condition must be checked in spatially multidimensional models.

Energy selectivity gives a second constraint on ratio tuning.
When the commutator identity \cref{eq:projected_momentum_identity}
holds, \(p_{nk}=\im m(E_n-E_k)x_{nk}/\hbar\). For indices
\(0\leq n<k<r\) with \(E_n<E_k\), the projected jump elements are
\begin{align}
 \langle\phi_n|L^{(r)}|\phi_k\rangle
 &=\left(\mu+\frac{\nu m(E_k-E_n)}{\hbar}\right)x_{nk},
 \nonumber\\
 \langle\phi_k|L^{(r)}|\phi_n\rangle
 &=\left(\mu-\frac{\nu m(E_k-E_n)}{\hbar}\right)x_{kn}.
 \label{eq:control_qld_energy_selectivity}
\end{align}
Starting from a pure energy eigenstate \(\ket{\phi_k}\), the
initial population gain in a different state \(\ket{\phi_n}\)
is twice the squared modulus of the corresponding jump element.
Consequently, \(\mu/\nu=m(E_k-E_n)/\hbar\) suppresses the
upward transition for that pair while retaining the downward one
when \(\nu>0\) and \(x_{nk}\ne0\). This recovers the balanced
two-state choice in \cref{eq:control_balanced_jump_coefficients}.
For an anharmonic potential \(V(x)\), different gaps \(E_k-E_n\)
generally require different ratios; a single pair \(\mu,\nu\)
cannot balance them all. The relevant transitions must be identified
from the occupied slow component, and the eventual target bias
must be checked. These instantaneous transfer formulas do not
replace coherent evolution by a population master equation.

More directly, the adjoint equation gives the control response
\begin{align}
 \frac{\mathrm d}{\mathrm dt}\operatorname{Tr}(F_s\rho_r(t))
 &=-\Gamma_s\operatorname{Tr}(F_s\rho_r(t))
   +\operatorname{Tr}\{F_s[\mathcal D_r(t)-\mathcal D_{r,\mathrm b}]
                              \rho_r(t)\}.
 \label{eq:control_qld_slow_response}
\end{align}
The second term is the effect of changing \(\mu,\nu\) on the
baseline slow expectation \(\operatorname{Tr}(F_s\rho_r(t))\).
Evaluating it for candidate coefficients connects the channel
analysis in \cref{eq:control_qld_initial_sector_transfer,eq:control_qld_energy_selectivity}
to the stopping observable; propagation verifies whether the
reduction persists. Conversely, if every allowed
generator satisfies \(\mathcal L_t^\dagger F_s=-g(t)F_s\), with
real \(g(t)\) and finite \(\int_0^{T_p}g(t)\,\mathrm dt\), then
\(\operatorname{Tr}(F_s\rho(T_p))
=e^{-\int_0^{T_p}g(t)\,\mathrm dt}\operatorname{Tr}(F_s\rho_0)\)
cannot vanish. This explains why access to a different dissipative
structure may be necessary. The sign condition is sufficient, not
necessary: the two-state pulse crosses zero by a transient rotation.

\paragraph{Residual overlap and finite switching errors.}
The following estimates justify \cref{rem:control_robustness} and
extend the exact-cancellation argument to nonzero residual overlap.
Write \(H=H_r\) for the fixed retained Hamiltonian and fix a relaxing baseline
generator \(\mathcal L=\mathcal K+\mathcal D_{\mathrm b}\), with
\(\mathcal K=-\im[H,\cdot]/\hbar\). Assume a simple real slow
eigenvalue \(\lambda_2=-\Gamma_s\), \(\Gamma_s>0\), is separated
from all other nonstationary decay rates. Choose its Hermitian left
eigenmatrix \(F_s\) and define the initial overlap \(w_s\) by
\begin{align}
 \mathcal L^\dagger(F_s)&=-\Gamma_sF_s,
 &\|F_s\|_{\mathrm{op}}&=1,
 &\operatorname{Tr}(F_s\rho_{\mathrm{ss}})&=0,
 \label{eq:general_control_slow_witness}\\
 w_s&=\tfrac12|\operatorname{Tr}(F_s\rho_0)|>0.
 \label{eq:general_control_initial_overlap}
\end{align}
The stationary expectation vanishes because \(\Gamma_s>0\).
Let \(\mathcal P_s\) be the spectral projector onto this simple
eigenvalue and \(\mathcal P_0(B)=\rho_{\mathrm{ss}}\operatorname{Tr}B\).
The complementary projector \(\mathcal Q_f=\mathcal I-\mathcal P_0-
\mathcal P_s\) contains the fast modes. These projectors need not
be orthogonal or positive maps.

An admissible stage changes the QLD coefficients \(\mu(t),\nu(t)\)
for \(0\leq t<T_p\). Its dissipator \(\mathcal D_t\) is
piecewise continuous and returns to
\(\mathcal D_{\mathrm b}\) at \(T_p\), without changing \(H\).
Its propagator \(\mathcal U(T_p,0)\) prepares
\(\rho_p=\mathcal U(T_p,0)\rho_0\). Define
\(A_p=\|\mathcal P_s(\rho_p-\rho_{\mathrm{ss}})\|_1/2\),
and verify, for \(C_p\geq0\) and \(f>\Gamma_s\), the bound
\begin{align}
 \tfrac12\|e^{t\mathcal L}\mathcal Q_f
                     (\rho_p-\rho_{\mathrm{ss}})\|_1
 &\leq C_pe^{-ft},\qquad t\geq0.
 \label{eq:control_endpoint_fast_bound}
\end{align}
For example, any \(f\) strictly below all fast decay rates permits
a finite prefactor \(C_p\), including in the presence of fast
Jordan blocks. The spectral decomposition then gives
\begin{align}
 D_{\mathrm{tr}}(e^{t\mathcal L}\rho_p,\rho_{\mathrm{ss}})
 &\leq A_pe^{-\Gamma_st}+C_pe^{-ft},\qquad t\geq0.
 \label{eq:general_control_prepared_error}
\end{align}
The pulse is useful if it makes \(A_p\) sufficiently small at an
acceptable cost \(T_p\); decreasing \(\chi_{\max}\) during the
pulse alone does not establish either fact. For a tolerance
\(\varepsilon>0\), define the comparison times
\begin{align}
 U_\varepsilon
 &=T_p+\max\left\{
 \frac{1}{\Gamma_s}\log\max\{1,2A_p/\varepsilon\},
 \frac{1}{f}\log\max\{1,2C_p/\varepsilon\}\right\},
 \label{eq:general_control_upper_certificate}\\
 L_\varepsilon
 &=\frac{1}{\Gamma_s}\log\max\{1,w_s/\varepsilon\}.
 \label{eq:general_control_lower_certificate}
\end{align}

Under \cref{eq:general_control_prepared_error}, one has
\(T_\varepsilon^{\mathrm{ctrl}}\leq U_\varepsilon\) and
\(T_\varepsilon^{\mathrm{base}}\geq L_\varepsilon\).
A strict speedup is certified by
\begin{align}
 2A_pe^{\Gamma_s T_p}&<w_s,
 &T_p+\frac1f\log\max\{1,2C_p/\varepsilon\}&<L_\varepsilon.
 \label{eq:general_control_time_comparison}
\end{align}
These conditions are equivalent to \(U_\varepsilon<L_\varepsilon\):
the slow-overlap reduction must outweigh the time cost \(T_p\),
and the fast remainder must relax before the baseline can finish.
If \(A_p=0\), the controlled upper bound improves to
\(T_p+f^{-1}\log\max\{1,C_p/\varepsilon\}\), and strict
acceleration holds for all sufficiently small \(\varepsilon>0\).
\begin{proof}[Verification of the time comparison]
The adjoint equation and trace duality give
\begin{align}
 \operatorname{Tr}[F_s(e^{t\mathcal L}\rho_0-\rho_{\mathrm{ss}})]
 &=e^{-\Gamma_st}\operatorname{Tr}(F_s\rho_0),
 \label{eq:general_control_witness_evolution}\\
 D_{\mathrm{tr}}(e^{t\mathcal L}\rho_0,\rho_{\mathrm{ss}})
 &\geq\tfrac12|\operatorname{Tr}[F_s(e^{t\mathcal L}\rho_0
                                      -\rho_{\mathrm{ss}})]|
 =w_se^{-\Gamma_st}.
 \label{eq:general_control_baseline_lower}
\end{align}
Distance at most \(\varepsilon\) therefore requires
\(t\geq L_\varepsilon\). Conversely, after the pulse, taking
\(t\) no smaller than both logarithmic terms in
\cref{eq:general_control_upper_certificate} makes each term in
\cref{eq:general_control_prepared_error} at most
\(\varepsilon/2\). Both terms remain below that value thereafter.
Including \(T_p\) proves the two time bounds. The second
condition in \cref{eq:general_control_time_comparison} implies
\(L_\varepsilon>T_p\geq0\), hence \(w_s>\varepsilon\) and
\(e^{\Gamma_s L_\varepsilon}=w_s/\varepsilon\).
The remaining condition for \(U_\varepsilon<L_\varepsilon\)
is then equivalent to
\(2A_p/\varepsilon<e^{\Gamma_s(L_\varepsilon-T_p)}\),
which gives the first condition in
\cref{eq:general_control_time_comparison}. This proves the
strict comparison, including \(A_p=0\).

If \(A_p=0<C_p\), only \(C_pe^{-ft}\) remains. For small
enough \(\varepsilon\),
\begin{align}
 L_\varepsilon-\left(T_p+\frac{1}{f}\log\frac{C_p}{\varepsilon}\right)
 &=\left(\frac{1}{\Gamma_s}-\frac{1}{f}\right)\log\frac{1}{\varepsilon}
   +\frac{\log w_s}{\Gamma_s}-\frac{\log C_p}{f}-T_p
 \longrightarrow+\infty.
 \label{eq:general_control_strict_speedup_limit}
\end{align}
The coefficient is positive since \(f>\Gamma_s\). If
\(A_p=C_p=0\), then \(\rho_p=\rho_{\mathrm{ss}}\), so the
controlled time is at most \(T_p\) while \(L_\varepsilon\to\infty\).
\end{proof}

\Cref{thm:two_state_dissipative_pulse,thm:qld_dissipative_pulse}
supply QLD constructions with \(A_p=0\); the time comparison
\cref{eq:general_control_time_comparison} also allows \(A_p>0\).
It makes the stopping rule quantitative: additional tuning is useful
only while the reduction of the residual component compensates for
the elapsed time. One should minimize the total bound
\(U_\varepsilon\), rather than the residual amplitude \(A_p\)
alone. When conservative modal prefactors make that bound
inconclusive, a converged trajectory calculation can assess the
specified finite model without claiming the analytic certificate.

The remaining question is how quickly the dissipator must be
switched. To compare an ideal schedule \(\mathcal D_t\) with an
implemented schedule \(\widetilde{\mathcal D}_t\), choose a common
end time \(T_f\geq T_p\), after which both use exactly
\(\mathcal D_{\mathrm b}\). The ideal schedule already uses
\(\mathcal D_{\mathrm b}\) for \(t\geq T_p\); the implemented
schedule may use this extra time for its return ramp. Define
\begin{align}
 \epsilon_{\mathrm{sw}}
 &=\frac12\int_0^{T_f}
    \|\widetilde{\mathcal D}_t-\mathcal D_t\|_{1\to1,\mathrm H}
       \,\mathrm dt,
 \label{eq:control_switching_error_budget}\\
 \|\mathcal A\|_{1\to1,\mathrm H}
 &:=\sup_{B=B^\dagger\ne0}\frac{\|\mathcal A(B)\|_1}{\|B\|_1}.
 \label{eq:control_hermitian_induced_norm}
\end{align}
The subscript \(\mathrm H\) denotes restriction to Hermitian
operators, not the Hamiltonian \(H\). The dimensionless error
\(\epsilon_{\mathrm{sw}}\) measures the integrated discrepancy
between two physical controls; it includes finite rise and fall
times, timing error, and coefficient calibration error.

Consider CPTP schedules with the same \(H,\rho_0\).
Combining \cref{eq:general_control_prepared_error} with the endpoint
estimate below gives \cref{eq:control_robust_postpulse_distance}, with
\(\mathcal L_b=\mathcal L\) and
\(\rho_{\mathrm{ss}}^{\mathrm b}=\rho_{\mathrm{ss}}\).
For \(\varepsilon>\epsilon_{\mathrm{sw}}\), a strict speedup is
certified whenever
\begin{align}
 T_f+\max\left\{
 \frac{1}{\Gamma_s}\log\max\left\{1,\frac{2A_p}{\varepsilon-\epsilon_{\mathrm{sw}}}\right\},
 \frac{1}{f}\log\max\left\{1,\frac{2C_p}{\varepsilon-\epsilon_{\mathrm{sw}}}\right\}
 \right\}<L_\varepsilon.
 \label{eq:control_robust_time_comparison}
\end{align}
If \(A_p=0\), replace the maximum by
\(f^{-1}\log\max\{1,C_p/(\varepsilon-\epsilon_{\mathrm{sw}})\}\).
To verify this robustness bound, let
\(\widetilde{\mathcal U}(t,u)\) be the implemented propagator.
Variation of constants, using the same Hamiltonian \(H\), gives
\begin{align}
 \widetilde\rho(T_f)-\rho(T_f)
 &=\int_0^{T_f}\widetilde{\mathcal U}(T_f,u)
      (\widetilde{\mathcal D}_u-\mathcal D_u)\rho(u)\,\mathrm du,
 \label{eq:control_switching_duhamel}\\
 D_{\mathrm{tr}}(\widetilde\rho(T_f),\rho(T_f))
 &\leq\frac12\int_0^{T_f}
      \|(\widetilde{\mathcal D}_u-\mathcal D_u)\rho(u)\|_1
        \,\mathrm du
 \leq\epsilon_{\mathrm{sw}}.
 \label{eq:control_switching_endpoint_bound}
\end{align}
The integrand before propagation is Hermitian; trace-norm
contractivity and \(\|\rho(u)\|_1=1\) justify the inequalities.
After \(T_f\), both schedules use \(\mathcal L\), so this
distance cannot increase. The ideal state at \(T_f+t\) has evolved
for \(T_f-T_p+t\) under \(\mathcal L\). Therefore
\begin{align}
 D_{\mathrm{tr}}(\widetilde\rho(T_f+t),\rho_{\mathrm{ss}})
 &\leq\epsilon_{\mathrm{sw}}
   +A_pe^{-\Gamma_s(T_f-T_p+t)}+C_pe^{-f(T_f-T_p+t)}
 \nonumber\\
 &\leq\epsilon_{\mathrm{sw}}+A_pe^{-\Gamma_st}+C_pe^{-ft}.
 \label{eq:control_robust_distance_derivation}
\end{align}
Allocating the remaining tolerance \(\varepsilon-\epsilon_{\mathrm{sw}}\)
to the decaying terms proves the time bound, while
\cref{eq:general_control_baseline_lower} supplies the unchanged
baseline lower bound \(L_\varepsilon\).

The bound gives a concrete switching-speed requirement. If the
schedules differ only on intervals of total duration
\(\tau_{\mathrm{sw}}\), and their generator discrepancy is at
most \(\Delta_{\mathrm{sw}}\) in the norm of
\cref{eq:control_hermitian_induced_norm}, then
\(\epsilon_{\mathrm{sw}}\leq\Delta_{\mathrm{sw}}\tau_{\mathrm{sw}}/2\).
For an ideally cancelled slow amplitude \(A_p=0\), the sufficient
conditions
\begin{align}
 \Delta_{\mathrm{sw}}\tau_{\mathrm{sw}}&\leq\varepsilon,
 &T_f+\frac{1}{f}\log\max\{1,2C_p/\varepsilon\}
 &<L_\varepsilon
 \label{eq:control_switching_speed_rule}
\end{align}
retain a strict speedup. Thus fast ramps are one way to preserve an
ideal pulse, but speed alone is not the criterion: a longer calibrated
schedule can also work if its actual residual amplitude \(A_p\)
passes \cref{eq:general_control_time_comparison}. For a projected QLD
jump \(L^{(r)}\) and its implemented value \(\widetilde L^{(r)}\),
the error can be bounded directly from the coefficient errors:
\begin{align}
 \|\widetilde{\mathcal D}_r-\mathcal D_r\|_{1\to1,\mathrm H}
 &\leq4(\|\widetilde L^{(r)}\|_{\mathrm{op}}+
           \|L^{(r)}\|_{\mathrm{op}})
           \|\widetilde L^{(r)}-L^{(r)}\|_{\mathrm{op}},
 \label{eq:control_jump_error_to_generator}\\
 \|\widetilde L^{(r)}-L^{(r)}\|_{\mathrm{op}}
 &\leq|\widetilde\mu-\mu|\,\|x_r\|_{\mathrm{op}}
      +|\widetilde\nu-\nu|\,\|p_r\|_{\mathrm{op}}.
 \label{eq:control_qld_coefficient_error}
\end{align}
To obtain the first inequality, expand the gain-term difference as
\(2(\widetilde L-L)B\widetilde L^\dagger+
2LB(\widetilde L^\dagger-L^\dagger)\), and bound each product by
\(\|XBY\|_1\leq\|X\|_{\mathrm{op}}\|B\|_1\|Y\|_{\mathrm{op}}\).
The anticommutator difference contributes the same bound because
\(\|\widetilde L^\dagger\widetilde L-L^\dagger L\|_{\mathrm{op}}
\leq(\|\widetilde L\|_{\mathrm{op}}+\|L\|_{\mathrm{op}})
\|\widetilde L-L\|_{\mathrm{op}}\).
Here the unadorned jumps in this calculation denote the projected
jumps in \cref{eq:control_jump_error_to_generator}.

The additive error \(\epsilon_{\mathrm{sw}}\) is a conservative
certificate, not a physical asymptotic error floor. The restored
generator \(\mathcal L\) still relaxes to \(\rho_{\mathrm{ss}}\),
but a nonzero residual slow component generally returns the late-time
rate \(\Gamma_s\). An asymptotic fast-rate claim therefore requires
exact cancellation, whereas a finite-accuracy speedup only requires
\cref{eq:control_robust_time_comparison}. A laboratory schedule must
also remain within the regime where its time-local Lindblad
description is valid; the mathematical allowance of jumps in the
coefficients is not a microscopic bandwidth guarantee.

For a different target \(\rho_{\mathrm{tar}}\), set
\(b_{\mathrm{tar}}=D_{\mathrm{tr}}(\rho_{\mathrm{ss}},
\rho_{\mathrm{tar}})\). The triangle inequalities give
\begin{align}
 D_{\mathrm{tr}}(e^{t\mathcal L}\rho_0,\rho_{\mathrm{tar}})
 &\geq w_se^{-\Gamma_st}-b_{\mathrm{tar}},
 \label{eq:general_control_biased_target_lower}\\
 D_{\mathrm{tr}}(e^{t\mathcal L}\rho_p,\rho_{\mathrm{tar}})
 &\leq b_{\mathrm{tar}}+A_pe^{-\Gamma_st}+C_pe^{-ft}.
 \label{eq:general_control_biased_target_upper}
\end{align}
Use \(\varepsilon+b_{\mathrm{tar}}\) in the baseline lower bound
and \(\varepsilon-b_{\mathrm{tar}}-\epsilon_{\mathrm{sw}}>0\)
in the implemented upper bound. This accounts for the accuracy
restriction found explicitly in the two-state model. For multiple
or nonreal slow eigenvalues, a pulse must suppress the entire
relevant slow spectral sector; a single expectation
\(\operatorname{Tr}(F_s\rho_p)\) no longer suffices.

To handle such a cluster, retain a conjugation-closed
slow projector \(\mathcal P_s\), excluding the stationary projector
\(\mathcal P_0\). Multiple-mode suppression is also relevant to
temporary-reset control \cite{bao2025temporary}; for a QLD pulse,
its achievement must be checked with the available coefficients
\(\mu,\nu\). The endpoint \(\rho_p\) then has the bound
\begin{align}
 D_{\mathrm{tr}}(e^{t\mathcal L}\rho_p,\rho_{\mathrm{ss}})
 &\leq \frac12\|e^{t\mathcal L}\mathcal P_s
                    (\rho_p-\rho_{\mathrm{ss}})\|_1
       +C_pe^{-ft}.
 \label{eq:control_multimode_residual_test}
\end{align}
Here the fast estimate uses \(\mathcal Q_f=\mathcal I-\mathcal P_0-
\mathcal P_s\); it must be verified for the chosen endpoint
\(\rho_p\). This follows by splitting the error into its two
invariant components and applying the triangle inequality.
The single-exponential test \cref{eq:general_control_time_comparison}
must not be applied to an unresolved slow cluster. In particular,
a smaller Hilbert--Schmidt norm of its initial projection is a
useful diagnostic but is not itself the trace-distance bound in
\cref{eq:control_multimode_residual_test}.

\subsection{Proof of \texorpdfstring{\cref{thm:qld_strength_control}}{Theorem 4.4}}
\label{app:proof_qld_strength_control}

\begin{proof}
For the traceless difference \(B(t)\) of two trajectories with
the same time-dependent generator \(\mathcal L_t\), the
Hamiltonian contribution has zero real quadratic form. Hence
\begin{align}
 \frac{\mathrm d}{\mathrm dt}\|B(t)\|_{\mathrm{HS}}^2
 &=2\operatorname{Re}\langle B(t),\mathcal K_r B(t)
       +\gamma_{\mathcal D}^{(r)}(t)\widehat{\mathcal D}_r B(t)\rangle
 \nonumber\\
 &\leq-2a_D\gamma_{\mathcal D}^{(r)}(t)\|B(t)\|_{\mathrm{HS}}^2,
 \label{eq:qld_control_hs_differential_bound}\\
 \|B(t)\|_{\mathrm{HS}}
 &\leq\exp\!\left[-a_D\int_u^t
       \gamma_{\mathcal D}^{(r)}(v)\,\mathrm dv\right]
       \|B(u)\|_{\mathrm{HS}},\qquad t\geq u\geq0.
 \label{eq:qld_control_hs_integrated_bound}
\end{align}
The second line follows by an integrating factor. It does not
differentiate \(\gamma_{\mathcal D}^{(r)}\) or an instantaneous
stationary state. Existence of \(\rho_{\mathrm{ss}}^{\mathrm h}\)
follows from finite-dimensional Lindblad dynamics; applying the
constant-stage estimate to two stationary states proves uniqueness.
Since \(\widehat{\mathcal D}_r(\sigma_D)=0\), their stationary
equations give
\begin{align}
 \gamma_{\mathrm h}\widehat{\mathcal D}_r
       (\rho_{\mathrm{ss}}^{\mathrm h}-\sigma_D)
 &=-\mathcal K_r\rho_{\mathrm{ss}}^{\mathrm h},
 \label{eq:qld_control_stationary_difference_equation}\\
 \|\rho_{\mathrm{ss}}^{\mathrm h}-\sigma_D\|_{\mathrm{HS}}
 &\leq\frac{R_D\|\mathcal K_r\|_{2\to2}}{\gamma_{\mathrm h}}
       \|\rho_{\mathrm{ss}}^{\mathrm h}\|_{\mathrm{HS}}
 \leq\frac{R_D\omega_{\mathrm{bw}}^{(r)}}{\gamma_{\mathrm h}}.
 \label{eq:qld_control_stationary_difference_bound}
\end{align}
The inverse acts on traceless operators, the norm of
\(\mathcal K_r\) is given by
\cref{eq:r_state_hamiltonian_bandwidth_norm}, and positivity gives
\(\|\rho_{\mathrm{ss}}^{\mathrm h}\|_{\mathrm{HS}}\leq1\).
The Schatten-norm inequality \(\|B\|_1\leq\sqrt r\|B\|_{\mathrm{HS}}\)
and the triangle inequality give
\begin{align}
 D_{\mathrm{tr}}(\rho_{\mathrm{ss}}^{\mathrm h},\rho_{\mathrm{tar}})
 &\leq D_{\mathrm{tr}}(\sigma_D,\rho_{\mathrm{tar}})
       +\frac{\sqrt r}{2}
          \|\rho_{\mathrm{ss}}^{\mathrm h}-\sigma_D\|_{\mathrm{HS}}
 \leq b_r.
 \label{eq:qld_control_stationary_accuracy}
\end{align}

The ramp produces a density operator \(\rho(T_r)\), irrespective
of its duration or profile. During the constant stage,
\cref{eq:qld_control_hs_integrated_bound} and
\(\|\rho(T_r)-\rho_{\mathrm{ss}}^{\mathrm h}\|_{\mathrm{HS}}
\leq\sqrt2\) imply
\begin{align}
 D_{\mathrm{tr}}(\rho(T_r+t),\rho_{\mathrm{ss}}^{\mathrm h})
 &\leq\frac{\sqrt r}{2}e^{-a_D\gamma_{\mathrm h}t}
       \|\rho(T_r)-\rho_{\mathrm{ss}}^{\mathrm h}\|_{\mathrm{HS}}
 \leq\sqrt{r/2}e^{-a_D\gamma_{\mathrm h}t}.
 \label{eq:qld_control_stationary_relaxation_bound}
\end{align}
Adding \(b_r\) proves \cref{eq:qld_control_convergence_bound}.
Its right-hand side decreases throughout the constant stage. For
\(b_r<\varepsilon\), solving for a value at most \(\varepsilon\)
gives
\begin{align}
 T_\varepsilon^{\mathrm{ctrl}}
 &\leq T_r+\frac{1}{a_D\gamma_{\mathrm h}}
   \log\max\left\{1,\frac{\sqrt{r/2}}{\varepsilon-b_r}\right\}.
 \label{eq:qld_control_settling_bound}
\end{align}

To compare times, stationarity and \(\Gamma_s>0\) give
\(\operatorname{Tr}(F_s\rho_{\mathrm{ss}}^{\mathrm b})=0\).
The adjoint equation and trace duality yield
\begin{align}
 \operatorname{Tr}\!\left[F_s
   (e^{t\mathcal L_b}\rho_0-\rho_{\mathrm{ss}}^{\mathrm b})\right]
 &=e^{-\Gamma_st}\operatorname{Tr}(F_s\rho_0),
 \label{eq:qld_speedup_baseline_witness}\\
 D_{\mathrm{tr}}(e^{t\mathcal L_b}\rho_0,\rho_{\mathrm{tar}})
 &\geq D_{\mathrm{tr}}(e^{t\mathcal L_b}\rho_0,
                         \rho_{\mathrm{ss}}^{\mathrm b})-b_{\mathrm b}
 \nonumber\\
 &\geq w_s e^{-\Gamma_st}-b_{\mathrm b}.
 \label{eq:qld_speedup_baseline_error}
\end{align}
Thus the necessary baseline time and sufficient controlled time obey
\begin{align}
 T_\varepsilon^{\mathrm{base}}
 &\geq\frac{1}{\Gamma_s}
     \log\max\left\{1,\frac{w_s}{\varepsilon+b_{\mathrm b}}\right\},
 \label{eq:qld_speedup_baseline_lower}\\
 T_\varepsilon^{\mathrm{ctrl}}
 &\leq T_r+\frac{1}{a_D\gamma_{\mathrm h}}
     \log\max\left\{1,\frac{\sqrt{r/2}}{\varepsilon-b_r}\right\}.
 \label{eq:qld_speedup_control_upper}
\end{align}
The second inequality is \cref{eq:qld_control_settling_bound},
so \cref{eq:qld_speedup_schedule_condition} proves a strict speedup.
For the restore protocol, denote the state at restoration by
\(\rho_p\). Contractivity and
\(\mathcal L_b(\rho_{\mathrm{ss}}^{\mathrm b})=0\) give
\begin{align}
 D_{\mathrm{tr}}(e^{t\mathcal L_b}\rho_p,
                  \rho_{\mathrm{ss}}^{\mathrm b})
 &\leq D_{\mathrm{tr}}(\rho_p,\rho_{\mathrm{ss}}^{\mathrm b})
 \leq\varepsilon,\qquad t\geq0.
 \label{eq:qld_control_restore_contractivity}
\end{align}
Consequently, restoration cannot spoil the attained tolerance
\(\varepsilon\) for this stationary target.
\end{proof}

\paragraph{Restoration with finite switching time.}
For a noninstantaneous
return to \(\mathcal L_b\), reserve an endpoint error allowance
\(\epsilon_{\mathrm{sw}}<\varepsilon\), reach
\(\varepsilon-\epsilon_{\mathrm{sw}}\) before returning, and
include the return duration in the time comparison. The
propagator estimate \cref{eq:control_switching_endpoint_bound}
bounds this error; \cref{eq:qld_control_restore_contractivity}
then controls all subsequent times.
If the target \(\rho_{\mathrm{tar}}\) is not stationary for
\(\mathcal L_b\), the restore claim does not apply: one keeps
\(\mathcal L_{\mathrm h}\) or verifies the final target bias
separately.

\addtocontents{toc}{\protect\endgroup}
\end{document}